\documentclass[a4paper,11pt]{article}
\pdfoutput=1 

\usepackage{jheppub} 

\usepackage[T1]{fontenc} 
\usepackage{bm}
\usepackage{slashed}

\title{\boldmath Gauge invariant canonical symplectic algorithms for real-time lattice strong-field quantum electrodynamics}

\author[a]{Qiang~Chen}
\author[b]{Jianyuan~Xiao}
\author[c,d]{and Peifeng~Fan}

\affiliation[a]{National Supercomputing Center in Zhengzhou, Zhengzhou University, Zhengzhou, Henan 450001, China}
\affiliation[b]{School of Nuclear Science and Technology, University of Science and Technology of China, Hefei, Anhui 230026, China}
\affiliation[c]{Key Laboratory of Optoelectronic Devices and Systems, College of Physics and Optoelectronic Engineering, Shenzhen University, Shenzhen, Guangdong 518060, China}
\affiliation[d]{Advanced Energy Research Center, Shenzhen University, Shenzhen, Guangdong 518060, China}

\emailAdd{cq0405@ustc.edu.cn}
\emailAdd{xiaojy@ustc.edu.cn}
\emailAdd{pffan@mail.ustc.edu.cn}

\abstract{A class of high-order canonical symplectic structure-preserving geometric algorithms are developed for high-quality 
simulations of the quantized Dirac-Maxwell theory based strong-field quantum electrodynamics (SFQED) and relativistic 
quantum plasmas (RQP) phenomena. With minimal coupling, the Lagrangian density of an interacting bispinor-gauge fields 
theory is constructed in a conjugate real fields form. The canonical symplectic form and canonical equations of this field 
theory are obtained by the general Hamilton's principle on cotangent bundle. Based on discrete exterior calculus, the gauge 
field components are discreted to form a cochain complex, and the bispinor components are naturally discreted on a staggered 
dual lattice as combinations of differential forms. With pull-back and push-forward gauge covariant derivatives, the discrete 
action is gauge invariant. A well-defined discrete canonical Poisson bracket generates a semi-discrete lattice canonical 
field theory (LCFT), which admits the canonical symplectic form, unitary property, gauge symmetry and discrete Poincar\'e 
subgroup, which are good approximations of the original continuous geometric structures. The Hamiltonian splitting method, 
Cayley transformation and symmetric composition technique are introduced to construct a class of high-order numerical schemes 
for the semi-discrete LCFT. These schemes involve two degenerate fermion flavors and are locally unconditional stable, which also 
preserve the geometric structures. Admitting Nielsen-Ninomiya theorem, the continuous chiral symmetry is partially broken on the 
lattice. As an extension, a pair of discrete chiral operators are introduced to reconstruct the lattice chirality. Equipped 
with statistically quantization-equivalent ensemble models of the Dirac vacuum and non-trivial plasma backgrounds, the schemes 
are expected to have excellent performance in secular simulations of relativistic quantum effects, where the numerical errors 
of conserved quantities are well bounded by very small values without coherent accumulation. The algorithms are verified in 
detail by numerical energy spectra. Real-time LCFT simulations are successfully implemented for the nonlinear Schwinger mechanism 
induced $e$-$e^+$ pairs creation and vacuum Kerr effect, where the nonlinear and non-perturbative features captured by the 
solutions provide a complete strong-field physical picture in a very wide range, which open a new door toward high-quality 
simulations in SFQED and RQP fields.}

\begin{document} 
\maketitle
\flushbottom

\section{Introduction}
\label{sec:1}

Quantum electrodynamics (QED) at extreme conditions is becoming more and more important, as the relativistic quantum 
effects are becoming dominant mechanism in many branches of modern physics. With the development of high power laser 
technology, e.g. chirped pulse amplification (CPA), the peak intensity above $10^{22}$ W$\cdot$cm$^{-2}$ is available by 
1$\sim$10 PW lasers, which is far stronger than the direct ionization threshold of $10^{16}\sim10^{18}$ W$\cdot$cm$^{-2}$ 
\cite{Mourou,Krausz}. When the matter is exposed in such intense laser beams, strong ionization can be generated and large 
relativistic quantum plasmas (RQP) will be produced \cite{EPLiang,Gahn,Nerush,Eliasson,Ridgers,Sarri}. Next generation 
10$\sim$PW laser projects, such as the extreme light infrastructure (ELI) and the high power laser energy research 
facility (HiPER), aimed to approach the Schwinger threshold of $10^{29}$ W$\cdot$cm$^{-2}$ or realize the fast ignition 
\cite{Mourou,Krausz,Nerush}. The Schwinger mechanism induced creation and following annihilation of fermion pairs play 
a fundamentally important role in modern high energy density physics (HEDP), astrophysics, and strong-field quantum 
electrodynamics (SFQED) \cite{Heisenberg,Schwinger1,EPLiang,Gahn,Nerush,Ridgers,Sarri}. Although the direct experimental 
verification of electron-positron ($e$-$e^{+}$) pair creation under the Schwinger limit in laboratory is still expected 
to realize in near future, the $e$-$e^{+}$ RQP is already an important target for astronomical observers, such as the magnetosphere 
of an X-ray pulsar \cite{Santangelo,Pottschmidt,Tsygankov}. The typical magnetic field of X-ray pulsars is $10^{12}$ G, 
and the effective temperature of X-ray pulsars is $10$ KeV. In such a environment, the magnetic energy approaches to the 
rest energies of electron and positron, and it is higher than the thermal energy. As a result, the relativistic quantum 
effects lead to anharmonic cyclotron absorption features observed in spectra of X-ray pulsars \cite{Santangelo,Pottschmidt,
Tsygankov,YShi1}. Effective and accurate non-perturbative methods are needed in understanding these SFQED and RQP phenomena. 
Among a group of semi-analytical and numerical methods, the lattice quantum field theory (LQFT) is an advanced theoretical 
tool to study relativistic quantum effects both in vacuum and plasmas.

As a quantum gauge field theory on the discrete lattice in Euclidean space-time, the LQFT first developed by Wilson has 
been widely used in quantum chromodynamics (QCD) to describe the strong interactions, such as the quark confinement 
and the quark-gluon plasmas (QGP) \cite{Wilson,Creutz,Satz,Yagi,Rothe}. Based on numerical path integrals and large-scale 
Monte Carlo (MC) simulations, the lattice quantum chromodynamics (LQCD) brings many significant results, such as the QCD 
phase transition and the hadron spectroscopy \cite{Yagi,Rothe}. By using the Schwinger-Keldysh time contours, the LQCD can 
even be expanded to simulate non-equilibrium statistical systems \cite{Schwinger2,Keldysh}. The LQCD can not only capture 
the basic quantum loop effects, but also provide us with a well-defined non-perturbative theory of QCD \cite{Rothe}. As 
a post-MC technique, the tensor network (TN) techniques provide an alternative approach to simulate the lattice gauge 
theories (LGT), which can be efficiently extended to real-time evolution of out-of-equilibrium systems \cite{Tagliacozzo,
Pichler,Buyens0,Carmen,Tilloy}. When it comes to phenomena with high occupation numbers and weak coupling, e.g. SFQED with 
non-trivial backgrounds and RQP, the classical relativistic field equations can be treated as good approximations to 
describe the dynamics of particles, where the quantum fluctuations can be introduced by constructing a statistically 
quantization-equivalent ensemble \cite{Aarts,Polkovnikov,Borsanyi,Hebenstreit1,Gelis,Hebenstreit2,Kasper1,Buyens1,Tanji1,
Hebenstreit3,MMuller,Buyens2,Tanji2,YShi2,Spitz}. Based on this real-time lattice quantum electrodynamics (LQED) method 
in classical statistic regime, some interesting phenomena have been numerically studied, such as the pair creation of 
fermions beyond the Schwinger limit \cite{Kasper1}, the real-time dynamics of string breaking \cite{Hebenstreit2}, the 
chiral magnetic effects \cite{MMuller}, and the $e$-$e^{+}$ pair production in laser-plasma interactions (LPI) \cite{YShi2}. 
The LGT simulations are even reconstructed to implement on optical lattice based quantum simulators in recent time, which 
show great vitality \cite{Kasper2,Zache}.

When implementing a real-time LQED simulation in classical statistic regime, a stable and high fidelity numerical algorithm 
is needed to obtain reliable and accurate results. When it comes to the $U(1)$ gauge field, there are many popular schemes 
for Maxwell's equations, such as the finite-difference time-domain (FDTD) method, the finite-element (FE) method, and the 
method of moments (MoM) etc., which are widely used in computational electrodynamics (CED) \cite{Yee,Harrington,Taflove,
GSun,QChen1}. When it comes to the bispinor field, the numerical calculations of Dirac equation may encounter more 
difficult, e.g. fermion doubling problem, which will bring pseudo-fermion modes on the lattice \cite{Nielsen}. There are 
several stable Dirac solvers, such as the time-splitting spectral (TSSM) method, quantum lattice Boltzmann (QLBM) technique, 
summation-by-parts-simultaneous approximation term (SBP-SAT) method, and time-dependent Galerkin (TDG) method etc 
\cite{Weizhu1,Zhongyi,Jialin,Gourdeau1,JianXu,Hammer1,Almquist,Hammer2,Gourdeau2,Beerwerth,Gourdeau3,Weizhu2}. Although 
these algorithms have different advantages in part, an unified scheme with almost perfect performance is still a beautiful 
goal. Because of the nonlinearity and the multi-scale nature of the Dirac-Maxwell equations, high-quality simulations of 
real-time LQED face challenges. For instance, the numerical errors of conserved quantities can coherently accumulate, though 
these errors may be very small in each numerical step. The breakdown of conservation laws over a long simulation time amounts to 
pseudophysics. The structure-preserving geometric algorithms first developed by Feng \emph{et. al.} for classical Hamiltonian 
systems have excellent performance in long-term simulations \cite{Feng1,Feng2,Benettin,Reich,Marsden,Lew,West,Hairer1,YWu,
Hairer2,SChin}, which are widely used in many complex systems, especially in geophysics and plasma physics \cite{Qin1,Qin2,
Qin3,Qin4,Qin5,Shadwick,Qin6,Qin7,Qin8,Morrison1,QChen2,Morrison2,QChen3,QChen4}. In this work, we construct a class of 
structure-preserving geometric algorithms for Dirac-Maxwell theory. The algorithms preserve the symplectic and unitary 
structures, which also admit the $U(1)$ gauge symmetry. The continuous Poincar\'e symmetry is reduced to a discrete subgroup, 
and there are only two degenerate fermion flavors exist on the lattice. The algorithms provide a powerful numerical tool for 
Dirac-Maxwell theory based real-time LQED simulations.

In Sec.\ref{sec:2}, a canonical field theory of Dirac-Maxwell systems is constructed in a conjugate real fields form 
to describe the fermion-photon interactions. The canonical symplectic form on cotangent bundle is obtained explicitly. 
In Sec.\ref{sec:3}, a semi-discrete lattice canonical field theory (LCFT) is constructed via the discrete exterior calculus 
(DEC) \cite{Hirani,Hiptmair,Arnold}, the pull-back and push-forward gauge covariant derivatives, and a discrete canonical 
Poisson bracket. The LCFT admits symplectic and unitary structures, and also admits $U(1)$ gauge symmetry on the lattice. 
In Sec.\ref{sec:4}, a class of high-order structure-preserving geometric algorithms are constructed for the LCFT. These schemes 
are locally unconditional stable, which also preserve the geometric structures and symmetries. The fermion doubling and chirality 
problems are discussed in detail. An ensemble model based field quantization procedure is reconstructed to simulate the Dirac 
vacuum and RQP. In Sec.\ref{sec:5}, the algorithms are detailedly verified and successfully used to simulate Schwinger 
mechanism induced $e$-$e^+$ pairs creation and vacuum Kerr effect. Numerical results show good properties in secular 
simulations. In Sec.\ref{sec:6}, we give a brief discussion about the advantages and attentions in implementing a real-time 
LCFT simulation by our algorithms, and show the outlook of applications in SFQED and RQP researches.

\section{Canonical field theory of the Dirac-Maxwell systems}
\label{sec:2}

\subsection{Lagrangian field theory}
\label{sec:2-1}
The Lagrangian density of the Dirac-Maxwell theory based QED can be written as \cite{Weinberg,Peskin,Zeidler,Dirac1,Dirac2},
\begin{eqnarray}
\mathcal{L}=\mathcal{L}_{\rm{D}}+\mathcal{L}_{\rm{M}},\label{eq:1}
\end{eqnarray}
\begin{eqnarray}
\mathcal{L}_{\rm{D}}=\bar{\psi}\left(i\hbar{c}\slashed{D}-mc^2\right)\psi,\label{eq:2}
\end{eqnarray}
\begin{eqnarray}
\mathcal{L}_{\rm{M}}=-\frac{1}{16\pi}\mathcal{F}^{\mu\nu}\mathcal{F}_{\mu\nu}.\label{eq:3}
\end{eqnarray}
Where the Dirac bispinor $\psi$ is a 4 components complex field. With the Minkowski metric 
$g^{\mu\nu}=g_{\mu\nu}=\rm{diag}(+,-,-,-)$, the contravariant 4-vectors of coordinate and $U(1)$ gauge field can be 
given by $x^{\mu}=(ct,\bm{x})$, $A^{\mu}=(\phi,\bm{A})$, and the Maxwell gauge field strength tensor 
$\mathcal{F}^{\mu\nu}=\partial^{\mu}A^{\nu}-\partial^{\nu}A^{\mu}$. The Dirac conjugate bispinor $\bar\psi=\psi^{+}\gamma^{0}$, 
where the superscript $+$ means Hermitian and the $4\times4$ matrices $\gamma^{\mu}$ belong to a Clifford algebra 
$\{\gamma^{\mu},\gamma^{\nu}\}=2g^{\mu\nu}$. With minimal coupling, the gauge covariant derivative 
$D_{\mu}=\partial_{\mu}+i\frac{e}{\hbar{c}}A_{\mu}$, and the Feynman dagger $\slashed{D}=\gamma^{\mu}D_{\mu}$ means Dirac 
contraction, where the charge $e$, reduced Planck constant $\hbar$ and light speed $c$ have their usual meanings. 
In the Dirac representation, the $\gamma^{\mu}$ matrices are given by \cite{Dirac1,Dirac2},
\begin{eqnarray}
\gamma^{0}=\left(\begin{array}{cc}
I_{2} & 0_{2}\\
0_{2} & -I_{2}
\end{array}\right),~~\gamma^{i}=\left(\begin{array}{cc}
0_{2} & \hat\sigma_{i}\\
-\hat\sigma_{i} & 0_{2}
\end{array}\right).\label{eq:4}
\end{eqnarray}
Where the Pauli matrices $\hat\sigma_{i}$ have their usual forms,
\begin{eqnarray}
\hat\sigma_{x}=\left(\begin{array}{cc}
0 & 1\\
1 & 0
\end{array}\right),~~\hat\sigma_{y}=\left(\begin{array}{cc}
0 & -i\\
i & 0
\end{array}\right),~~\hat\sigma_{z}=\left(\begin{array}{cc}
1 & 0\\
0 & -1
\end{array}\right).\label{eq:5}
\end{eqnarray}

The Dirac bispinor and it's Hermitian can be rewritten in a conjugate real fields form as,
\begin{eqnarray}
\psi=\frac{1}{\sqrt{2\hbar}}\left(\psi_{R}+i\psi_{I}\right)=\frac{1}{\sqrt{2\hbar}}\left(\begin{array}{c}
\psi_{1R}+i\psi_{1I}\\
\psi_{2R}+i\psi_{2I}\\
\psi_{3R}+i\psi_{3I}\\
\psi_{4R}+i\psi_{4I}
\end{array}\right),\label{eq:6}
\end{eqnarray}
\begin{eqnarray}
\psi^{+}=\frac{1}{\sqrt{2\hbar}}\left(\psi_{R}^{T}-i\psi_{I}^{T}\right)=\frac{1}{\sqrt{2\hbar}}\left(\begin{array}{c}
\psi_{1R}-i\psi_{1I}\\
\psi_{2R}-i\psi_{2I}\\
\psi_{3R}-i\psi_{3I}\\
\psi_{4R}-i\psi_{4I}
\end{array}\right)^{T}.\label{eq:7}
\end{eqnarray}
By introducing the Dirac matrices \cite{Dirac1,Dirac2},
\begin{eqnarray}
\alpha_{i}=\gamma^{0}\gamma^{i}=\left(\begin{array}{cc}
0_{2} & \hat\sigma_{i}\\
\hat\sigma_{i} & 0_{2}
\end{array}\right),~~\beta=\gamma^{0}I_{4}=\gamma^{0},\label{eq:8}
\end{eqnarray}
we can rewrite the Lagrangian density \eqref{eq:1} as,
\begin{eqnarray}
\mathcal{L}&=&\frac{1}{2\hbar}\left(\psi_{R}^{T}-i\psi_{I}^{T}\right)\left(i\hbar\frac{\partial}{\partial{t}}+i\hbar{c}\bm{\alpha}\cdot\bigtriangledown-e\phi+e\bm{\alpha}\cdot\bm{A}-\beta{mc^2}\right)\left(\psi_{R}+i\psi_{I}\right)+\frac{1}{8\pi}\left(\bm{E}^2-\bm{B}^2\right)\nonumber\\
&=&\frac{1}{2}\left(\psi_{I}^{T}\frac{\partial}{\partial{t}}\psi_{R}-\psi_{R}^{T}\frac{\partial}{\partial{t}}\psi_{I}\right)+\frac{i}{2}\left(\psi_{R}^{T}\frac{\partial}{\partial{t}}\psi_{R}+\psi_{I}^{T}\frac{\partial}{\partial{t}}\psi_{I}\right)-\frac{1}{2\hbar}\left(\psi_{R}^{T}-i\psi_{I}^{T}\right)\hat{H}\left(\psi_{R}+i\psi_{I}\right)\nonumber\\
& &+\frac{1}{8\pi}\left(\bm{E}^2-\bm{B}^2\right).\label{eq:9}
\end{eqnarray}
Where $\bm{E}=-\dot{\bm{A}}/c-\bigtriangledown{\phi}$ and $\bm{B}=\bigtriangledown\times\bm{A}$ are electric and magnetic 
strengths of the $U(1)$ gauge field, and the superscript $\cdot$ means derivative with respect to time. 
$\hat{H}=\hat{H}_{R}+i\hat{H}_{I}=-i\hbar{c}\bm{\alpha}\cdot\bigtriangledown-e\bm{\alpha}\cdot\bm{A}+e\phi+\beta{mc^2}$ 
is the Hamiltonian operator of the Dirac equation, which can be given by,
\begin{eqnarray}
\hat{H}_{R}=\left(\begin{array}{cccc}
e\phi+mc^2 & 0 & -eA_{z} & -eA_{x}-\hbar{c}\partial_{y}\\
0 & e\phi+mc^2 & -eA_{x}+\hbar{c}\partial_{y} & eA_{z}\\
-eA_{z} & -eA_{x}-\hbar{c}\partial_{y} & e\phi-mc^2 & 0\\
-eA_{x}+\hbar{c}\partial_{y} & eA_{z} & 0 & e\phi-mc^2
\end{array}\right),\label{eq:10}
\end{eqnarray}
\begin{eqnarray}
\hat{H}_{I}=\left(\begin{array}{cccc}
0 & 0 & -\hbar{c}\partial_{z} & eA_{y}-\hbar{c}\partial_{x}\\
0 & 0 & -eA_{y}-\hbar{c}\partial_{x} & \hbar{c}\partial_{z}\\
-\hbar{c}\partial_{z} & eA_{y}-\hbar{c}\partial_{x} & 0 & 0\\
-eA_{y}-\hbar{c}\partial_{x} & \hbar{c}\partial_{z} & 0 & 0
\end{array}\right).\label{eq:11}
\end{eqnarray}
By substituting Eqs.~\eqref{eq:10}-\eqref{eq:11} into Eq.~\eqref{eq:9} and integrating it in full Minkowski space-time 
manifold, we obtain the action functional $S=\int_{T}L{\rm{d}}t=\int_{T}\int_{V}\mathcal{L}{\rm{d}}^{4}x$. Where the 
Lagrangian functional can be given by,
\begin{eqnarray}
L&=&\int_{V}\mathcal{L}{\rm{d}}^{3}x\nonumber\\
&=&\int_{V}\left[\frac{1}{2}\left(\psi_{I}^{T}\dot{\psi_{R}}-\psi_{R}^{T}\dot{\psi_{I}}\right)-\frac{1}{2\hbar}\left(\psi_{R}^{T}\hat{H}_{R}\psi_{R}+\psi_{I}^{T}\hat{H}_{R}\psi_{I}-\psi_{R}^{T}\hat{H}_{I}\psi_{I}+\psi_{I}^{T}\hat{H}_{I}\psi_{R}\right)\right.\nonumber\\
& &\left.+\frac{1}{8\pi}\left(\bm{E}^2-\bm{B}^2\right)\right]{\rm{d}}^{3}x.\label{eq:12}
\end{eqnarray}
On the tangent bundle $TG$ of the configuration manifold $G=(\psi_{R},\psi_{I},\bm{A},\phi)$, the Hamilton's principle 
$\delta{S}=0$ gives rise to the classical dynamical equations of the fields,
\begin{eqnarray}
\hbar\frac{\partial}{\partial{t}}\left(\begin{array}{c}
\psi_{R}\\
\psi_{I}
\end{array}\right)=\left(\begin{array}{cc}
\hat{H}_{I} & \hat{H}_{R}\\
-\hat{H}_{R} & \hat{H}_{I}
\end{array}\right)\left(\begin{array}{c}
\psi_{R}\\
\psi_{I}
\end{array}\right),\label{eq:13}
\end{eqnarray}
\begin{eqnarray}
\frac{1}{c^2}\ddot{\bm{A}}+\bigtriangledown\times\bigtriangledown\times\bm{A}+\frac{1}{c}\bigtriangledown\dot\phi=\frac{4\pi}{c}\bm{J}.\label{eq:14}
\end{eqnarray}
Where the bilinear form $\bm{J}=ec\psi^{+}\bm{\alpha}\psi$ is the Dirac current density \cite{Dirac2}, and the gauge 
field components are restricted by the Gauss's law $c\bigtriangledown^2\phi+\bigtriangledown\cdot\dot{\bm{A}}=-4{\pi}ec\psi^{+}\psi$.

\subsection{Hamiltonian field theory}
\label{sec:2-2}
The cotangent bundle of the configuration manifold $G$ can be defined as $T^{*}G=(\psi_{R},\psi_{I},\bm{A},$ $\phi,
\delta{L}/\delta\dot{\psi_{R}},\delta{L}/\delta\dot{\psi_{I}},\delta{L}/\delta\dot{\bm{A}},\delta{L}/\delta\dot{\phi})$, 
where the variational derivatives of Lagrangian functional with respect to field components are given by,
\begin{eqnarray}
\frac{\delta{L}}{\delta\dot{\psi_{R}}}=\frac{1}{2}\psi_{I},~~\frac{\delta{L}}{\delta\dot{\psi_{I}}}=-\frac{1}{2}\psi_{R},\label{eq:15}
\end{eqnarray}
\begin{eqnarray}
\bm{Y}=\frac{\delta{L}}{\delta\dot{\bm{A}}}=\frac{1}{4\pi{c^2}}\dot{\bm{A}}+\frac{1}{4\pi{c}}\bigtriangledown\phi,~~\frac{\delta{L}}{\delta\dot\phi}=0.\label{eq:16}
\end{eqnarray}
The Hamiltonian functional $H$ can be obtained via the Legendre transformation of $L$, which is a map $TG \to T^{*}G$,
\begin{eqnarray}
H&=&\int_{V}\left[\left(\frac{\delta{L}}{\delta\dot{\psi_{R}}}\right)^{T}\dot{\psi_{R}}+\left(\frac{\delta{L}}{\delta\dot{\psi_{I}}}\right)^{T}\dot{\psi_{I}}+\frac{\delta{L}}{\delta\dot{\bm{A}}}\cdot\dot{\bm{A}}\right]{\rm{d}}^{3}x-L\nonumber\\
&=&\int_{V}\left\{\frac{1}{2\hbar}\left(\psi_{R}^{T}\hat{H}_{R}\psi_{R}+\psi_{I}^{T}\hat{H}_{R}\psi_{I}-\psi_{R}^{T}\hat{H}_{I}\psi_{I}+\psi_{I}^{T}\hat{H}_{I}\psi_{R}\right)\right.\nonumber\\
& &\left.+\frac{1}{8\pi}\left[16\pi^2c^2\bm{Y}^2+\left(\bigtriangledown\times\bm{A}\right)^2-8\pi{c}\bm{Y}\cdot\bigtriangledown\phi\right]\right\}{\rm{d}}^{3}x.\label{eq:17}
\end{eqnarray}

On the cotangent bundle $T^{*}G$, we can obtain a 2-form field,
\begin{eqnarray}
\Omega&=&\bm{{\rm{d}}}\left[\left(\frac{\delta{L}}{\delta\dot{\psi_{R}}}\right)^{T}\bm{{\rm{d}}}\psi_{R}+\left(\frac{\delta{L}}{\delta\dot{\psi_{I}}}\right)^{T}\bm{{\rm{d}}}\psi_{I}+\frac{\delta{L}}{\delta\dot{\bm{A}}}\cdot\bm{{\rm{d}}}\bm{A}\right]\nonumber\\
&=&\sum^{4}_{i=1}\bm{{\rm{d}}}\psi_{iI}\wedge\bm{{\rm{d}}}\psi_{iR}+\sum^{3}_{i=1}\bm{{\rm{d}}}Y_{i}\wedge\bm{{\rm{d}}}A_{i},\label{eq:18}
\end{eqnarray}
which is obviously exact and closed, where $\bm{{\rm{d}}}$ is the exterior derivative operator. The 2-form $\Omega$ is a 
canonical symplectic form which can be used to construct a Poisson algebra,
\begin{eqnarray}
\left\{F,G\right\}&=&\int_{V}\left[\left(\frac{\delta{F}}{\delta\psi_{R}}\right)^{T},\frac{\delta{F}}{\delta\bm{A}},\left(\frac{\delta{F}}{\delta\psi_{I}}\right)^{T},\frac{\delta{F}}{\delta\bm{Y}}\right]\Omega^{-1}\left[\left(\frac{\delta{G}}{\delta\psi_{R}}\right)^{T},\frac{\delta{G}}{\delta\bm{A}},\left(\frac{\delta{G}}{\delta\psi_{I}}\right)^{T},\frac{\delta{G}}{\delta\bm{Y}}\right]^{T}{\rm{d}}^{3}x\nonumber\\
&=&\int_{V}\left[\sum^{4}_{i=1}\left(\frac{\delta{F}}{\delta\psi_{iR}}\frac{\delta{G}}{\delta\psi_{iI}}-\frac{\delta{G}}{\delta\psi_{iR}}\frac{\delta{F}}{\delta\psi_{iI}}\right)+\frac{\delta{F}}{\delta\bm{A}}\cdot\frac{\delta{G}}{\delta\bm{Y}}-\frac{\delta{G}}{\delta\bm{A}}\cdot\frac{\delta{F}}{\delta\bm{Y}}\right]{\rm{d}}^{3}x.\label{eq:19}
\end{eqnarray}
Where $F$ and $G$ are arbitary functionals on $T^{*}G$. The dynamical equations of a field theory with canonical symplectic 
structure can be generated by the Hamiltonian functional of this field theory,
\begin{eqnarray}
\dot{F}=\left\{F,H\right\}.\label{eq:20}
\end{eqnarray}

By taking the total variation of the Hamiltonian functional \eqref{eq:17} with fixed boundary, we obtain,
\begin{eqnarray}
\delta{H}&=&\int_{V}\left\{\frac{1}{\hbar}\left[\left(e\phi+mc^2\right)\psi_{1R}-eA_{x}\psi_{4R}-eA_{y}\psi_{4I}-eA_{z}\psi_{3R}-\hbar{c}\left(-\partial_{x}\psi_{4I}+\partial_{y}\psi_{4R}-\partial_{z}\psi_{3I}\right)\right]\delta\psi_{1R}\right.\nonumber\\
& &+\frac{1}{\hbar}\left[\left(e\phi+mc^2\right)\psi_{2R}-eA_{x}\psi_{3R}+eA_{y}\psi_{3I}+eA_{z}\psi_{4R}-\hbar{c}\left(-\partial_{x}\psi_{3I}-\partial_{y}\psi_{3R}+\partial_{z}\psi_{4I}\right)\right]\delta\psi_{2R}\nonumber\\
& &+\frac{1}{\hbar}\left[\left(e\phi-mc^2\right)\psi_{3R}-eA_{x}\psi_{2R}-eA_{y}\psi_{2I}-eA_{z}\psi_{1R}-\hbar{c}\left(-\partial_{x}\psi_{2I}+\partial_{y}\psi_{2R}-\partial_{z}\psi_{1I}\right)\right]\delta\psi_{3R}\nonumber\\
& &+\frac{1}{\hbar}\left[\left(e\phi-mc^2\right)\psi_{4R}-eA_{x}\psi_{1R}+eA_{y}\psi_{1I}+eA_{z}\psi_{2R}-\hbar{c}\left(-\partial_{x}\psi_{1I}-\partial_{y}\psi_{1R}+\partial_{z}\psi_{2I}\right)\right]\delta\psi_{4R}\nonumber\\
& &+\frac{1}{\hbar}\left[\left(e\phi+mc^2\right)\psi_{1I}-eA_{x}\psi_{4I}+eA_{y}\psi_{4R}-eA_{z}\psi_{3I}-\hbar{c}\left(\partial_{x}\psi_{4R}+\partial_{y}\psi_{4I}+\partial_{z}\psi_{3R}\right)\right]\delta\psi_{1I}\nonumber\\
& &+\frac{1}{\hbar}\left[\left(e\phi+mc^2\right)\psi_{2I}-eA_{x}\psi_{3I}-eA_{y}\psi_{3R}+eA_{z}\psi_{4I}-\hbar{c}\left(\partial_{x}\psi_{3R}-\partial_{y}\psi_{3I}-\partial_{z}\psi_{4R}\right)\right]\delta\psi_{2I}\nonumber\\
& &+\frac{1}{\hbar}\left[\left(e\phi-mc^2\right)\psi_{3I}-eA_{x}\psi_{2I}+eA_{y}\psi_{2R}-eA_{z}\psi_{1I}-\hbar{c}\left(\partial_{x}\psi_{2R}+\partial_{y}\psi_{2I}+\partial_{z}\psi_{1R}\right)\right]\delta\psi_{3I}\nonumber\\
& &+\frac{1}{\hbar}\left[\left(e\phi-mc^2\right)\psi_{4I}-eA_{x}\psi_{1I}-eA_{y}\psi_{1R}+eA_{z}\psi_{2I}-\hbar{c}\left(\partial_{x}\psi_{1R}-\partial_{y}\psi_{1I}-\partial_{z}\psi_{2R}\right)\right]\delta\psi_{4I}\nonumber\\
& &\left.+\left[-\frac{1}{c}\bm{J}+\frac{1}{4\pi}\bigtriangledown\times\bigtriangledown\times\bm{A}\right]\cdot\delta\bm{A}+\left(4{\pi}c^2\bm{Y}-c\bigtriangledown\phi\right)\cdot\delta\bm{Y}-c\bigtriangledown\cdot\bm{Y}\delta{\phi}\right\}{\rm{d}}^{3}x.\label{eq:21}
\end{eqnarray}
Where the Dirac current density can be expanded as,
\begin{eqnarray}
\bm{J}=\frac{ec}{\hbar}\left(\begin{array}{c}
\psi_{1R}\psi_{4R}+\psi_{1I}\psi_{4I}+\psi_{2R}\psi_{3R}+\psi_{2I}\psi_{3I}\\
\psi_{1R}\psi_{4I}-\psi_{1I}\psi_{4R}+\psi_{2I}\psi_{3R}-\psi_{2R}\psi_{3I}\\
\psi_{1R}\psi_{3R}+\psi_{1I}\psi_{3I}-\psi_{2R}\psi_{4R}-\psi_{2I}\psi_{4I}
\end{array}\right)=ec\psi^{+}\bm{\alpha}\psi.\label{eq:22}
\end{eqnarray}

By substituting Eqs.~\eqref{eq:21}-\eqref{eq:22} into Eq.~\eqref{eq:20}, we obtain the canonical equations of the 
Dirac-Maxwell fields theory as,
\begin{eqnarray}
\dot\psi_{1R}&=&\left\{\psi_{1R},H\right\}\nonumber\\
&=&\frac{1}{\hbar}\left[\left(e\phi+mc^2\right)\psi_{1I}-eA_{x}\psi_{4I}+eA_{y}\psi_{4R}-eA_{z}\psi_{3I}-\hbar{c}\left(\partial_{x}\psi_{4R}+\partial_{y}\psi_{4I}+\partial_{z}\psi_{3R}\right)\right],\label{eq:23}
\end{eqnarray}
\begin{eqnarray}
\dot\psi_{2R}&=&\left\{\psi_{2R},H\right\}\nonumber\\
&=&\frac{1}{\hbar}\left[\left(e\phi+mc^2\right)\psi_{2I}-eA_{x}\psi_{3I}-eA_{y}\psi_{3R}+eA_{z}\psi_{4I}-\hbar{c}\left(\partial_{x}\psi_{3R}-\partial_{y}\psi_{3I}-\partial_{z}\psi_{4R}\right)\right],\label{eq:24}
\end{eqnarray}
\begin{eqnarray}
\dot\psi_{3R}&=&\left\{\psi_{3R},H\right\}\nonumber\\
&=&\frac{1}{\hbar}\left[\left(e\phi-mc^2\right)\psi_{3I}-eA_{x}\psi_{2I}+eA_{y}\psi_{2R}-eA_{z}\psi_{1I}-\hbar{c}\left(\partial_{x}\psi_{2R}+\partial_{y}\psi_{2I}+\partial_{z}\psi_{1R}\right)\right],\label{eq:25}
\end{eqnarray}
\begin{eqnarray}
\dot\psi_{4R}&=&\left\{\psi_{4R},H\right\}\nonumber\\
&=&\frac{1}{\hbar}\left[\left(e\phi-mc^2\right)\psi_{4I}-eA_{x}\psi_{1I}-eA_{y}\psi_{1R}+eA_{z}\psi_{2I}-\hbar{c}\left(\partial_{x}\psi_{1R}-\partial_{y}\psi_{1I}-\partial_{z}\psi_{2R}\right)\right],\label{eq:26}
\end{eqnarray}
\begin{eqnarray}
\dot\psi_{1I}&=&\left\{\psi_{1I},H\right\}\nonumber\\
&=&-\frac{1}{\hbar}\left[\left(e\phi+mc^2\right)\psi_{1R}-eA_{x}\psi_{4R}-eA_{y}\psi_{4I}-eA_{z}\psi_{3R}-\hbar{c}\left(-\partial_{x}\psi_{4I}+\partial_{y}\psi_{4R}-\partial_{z}\psi_{3I}\right)\right],\label{eq:27}
\end{eqnarray}
\begin{eqnarray}
\dot\psi_{2I}&=&\left\{\psi_{2I},H\right\}\nonumber\\
&=&-\frac{1}{\hbar}\left[\left(e\phi+mc^2\right)\psi_{2R}-eA_{x}\psi_{3R}+eA_{y}\psi_{3I}+eA_{z}\psi_{4R}-\hbar{c}\left(-\partial_{x}\psi_{3I}-\partial_{y}\psi_{3R}+\partial_{z}\psi_{4I}\right)\right],\label{eq:28}
\end{eqnarray}
\begin{eqnarray}
\dot\psi_{3I}&=&\left\{\psi_{3I},H\right\}\nonumber\\
&=&-\frac{1}{\hbar}\left[\left(e\phi-mc^2\right)\psi_{3R}-eA_{x}\psi_{2R}-eA_{y}\psi_{2I}-eA_{z}\psi_{1R}-\hbar{c}\left(-\partial_{x}\psi_{2I}+\partial_{y}\psi_{2R}-\partial_{z}\psi_{1I}\right)\right],\label{eq:29}
\end{eqnarray}
\begin{eqnarray}
\dot\psi_{4I}&=&\left\{\psi_{4I},H\right\}\nonumber\\
&=&-\frac{1}{\hbar}\left[\left(e\phi-mc^2\right)\psi_{4R}-eA_{x}\psi_{1R}+eA_{y}\psi_{1I}+eA_{z}\psi_{2R}-\hbar{c}\left(-\partial_{x}\psi_{1I}-\partial_{y}\psi_{1R}+\partial_{z}\psi_{2I}\right)\right],\label{eq:30}
\end{eqnarray}
\begin{eqnarray}
\dot{\bm{A}}&=&\left\{\bm{A},H\right\}=4\pi{c}^2\bm{Y}-c\bigtriangledown\phi,\label{eq:31}
\end{eqnarray}
\begin{eqnarray}
\dot{\bm{Y}}&=&\left\{\bm{Y},H\right\}=-\frac{1}{4\pi}\bigtriangledown\times\bigtriangledown\times\bm{A}+\frac{1}{c}\bm{J}.\label{eq:32}
\end{eqnarray}

The canonical equations \eqref{eq:23}-\eqref{eq:32} equal to the dynamical equations \eqref{eq:13}-\eqref{eq:14}, which 
means that the Hamiltonian field theory on $T^{*}G$ is an equivalent theory to the Lagrangian field theory on $TG$, both 
of which describe the intrinsic geometric structures of the interacting particles.

\subsection{Gauge and Poincar\'e invariances}
\label{sec:2-3}
The QED is $U(1)$ gauge and Poincar\'e invariant. Based on the Noether's theorem, the canonical field theory constructed 
in Sec.\ref{sec:2} for Dirac-Maxwell systems admits charge, energy-momentum and angular momentum conservation laws 
\cite{Weinberg,Peskin}.

The $U(1)$ gauge symmetry means that the action and dynamical equations are invariant under the $U(1)$ gauge transformation,
\begin{eqnarray}
\left(A^{\mu},\psi,\psi^{+}\right)\to\left(A^{\mu}+\partial^{\mu}\theta,\psi{\rm{e}}^{i\frac{e}{\hbar{c}}\theta},\psi^{+}{\rm{e}}^{-i\frac{e}{\hbar{c}}\theta}\right),\label{eq:33}
\end{eqnarray}
where the gauge parameter $\theta$ is an arbitrary scalar field. It is convenient to verify the $U(1)$ gauge symmetry of 
the canonical field theory by substituting Eq.~\eqref{eq:33} into the Lagrangian density \eqref{eq:9} or canonical 
equations \eqref{eq:23}-\eqref{eq:32}. With an infinitesimal gauge transformation $\delta(A^{\mu},\psi,\psi^{+})
=(\partial^{\mu}\theta,i\frac{e}{\hbar{c}}\theta\psi,-i\frac{e}{\hbar{c}}\theta\psi^{+})$ and using the dynamical 
equations \eqref{eq:13}-\eqref{eq:14}, we can obtain the charge conservation law via $\delta{S}=0$,
\begin{eqnarray}
\left(\frac{\delta{S}}{\delta\psi}\right)^{T}\delta\psi+\frac{\delta{S}}{\delta\psi^{+}}\delta\psi^{+T}=0,\label{eq:34}
\end{eqnarray}
which can be explicitly written as,
\begin{eqnarray}
\partial_{\mu}J^{\mu}=0,~~J^{\mu}=\left(e\psi^{+}\psi,\frac{1}{c}\bm{J}\right).\label{eq:35}
\end{eqnarray}

The Poincar\'e symmetry consists of two parts, which are translation symmetry and Lorentz covariance. The translation 
symmetry means that the action and dynamical equations are invariant under the space-time translation,
\begin{eqnarray}
\left(x^{\mu},A^{\mu},\psi,\psi^{+}\right)\to\left(x^{\mu}+\epsilon^{\mu},A^{'\mu},\psi',\psi^{'+}\right).\label{eq:36}
\end{eqnarray}
Where $\epsilon^{\mu}$ is an arbitrary translation parameter, and the local field components admit 
$A^{'\mu}(x^{\mu}+\epsilon^{\mu})=A^{\mu}(x^{\mu})$, $\psi'(x^{\mu}+\epsilon^{\mu})=\psi(x^{\mu})$, 
$\psi^{'+}(x^{\mu}+\epsilon^{\mu})=\psi^{+}(x^{\mu})$. By substituting an infinitesimal translations 
$\delta(x^{\mu},A^{\mu},\psi,\psi^{+})=(\epsilon^{\mu},0,0,0)$ into $\delta{S}=0$ and using the dynamical equations 
\eqref{eq:13}-\eqref{eq:14}, we can obtain the energy-momentum conservation law as,
\begin{eqnarray}
\partial_{\mu}\mathcal{T}^{\mu\nu}=0.\label{eq:37}
\end{eqnarray}
Where the energy-momentum tensor is given by,
\begin{eqnarray}
\mathcal{T}^{\mu\nu}=i\hbar{c}\bar\psi\gamma^{\mu}\partial^{\nu}\psi+\frac{1}{16\pi}\mathcal{F}^2g^{\mu\nu}-\frac{1}{4\pi}\mathcal{F}^{\mu\rho}\partial^{\nu}A_{\rho}.\label{eq:38}
\end{eqnarray}
The Lorentz covariance means that the action and dynamical equations are invariant under the Lorentz transformation,
\begin{eqnarray}
\left(x^{\mu},A^{\mu},\psi,\psi^{+}\right)\to\left(\Lambda^{\mu}_{\nu}x^{\nu},\Lambda^{\mu}_{\nu}A^{\nu},S(\Lambda)\psi,\psi^{+}S^{-1}(\Lambda)\right).\label{eq:39}
\end{eqnarray}
Where $S(\Lambda)$ admits $S(\Lambda)\gamma^{\mu}S^{-1}(\Lambda)=\Lambda^{-1\mu}_{\nu}\gamma^{\nu}$, and $\Lambda^{\mu}_{\nu}$ 
is an arbitrary Lorentz transformation parameter. It is convenient to verify the Lorentz covariance of the canonical 
field theory by substituting Eq.~\eqref{eq:39} into the Lagrangian density \eqref{eq:9} or canonical equations 
\eqref{eq:23}-\eqref{eq:32}. By substituting an infinitesimal Lorentz transformation $\delta(x^{\mu},A^{\mu},\psi,
\psi^{+})=(\epsilon^{\mu\nu}x_{\nu},\epsilon^{\mu\nu}A_{\nu},-\frac{i}{4}\epsilon^{\eta\xi}\sigma_{\eta\xi}\psi,
\frac{i}{4}\psi^{+}\epsilon^{\eta\xi}\sigma_{\eta\xi})$ into $\delta{S}=0$ and using the dynamical equations 
\eqref{eq:13}-\eqref{eq:14}, we can obtain the angular momentum conservation law as,
\begin{eqnarray}
\partial_{\mu}\mathcal{M}^{\mu\nu\rho}=0.\label{eq:40}
\end{eqnarray}
Where the general angular momentum tensor is given by,
\begin{eqnarray}
\mathcal{M}^{\mu\nu\rho}=x^{\nu}\mathcal{T}^{\mu\rho}-x^{\rho}\mathcal{T}^{\mu\nu}+\frac{\hbar{c}}{2}\bar\psi\gamma^{\mu}\sigma^{\nu\rho}\psi+\frac{1}{4\pi}\left(A^{\nu}\mathcal{F}^{\mu\rho}-A^{\rho}\mathcal{F}^{\mu\nu}\right).\label{eq:41}
\end{eqnarray}
Here $\epsilon^{\mu\nu}=-\epsilon^{\nu\mu}$ is an arbitrary infinitesimal Lorentz parameter, and the Lorentz generator 
on bispinor field is defined as $\sigma^{\mu\nu}=\frac{i}{2}[\gamma^{\mu},\gamma^{\nu}]$ \cite{Weinberg,Peskin}.

\section{Lattice canonical field theory of the Dirac-Maxwell systems}
\label{sec:3}

\subsection{DEC Based Discretization}
\label{sec:3-1}
The first step to construct a lattice field theory is discretization. As a differential geometry based numerical framework, 
DEC defines a class of complete operational rules and differential forms on a discrete differential manifold, which form 
a cochain complex \cite{Hirani,Hiptmair,Arnold}. To construct a semi-discrete LCFT for Dirac-Maxwell systems, the 
space-like submanifold of the Minkowski space-time manifold is discretized by using a rectangular lattice (other lattices 
are also viable). Then the scalar field $A^{0}=\phi$, which is a 0-form on the space-like submanifold, naturally lives on 
the vertex of the lattice,
\begin{eqnarray}
\phi_{J}\left(t\right):\phi\left(t,x_{i},y_{j},z_{k}\right).\label{eq:42}
\end{eqnarray}
Where the subscript $J$ indicates lattice label which traverses all lattice points, and $(x_{i},y_{j},z_{k})$ is the 
coordinate of the lattice vertex. The $U(1)$ gauge and electric field 1-forms $\bm{A}=A_{i}dx^{i}$ and $\bm{Y}=Y_{i}dx^{i}$ 
naturally live along the edges of the lattice,
\begin{eqnarray}
A/Y_{xJ}\left(t\right):A/Y_{x}\left(t,x_{i}+\frac{\Delta{x}}{2},y_{j},z_{k}\right),\label{eq:43}\\
A/Y_{yJ}\left(t\right):A/Y_{y}\left(t,x_{i},y_{j}+\frac{\Delta{y}}{2},z_{k}\right),\label{eq:44}\\
A/Y_{zJ}\left(t\right):A/Y_{z}\left(t,x_{i},y_{j},z_{k}+\frac{\Delta{z}}{2}\right).\label{eq:45}
\end{eqnarray}
In the above discretization, a half integer index indicates along which edge does the field resides, where $\Delta{x}$, 
$\Delta{y}$ and $\Delta{z}$ are lattice periods. In DEC framework, the magnetic field 2-form $\bm{{\rm{d}}}\bm{A}$ lives 
on the face center of the lattice. By using the Hodge dual operator $*$, we obtain the discrete charge 3-form 
$-\bm{{\rm{d}}}*\bm{Y}$ and current 2-form $\bm{{\rm{d}}}*\bm{{\rm{d}}}\bm{A}$ on the volume and face centers of the 
dual lattice respectively. Where the coordinate of a form on dual lattice is translated by $(\Delta{x},\Delta{y},\Delta{z})/2$ 
after the Hodge operation, which means the primary-dual lattice generated by the Hodge star is a staggered lattice. The 
discrete gradient $\bigtriangledown_{d}$, curl $\bigtriangledown_{d}\times$, and divergence $\bigtriangledown_{d}\cdot$ 
operators in DEC framework can be defined as \cite{Hirani,QChen3},
\begin{eqnarray}
\bm{{\rm{d}}}\phi_{J}=\left(\bigtriangledown_{d}\phi\right)_{J}=\left(\begin{array}{c}
\frac{\phi_{i+1,j,k}-\phi_{i,j,k}}{\bigtriangleup{x}}\\
\frac{\phi_{i,j+1,k}-\phi_{i,j,k}}{\bigtriangleup{y}}\\
\frac{\phi_{i,j,k+1}-\phi_{i,j,k}}{\bigtriangleup{z}}
\end{array}\right),\label{eq:46}
\end{eqnarray}
\begin{eqnarray}
\bm{{\rm{d}}}\bm{A}_{J}=\left(\bigtriangledown_{d}\times\bm{A}\right)_{J}=\left(\begin{array}{c}
\frac{Az_{i,j+1,k+\frac{1}{2}}-Az_{i,j,k+\frac{1}{2}}}{\bigtriangleup{y}}-\frac{Ay_{i,j+\frac{1}{2},k+1}-Ay_{i,j+\frac{1}{2},k}}{\bigtriangleup{z}}\\
\frac{Ax_{i+\frac{1}{2},j,k+1}-Ax_{i+\frac{1}{2},j,k}}{\bigtriangleup{z}}-\frac{Az_{i+1,j,k+\frac{1}{2}}-Az_{i,j,k+\frac{1}{2}}}{\bigtriangleup{x}}\\
\frac{Ay_{i+1,j+\frac{1}{2},k}-Ay_{i,j+\frac{1}{2},k}}{\bigtriangleup{x}}-\frac{Ax_{i+\frac{1}{2},j+1,k}-Ax_{i+\frac{1}{2},j,k}}{\bigtriangleup{y}}
\end{array}\right),\label{eq:47}
\end{eqnarray}
\begin{eqnarray}
\bm{{\rm{d}}}*\bm{Y}_{J}=\left(\bigtriangledown_{d}\cdot\bm{Y}\right)_{J}=\frac{Yx_{i+\frac{1}{2},j,k}-Yx_{i-\frac{1}{2},j,k}}{\bigtriangleup{x}}+\frac{Yy_{i,j+\frac{1}{2},k}-Yy_{i,j-\frac{1}{2},k}}{\bigtriangleup{y}}+\frac{Yz_{i,j,k+\frac{1}{2}}-Yz_{i,j,k-\frac{1}{2}}}{\bigtriangleup{z}}.\label{eq:48}
\end{eqnarray}

\begin{figure}[htbp]
\centerline{\includegraphics[width=8.8cm,height=4.2cm]{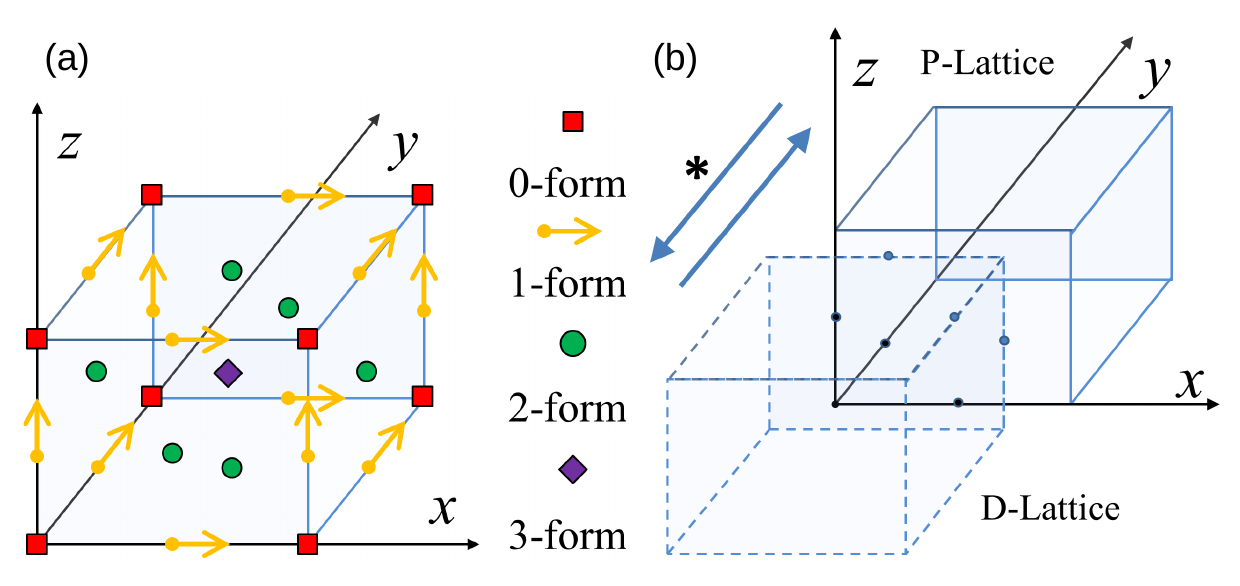}}
\caption{DEC based discretization of the LCFT for Dirac-Maxwell systems on a rectangular lattice. (a) Discrete forms 
on the space-like submanifold. (b) Staggered lattice generated by the Hodge dual operator *, where P- indicates primary 
lattice and D- indiactes dual lattice.}
\label{fig:1}
\end{figure}

When it comes to the bispinor field, the fermion doubling is a serious problem in LGT, especially in LQCD simulations. 
Nielsen-Ninomiya no-go theorem states that the discretization of the Dirac equation on a regular space lattice forbids 
a single chirally invariant fermion flavor without breaking one or more of the following assumptions: translation 
invariance, locality, and Hermiticity \cite{Nielsen}. There are several strategies to solve this problem, e.g. Wilson's 
momentum-dependent mass term and Kogut-Susskind staggered fermion \cite{Kogut,Rothe}. In particular, the K\"ahler fermion 
constructed by P. Becher and H. Joos is the first geometric theory based lattice fermion, which is proved equivalent to a 
staggered fermion \cite{Becher}. Here, in the unified DEC framework, we treat the bispinor components as different differential forms on 
the space-like submanifold, which are $\psi_{1}$, $\psi_{2}dx{\wedge}dz$, $\psi_{3}dz$, $\psi_{4}dx$, $\tilde\psi_{1}dx{\wedge}dy$, 
$\tilde\psi_{2}dy{\wedge}dz$, $\tilde\psi_{3}dx{\wedge}dy{\wedge}dz$ and $\tilde\psi_{4}dy$,
\begin{eqnarray}
\psi_{1J}\left(t\right):\psi_{1}\left(t,x_{i},y_{j},z_{k}\right),~\tilde\psi_{1J}\left(t\right):\psi_{1}\left(t,x_{i}+\frac{\Delta{x}}{2},y_{j}+\frac{\Delta{y}}{2},z_{k}\right),\label{eq:49}\\
\psi_{2J}\left(t\right):\psi_{2}\left(t,x_{i}+\frac{\Delta{x}}{2},y_{j},z_{k}+\frac{\Delta{z}}{2}\right),~\tilde\psi_{2J}\left(t\right):\psi_{2}\left(t,x_{i},y_{j}+\frac{\Delta{y}}{2},z_{k}+\frac{\Delta{z}}{2}\right),\label{eq:50}\\
\psi_{3J}\left(t\right):\psi_{3}\left(t,x_{i},y_{j},z_{k}+\frac{\Delta{z}}{2}\right),~\tilde\psi_{3J}\left(t\right):\psi_{3}\left(t,x_{i}+\frac{\Delta{x}}{2},y_{j}+\frac{\Delta{y}}{2},z_{k}+\frac{\Delta{z}}{2}\right),\label{eq:51}\\
\psi_{4J}\left(t\right):\psi_{4}\left(t,x_{i}+\frac{\Delta{x}}{2},y_{j},z_{k}\right),~\tilde\psi_{4J}\left(t\right):\psi_{4}\left(t,x_{i},y_{j}+\frac{\Delta{y}}{2},z_{k}\right).\label{eq:52}
\end{eqnarray}
Where the double sampled bispinor components live on the vertex, edge, faces and volume centers of the lattice respectively. 
By using the Hodge dual operator, we find the dual relations $\psi_{1}\stackrel{*}\leftrightarrows\tilde\psi_{3}$, 
$\psi_{2}\stackrel{*}\leftrightarrows\tilde\psi_{4}$, $\psi_{3}\stackrel{*}\leftrightarrows\tilde\psi_{1}$ and 
$\psi_{4}\stackrel{*}\leftrightarrows\tilde\psi_{2}$, which naturally generate a staggered checkerboard-like lattice. The DEC 
dual relations could lead to a lattice bispinor field which involves two degenerate flavors. The DEC based discretization 
of the LCFT on the space-like submanifold is shown in Fig.\ref{fig:1}.

Based on the discretization of fields, we can construct a discrete Poisson bracket, which admits bilinearity, 
anticommutativity, product rule, and Jacobi identity. The fields can be reconstructed as,
\begin{eqnarray}
\bm{A}\left(\bm{x},t\right)=\sum_{J=1}^{M}\bm{A}_{J}\left(t\right)W\left(\bm{x}-\bm{x}_{J}\right),~~~\bm{Y}\left(\bm{x},t\right)=\sum_{J=1}^{M}\bm{Y}_{J}\left(t\right)W\left(\bm{x}-\bm{x}_{J}\right),\label{eq:53}
\end{eqnarray}
\begin{eqnarray}
\psi_{iR}/\tilde\psi_{iR}\left(\bm{x},t\right)=\sum_{J=1}^{M}\psi_{iRJ}/\tilde\psi_{iRJ}\left(t\right)W\left(\bm{x}-\bm{x}_{J}\right),~~~\psi_{iI}/\tilde\psi_{iI}\left(\bm{x},t\right)=\sum_{J=1}^{M}\psi_{iIJ}/\tilde\psi_{iIJ}\left(t\right)W\left(\bm{x}-\bm{x}_{J}\right).\label{eq:54}
\end{eqnarray} 
Where the distribution function $W\left(\bm{x}-\bm{x}_{J}\right)$ is defined as,
\begin{eqnarray}
W\left(\bm{x}-\bm{x}_{J}\right)=\left\{ \begin{array}{cc}
1, & |x-x_{J}|<\frac{\bigtriangleup{x}}{2},|y-y_{J}|<\frac{\bigtriangleup{y}}{2},|z-z_{J}|<\frac{\bigtriangleup{z}}{2}\\
0, & \rm{elsewhere}
\end{array}\right..\label{eq:55}
\end{eqnarray}
Then, the variational derivative with respect to $\bm{A}$ is \cite{Qin8,QChen2},
\begin{eqnarray}
\frac{\delta{F}}{\delta\bm{A}}=\sum_{J=1}^{M}\frac{\delta\bm{A}_{J}}{\delta\bm{A}}\frac{\partial{F}}{\partial\bm{A}_{J}}=\sum_{J=1}^{M}\frac{1}{\bigtriangleup{V}}W\left(\bm{x}-\bm{x}_{J}\right)\frac{\partial{F}}{\partial\bm{A}_{J}},\label{eq:56}
\end{eqnarray}
and the variational derivatives with respect to $\bm{Y}$, $\psi_{iR}/\tilde\psi_{iR}$ and $\psi_{iI}/\tilde\psi_{iI}$ 
have similar expressions. Here, $\bigtriangleup{V}=\bigtriangleup{x}\wedge\bigtriangleup{y}\wedge\bigtriangleup{z}$ is 
the volume form on the lattice. Based on Eq.~\eqref{eq:56}, the canonical Poisson bracket \eqref{eq:19} is discretized as,
\begin{eqnarray}
\left\{F,G\right\} _{d}&=&\sum_{J=1}^{M}\left[2\sum_{\sim}\sum_{i=1}^{4}\left(\frac{\partial{F}}{\partial\psi_{iRJ}}\frac{\partial{G}}{\partial\psi_{iIJ}}-\frac{\partial{G}}{\partial\psi_{iRJ}}\frac{\partial{F}}{\partial\psi_{iIJ}}\right)+\sum_{i=1}^{3}\left(\frac{\partial{F}}{\partial{A}_{iJ}}\frac{\partial{G}}{\partial{Y}_{iJ}}-\frac{\partial{G}}{\partial{A}_{iJ}}\frac{\partial{F}}{\partial{Y}_{iJ}}\right)\right]\frac{1}{\bigtriangleup{V}}\nonumber\\
&=&\sum_{J=1}^{M}\left\{2\sum_{\sim}\left[\left(\frac{\partial{F}}{\partial\psi_{RJ}}\right)^{T}\frac{\partial{G}}{\partial\psi_{IJ}}-\left(\frac{\partial{G}}{\partial\psi_{RJ}}\right)^{T}\frac{\partial{F}}{\partial\psi_{IJ}}\right]+\frac{\partial{F}}{\partial\bm{A}_{J}}\cdot\frac{\partial{G}}{\partial\bm{Y}_{J}}-\frac{\partial{G}}{\partial\bm{A}_{J}}\cdot\frac{\partial{F}}{\partial\bm{Y}_{J}}\right\}\frac{1}{\bigtriangleup{V}}.\label{eq:57}
\end{eqnarray}
With the discrete canonical Poisson bracket \eqref{eq:57}, the functionals on the discrete cotangent bundle 
$T^{*}G_{d}=(\psi_{RJ},\tilde\psi_{RJ},\bm{A}_{J},\psi_{IJ},\tilde\psi_{IJ},\bm{Y}_{J})$ form a complete Poisson algebra. 
Then a semi-discrete LCFT can be generated by this discrete canonical Poisson bracket with a proper Hamiltonian functional 
on $T^{*}G_{d}$.

\subsection{Pull-back and push-forward gauge covariant derivatives}
\label{sec:3-2}
The guage 1-form defines the guage connection on the $U(1)$ bundle, which enables parallel transport bispinor on the 
Minkowski manifold. In order to construct a gauge invariant semi-discrete LCFT, we introduce a pair of discrete gauge 
covariant derivatives for different bispinor components, which can be recognized as Wilson lines in the DEC framework \cite{YShi2}. 
When it comes to $D_{x}\psi_{1}$, $D_{y}\psi_{1}$, $D_{z}\psi_{1}$, $D_{y}\psi_{2}$, $D_{x}\psi_{3}$, $D_{y}\psi_{3}$, $D_{y}\psi_{4}$, 
$D_{z}\psi_{4}$, $D_{z}\tilde\psi_{1}$, $D_{x}\tilde\psi_{2}$, $D_{x}\tilde\psi_{4}$, and $D_{z}\tilde\psi_{4}$,
the pull-back gauge covariant derivative $D^{<}$ is used along the relative gauge connections, e.g.,
\begin{eqnarray}
\left(D^{<}_{x}\psi_{1}\right)_{J}&=&\frac{1}{\Delta{x}}\left(\psi_{1i+1,j,k}{\rm{e}}^{-i\frac{e}{\hbar{c}}A_{xi+\frac{1}{2},j,k}\Delta{x}}-\psi_{1i,j,k}\right)\nonumber\\
&=&\frac{1}{\sqrt{2\hbar}\Delta{x}}\left\{\psi_{1Ri+1,j,k}\cos\left(\frac{e}{\hbar{c}}A_{xi+\frac{1}{2},j,k}\Delta{x}\right)+\psi_{1Ii+1,j,k}\sin\left(\frac{e}{\hbar{c}}A_{xi+\frac{1}{2},j,k}\Delta{x}\right)-\psi_{1Ri,j,k}\right.\nonumber\\
& &\left.+i\left[\psi_{1Ii+1,j,k}\cos\left(\frac{e}{\hbar{c}}A_{xi+\frac{1}{2},j,k}\Delta{x}\right)-\psi_{1Ri+1,j,k}\sin\left(\frac{e}{\hbar{c}}A_{xi+\frac{1}{2},j,k}\Delta{x}\right)-\psi_{1Ii,j,k}\right]\right\}.\label{eq:58}
\end{eqnarray}
The other $D^{<}$ components can be given in a similar form.

When it comes to $D_{x}\psi_{2}$, $D_{z}\psi_{2}$, $D_{z}\psi_{3}$, $D_{x}\psi_{4}$, $D_{x}\tilde\psi_{1}$, $D_{y}\tilde\psi_{1}$, 
$D_{y}\tilde\psi_{2}$, $D_{z}\tilde\psi_{2}$, $D_{x}\tilde\psi_{3}$, $D_{y}\tilde\psi_{3}$, $D_{z}\tilde\psi_{3}$, and $D_{y}\tilde\psi_{4}$,
the push-forward gauge covariant derivative $D^{>}$ is used along the relative gauge connections, e.g.,
\begin{eqnarray}
\left(D^{>}_{z}\psi_{3}\right)_{J}&=&\frac{1}{\Delta{z}}\left(\psi_{3i,j,k+\frac{1}{2}}-\psi_{3i,j,k-\frac{1}{2}}{\rm{e}}^{i\frac{e}{\hbar{c}}A_{zi,j,k-\frac{1}{2}}\Delta{z}}\right)\nonumber\\
&=&\frac{1}{\sqrt{2\hbar}\Delta{z}}\left\{\psi_{3Ri,j,k+\frac{1}{2}}-\psi_{3Ri,j,k-\frac{1}{2}}\cos\left(\frac{e}{\hbar{c}}A_{zi,j,k-\frac{1}{2}}\Delta{z}\right)+\psi_{3Ii,j,k-\frac{1}{2}}\sin\left(\frac{e}{\hbar{c}}A_{zi,j,k-\frac{1}{2}}\Delta{z}\right)\right.\nonumber\\
& &\left.+i\left[\psi_{3Ii,j,k+\frac{1}{2}}-\psi_{3Ii,j,k-\frac{1}{2}}\cos\left(\frac{e}{\hbar{c}}A_{zi,j,k-\frac{1}{2}}\Delta{z}\right)-\psi_{3Ri,j,k-\frac{1}{2}}\sin\left(\frac{e}{\hbar{c}}A_{zi,j,k-\frac{1}{2}}\Delta{z}\right)\right]\right\}.\label{eq:59}
\end{eqnarray}
The other $D^{>}$ components can be given in a similar form.

By using $\bigtriangledown_{d}$ operator, the semi-discrete gauge transformation can be defined as,
\begin{eqnarray}
\phi'_{J}=\phi_{J}-\frac{1}{c}\dot{\theta}_{J},\label{eq:60}\\
\bm{A}'_{J}=\bm{A}_{J}+\left(\bigtriangledown_{d}\theta\right)_{J},\label{eq:61}\\
\psi'_{J}=\psi_{J}{\rm{e}}^{i\frac{e}{\hbar{c}}\theta_{J}}.\label{eq:62}
\end{eqnarray}
Where $\theta_{J}$ is an arbitrary discrete 0-form.

By substituting Eqs.~\eqref{eq:60}-\eqref{eq:62} into Eqs.~\eqref{eq:58}-\eqref{eq:59}, we obtain the gauge property of 
pull-back and push-forward gauge covariant derivatives,
\begin{eqnarray}
\left(D^{<}_{x}\psi_{1}\right)_{J}\xrightarrow{\psi_{1J}{\rm{e}}^{i\frac{e}{\hbar{c}}\theta_{J}},~A_{xJ}+\left(\bigtriangledown_{d}\theta\right)_{xJ}}\left(D^{<}_{x}\psi_{1}\right)_{J}{\rm{e}}^{i\frac{e}{\hbar{c}}\theta_{J}},\label{eq:63}\\
\left(D^{>}_{z}\psi_{3}\right)_{J}\xrightarrow{\psi_{3J}{\rm{e}}^{i\frac{e}{\hbar{c}}\theta_{J}},~A_{zJ}+\left(\bigtriangledown_{d}\theta\right)_{zJ}}\left(D^{>}_{z}\psi_{3}\right)_{J}{\rm{e}}^{i\frac{e}{\hbar{c}}\theta_{J}}.\label{eq:64}
\end{eqnarray}
It shows that after a gauge transformation, the pull-back and push-forward gauge covariant derivatives get an unified 
phase, which ensures the semi-discrete Lagrangian density of the bispinor is gauge invariant. 

The semi-discrete Lagrangian density of the $U(1)$ gauge field is also gauge invariant in the DEC framework, which can 
be directly verified. As a result, the semi-discrete action functional admits gauge symmetry.

\subsection{Semi-discrete canonical field theory}
\label{sec:3-3}
With the DEC and discrete gauge covariant derivatives, the Hamiltonian functional \eqref{eq:17} is discreted as,
\begin{eqnarray}
H_{d}=H^{(1)}_{d}+H^{(2)}_{d}+H^{(3)}_{d}.\label{eq:65}
\end{eqnarray}
Where the superscript means 1- bispinor momentum, 2- bispinor mass-energy, and 3- $U(1)$ gauge field respectively. The 
discrete Hamiltonian functionals are given by,
\begin{eqnarray}
H^{(1)}_{d}=-i\hbar{c}\sum^{M}_{J=1}\left(\begin{array}{c}
\psi_{1J}\\
\psi_{2J}\\
\psi_{3J}\\
\psi_{4J}\\
\tilde\psi_{1J}\\
\tilde\psi_{2J}\\
\tilde\psi_{3J}\\
\tilde\psi_{4J}
\end{array}\right)^{+}\left(\begin{array}{cccccccc}
0 & 0 & D^{>}_{z} & D^{>}_{x} & 0 & 0 & 0 & -iD^{>}_{y}\\
0 & 0 & D^{<}_{x} & -D^{<}_{z} & 0 & 0 & iD^{>}_{y} & 0\\
D^{<}_{z} & D^{>}_{x} & 0 & 0 & 0 & -iD^{>}_{y} & 0 & 0\\
D^{<}_{x} & -D^{>}_{z} & 0 & 0 & iD^{>}_{y} & 0 & 0 & 0\\
0 & 0 & 0 & -iD^{<}_{y} & 0 & 0 & D^{>}_{z} & D^{<}_{x}\\
0 & 0 & iD^{<}_{y} & 0 & 0 & 0 & D^{>}_{x} & -D^{<}_{z}\\
0 & -iD^{<}_{y} & 0 & 0 & D^{<}_{z} & D^{<}_{x} & 0 & 0\\
iD^{<}_{y} & 0 & 0 & 0 & D^{>}_{x} & -D^{>}_{z} & 0 & 0
\end{array}\right)\left(\begin{array}{c}
\psi_{1J}\\
\psi_{2J}\\
\psi_{3J}\\
\psi_{4J}\\
\tilde\psi_{1J}\\
\tilde\psi_{2J}\\
\tilde\psi_{3J}\\
\tilde\psi_{4J}
\end{array}\right)\frac{\Delta{V}}{2},\label{eq:66}
\end{eqnarray}
\begin{eqnarray}
H^{(2)}_{d}&=&\sum^{M}_{J=1}\left(\begin{array}{c}
\psi_{1J}\\
\psi_{2J}\\
\psi_{3J}\\
\psi_{4J}
\end{array}\right)^{+}\left(\begin{array}{cccc}
e\phi_{J}+mc^2 & 0 & 0 & 0\\
0 & e\phi_{J}+mc^2 & 0 & 0\\
0 & 0 & e\phi_{J}-mc^2 & 0\\
0 & 0 & 0 & e\phi_{J}-mc^2
\end{array}\right)\left(\begin{array}{c}
\psi_{1J}\\
\psi_{2J}\\
\psi_{3J}\\
\psi_{4J}
\end{array}\right)\frac{\Delta{V}}{2}\nonumber\\
& &+\sum^{M}_{J=1}\left(\begin{array}{c}
\tilde\psi_{1J}\\
\tilde\psi_{2J}\\
\tilde\psi_{3J}\\
\tilde\psi_{4J}
\end{array}\right)^{+}\left(\begin{array}{cccc}
e\phi_{J}+mc^2 & 0 & 0 & 0\\
0 & e\phi_{J}+mc^2 & 0 & 0\\
0 & 0 & e\phi_{J}-mc^2 & 0\\
0 & 0 & 0 & e\phi_{J}-mc^2
\end{array}\right)\left(\begin{array}{c}
\tilde\psi_{1J}\\
\tilde\psi_{2J}\\
\tilde\psi_{3J}\\
\tilde\psi_{4J}
\end{array}\right)\frac{\Delta{V}}{2},\label{eq:67}
\end{eqnarray}
\begin{eqnarray}
H^{(3)}_{d}=\sum^{M}_{J=1}\frac{1}{8\pi}\left[16\pi^2c^2\bm{Y}^{2}_{J}+\left(\bigtriangledown_{d}\times\bm{A}\right)^{2}_{J}-8\pi{c}\bm{Y}_{J}\cdot\left(\bigtriangledown_{d}\phi\right)_{J}\right]\Delta{V}.\label{eq:68}
\end{eqnarray}

By substituting the discrete Hamiltonian functional \eqref{eq:65} into the discrete Poisson bracket \eqref{eq:57}, we 
obtain the canonical equations of the semi-discrete LCFT for Dirac-Maxwell systems. Here, we introduce the Hamiltonian 
splitting method and generate three linear canonical subsystems \cite{QChen2}.

The subsystem generated by $H^{(1)}_{d}$ is given by,
\begin{eqnarray}
\dot{\psi}_{1RJ}&=&\left\{\psi_{1RJ},H^{(1)}_{d}\right\}\nonumber\\
&=&\frac{c}{\Delta{x}}\left(\psi_{4RJ-1}\cos^{x}_{J-1}-\psi_{4IJ-1}\sin^{x}_{J-1}-\psi_{4RJ}\right)+\frac{c}{\Delta{y}}\left(\tilde\psi_{4IJ-1}\cos^{y}_{J-1}+\tilde\psi_{4RJ-1}\sin^{y}_{J-1}-\tilde\psi_{4IJ}\right)\nonumber\\
& &+\frac{c}{\Delta{z}}\left(\psi_{3RJ-1}\cos^{z}_{J-1}-\psi_{3IJ-1}\sin^{z}_{J-1}-\psi_{3RJ}\right),\label{eq:69}
\end{eqnarray}
\begin{eqnarray}
\dot{\psi}_{2RJ}&=&\left\{\psi_{2RJ},H^{(1)}_{d}\right\}\nonumber\\
&=&\frac{c}{\Delta{x}}\left(\psi_{3RJ}-\psi_{3RJ+1}\cos^{x}_{J}-\psi_{3IJ+1}\sin^{x}_{J}\right)+\frac{c}{\Delta{y}}\left(\tilde\psi_{3IJ}-\tilde\psi_{3IJ-1}\cos^{y}_{J-1}-\tilde\psi_{3RJ-1}\sin^{y}_{J-1}\right)\nonumber\\
& &+\frac{c}{\Delta{z}}\left(\psi_{4RJ+1}\cos^{z}_{J}+\psi_{4IJ+1}\sin^{z}_{J}-\psi_{4RJ}\right),\label{eq:70}
\end{eqnarray}
\begin{eqnarray}
\dot{\psi}_{3RJ}&=&\left\{\psi_{3RJ},H^{(1)}_{d}\right\}\nonumber\\
&=&\frac{c}{\Delta{x}}\left(\psi_{2RJ-1}\cos^{x}_{J-1}-\psi_{2IJ-1}\sin^{x}_{J-1}-\psi_{2RJ}\right)+\frac{c}{\Delta{y}}\left(\tilde\psi_{2IJ-1}\cos^{y}_{J-1}+\tilde\psi_{2RJ-1}\sin^{y}_{J-1}-\tilde\psi_{2IJ}\right)\nonumber\\
& &+\frac{c}{\Delta{z}}\left(\psi_{1RJ}-\psi_{1RJ+1}\cos^{z}_{J}-\psi_{1IJ+1}\sin^{z}_{J}\right),\label{eq:71}
\end{eqnarray}
\begin{eqnarray}
\dot{\psi}_{4RJ}&=&\left\{\psi_{4RJ},H^{(1)}_{d}\right\}\nonumber\\
&=&\frac{c}{\Delta{x}}\left(\psi_{1RJ}-\psi_{1RJ+1}\cos^{x}_{J}-\psi_{1IJ+1}\sin^{x}_{J}\right)+\frac{c}{\Delta{y}}\left(\tilde\psi_{1IJ}-\tilde\psi_{1IJ-1}\cos^{y}_{J-1}-\tilde\psi_{1RJ-1}\sin^{y}_{J-1}\right)\nonumber\\
& &+\frac{c}{\Delta{z}}\left(\psi_{2RJ}-\psi_{2RJ-1}\cos^{z}_{J-1}+\psi_{2IJ-1}\sin^{z}_{J-1}\right),\label{eq:72}
\end{eqnarray}
\begin{eqnarray}
\dot{\psi}_{1IJ}&=&\left\{\psi_{1IJ},H^{(1)}_{d}\right\}\nonumber\\
&=&\frac{c}{\Delta{x}}\left(\psi_{4IJ-1}\cos^{x}_{J-1}+\psi_{4RJ-1}\sin^{x}_{J-1}-\psi_{4IJ}\right)+\frac{c}{\Delta{y}}\left(\tilde\psi_{4RJ}-\tilde\psi_{4RJ-1}\cos^{y}_{J-1}+\tilde\psi_{4IJ-1}\sin^{y}_{J-1}\right)\nonumber\\
& &+\frac{c}{\Delta{z}}\left(\psi_{3IJ-1}\cos^{z}_{J-1}+\psi_{3RJ-1}\sin^{z}_{J-1}-\psi_{3IJ}\right),\label{eq:73}
\end{eqnarray}
\begin{eqnarray}
\dot{\psi}_{2IJ}&=&\left\{\psi_{2IJ},H^{(1)}_{d}\right\}\nonumber\\
&=&\frac{c}{\Delta{x}}\left(\psi_{3IJ}-\psi_{3IJ+1}\cos^{x}_{J}+\psi_{3RJ+1}\sin^{x}_{J}\right)+\frac{c}{\Delta{y}}\left(\tilde\psi_{3RJ-1}\cos^{y}_{J-1}-\tilde\psi_{3IJ-1}\sin^{y}_{J-1}-\tilde\psi_{3RJ}\right)\nonumber\\
& &+\frac{c}{\Delta{z}}\left(\psi_{4IJ+1}\cos^{z}_{J}-\psi_{4RJ+1}\sin^{z}_{J}-\psi_{4IJ}\right),\label{eq:74}
\end{eqnarray}
\begin{eqnarray}
\dot{\psi}_{3IJ}&=&\left\{\psi_{3IJ},H^{(1)}_{d}\right\}\nonumber\\
&=&\frac{c}{\Delta{x}}\left(\psi_{2IJ-1}\cos^{x}_{J-1}+\psi_{2RJ-1}\sin^{x}_{J-1}-\psi_{2IJ}\right)+\frac{c}{\Delta{y}}\left(\tilde\psi_{2RJ}-\tilde\psi_{2RJ-1}\cos^{y}_{J-1}+\tilde\psi_{2IJ-1}\sin^{y}_{J-1}\right)\nonumber\\
& &+\frac{c}{\Delta{z}}\left(\psi_{1IJ}-\psi_{1IJ+1}\cos^{z}_{J}+\psi_{1RJ+1}\sin^{z}_{J}\right),\label{eq:75}
\end{eqnarray}
\begin{eqnarray}
\dot{\psi}_{4IJ}&=&\left\{\psi_{4IJ},H^{(1)}_{d}\right\}\nonumber\\
&=&\frac{c}{\Delta{x}}\left(\psi_{1IJ}-\psi_{1IJ+1}\cos^{x}_{J}+\psi_{1RJ+1}\sin^{x}_{J}\right)+\frac{c}{\Delta{y}}\left(\tilde\psi_{1RJ-1}\cos^{y}_{J-1}-\tilde\psi_{1IJ-1}\sin^{y}_{J-1}-\tilde\psi_{1RJ}\right)\nonumber\\
& &+\frac{c}{\Delta{z}}\left(\psi_{2IJ}-\psi_{2IJ-1}\cos^{z}_{J-1}-\psi_{2RJ-1}\sin^{z}_{J-1}\right),\label{eq:76}
\end{eqnarray}

\begin{eqnarray}
\dot{\tilde\psi}_{1RJ}&=&\left\{\tilde\psi_{1RJ},H^{(1)}_{d}\right\}\nonumber\\
&=&\frac{c}{\Delta{x}}\left(\tilde\psi_{4RJ}-\tilde\psi_{4RJ+1}\cos^{x}_{J}-\tilde\psi_{4IJ+1}\sin^{x}_{J}\right)+\frac{c}{\Delta{y}}\left(\psi_{4IJ}-\psi_{4IJ+1}\cos^{y}_{J}+\psi_{4RJ+1}\sin^{y}_{J}\right)\nonumber\\
& &+\frac{c}{\Delta{z}}\left(\tilde\psi_{3RJ-1}\cos^{z}_{J-1}-\tilde\psi_{3IJ-1}\sin^{z}_{J-1}-\tilde\psi_{3RJ}\right),\label{eq:d69}
\end{eqnarray}
\begin{eqnarray}
\dot{\tilde\psi}_{2RJ}&=&\left\{\tilde\psi_{2RJ},H^{(1)}_{d}\right\}\nonumber\\
&=&\frac{c}{\Delta{x}}\left(\tilde\psi_{3RJ-1}\cos^{x}_{J-1}-\tilde\psi_{3IJ-1}\sin^{x}_{J-1}-\tilde\psi_{3RJ}\right)+\frac{c}{\Delta{y}}\left(\psi_{3IJ+1}\cos^{y}_{J}-\psi_{3RJ+1}\sin^{y}_{J}-\psi_{3IJ}\right)\nonumber\\
& &+\frac{c}{\Delta{z}}\left(\tilde\psi_{4RJ+1}\cos^{z}_{J}+\tilde\psi_{4IJ+1}\sin^{z}_{J}-\tilde\psi_{4RJ}\right),\label{eq:d70}
\end{eqnarray}
\begin{eqnarray}
\dot{\tilde\psi}_{3RJ}&=&\left\{\tilde\psi_{3RJ},H^{(1)}_{d}\right\}\nonumber\\
&=&\frac{c}{\Delta{x}}\left(\tilde\psi_{2RJ}-\tilde\psi_{2RJ+1}\cos^{x}_{J}-\tilde\psi_{2IJ+1}\sin^{x}_{J}\right)+\frac{c}{\Delta{y}}\left(\psi_{2IJ}-\psi_{2IJ+1}\cos^{y}_{J}+\psi_{2RJ+1}\sin^{y}_{J}\right)\nonumber\\
& &+\frac{c}{\Delta{z}}\left(\tilde\psi_{1RJ}-\tilde\psi_{1RJ+1}\cos^{z}_{J}-\tilde\psi_{1IJ+1}\sin^{z}_{J}\right),\label{eq:d71}
\end{eqnarray}
\begin{eqnarray}
\dot{\tilde\psi}_{4RJ}&=&\left\{\tilde\psi_{4RJ},H^{(1)}_{d}\right\}\nonumber\\
&=&\frac{c}{\Delta{x}}\left(\tilde\psi_{1RJ-1}\cos^{x}_{J-1}-\tilde\psi_{1IJ-1}\sin^{x}_{J-1}-\tilde\psi_{1RJ}\right)+\frac{c}{\Delta{y}}\left(\psi_{1IJ+1}\cos^{y}_{J}-\psi_{1RJ+1}\sin^{y}_{J}-\psi_{1IJ}\right)\nonumber\\
& &+\frac{c}{\Delta{z}}\left(\tilde\psi_{2RJ}-\tilde\psi_{2RJ-1}\cos^{z}_{J-1}+\tilde\psi_{2IJ-1}\sin^{z}_{J-1}\right),\label{eq:d72}
\end{eqnarray}
\begin{eqnarray}
\dot{\tilde\psi}_{1IJ}&=&\left\{\tilde\psi_{1IJ},H^{(1)}_{d}\right\}\nonumber\\
&=&\frac{c}{\Delta{x}}\left(\tilde\psi_{4IJ}-\tilde\psi_{4IJ+1}\cos^{x}_{J}+\tilde\psi_{4RJ+1}\sin^{x}_{J}\right)+\frac{c}{\Delta{y}}\left(\psi_{4RJ+1}\cos^{y}_{J}+\psi_{4IJ+1}\sin^{y}_{J}-\psi_{4RJ}\right)\nonumber\\
& &+\frac{c}{\Delta{z}}\left(\tilde\psi_{3IJ-1}\cos^{z}_{J-1}+\tilde\psi_{3RJ-1}\sin^{z}_{J-1}-\tilde\psi_{3IJ}\right),\label{eq:d73}
\end{eqnarray}
\begin{eqnarray}
\dot{\tilde\psi}_{2IJ}&=&\left\{\tilde\psi_{2IJ},H^{(1)}_{d}\right\}\nonumber\\
&=&\frac{c}{\Delta{x}}\left(\tilde\psi_{3IJ-1}\cos^{x}_{J-1}+\tilde\psi_{3RJ-1}\sin^{x}_{J-1}-\tilde\psi_{3IJ}\right)+\frac{c}{\Delta{y}}\left(\psi_{3RJ}-\psi_{3RJ+1}\cos^{y}_{J}-\psi_{3IJ+1}\sin^{y}_{J}\right)\nonumber\\
& &+\frac{c}{\Delta{z}}\left(\tilde\psi_{4IJ+1}\cos^{z}_{J}-\tilde\psi_{4RJ+1}\sin^{z}_{J}-\tilde\psi_{4IJ}\right),\label{eq:d74}
\end{eqnarray}
\begin{eqnarray}
\dot{\tilde\psi}_{3IJ}&=&\left\{\tilde\psi_{3IJ},H^{(1)}_{d}\right\}\nonumber\\
&=&\frac{c}{\Delta{x}}\left(\tilde\psi_{2IJ}-\tilde\psi_{2IJ+1}\cos^{x}_{J}+\tilde\psi_{2RJ+1}\sin^{x}_{J}\right)+\frac{c}{\Delta{y}}\left(\psi_{2RJ+1}\cos^{y}_{J}+\psi_{2IJ+1}\sin^{y}_{J}-\psi_{2RJ}\right)\nonumber\\
& &+\frac{c}{\Delta{z}}\left(\tilde\psi_{1IJ}-\tilde\psi_{1IJ+1}\cos^{z}_{J}+\tilde\psi_{1RJ+1}\sin^{z}_{J}\right),\label{eq:d75}
\end{eqnarray}
\begin{eqnarray}
\dot{\tilde\psi}_{4IJ}&=&\left\{\tilde\psi_{4IJ},H^{(1)}_{d}\right\}\nonumber\\
&=&\frac{c}{\Delta{x}}\left(\tilde\psi_{1IJ-1}\cos^{x}_{J-1}+\tilde\psi_{1RJ-1}\sin^{x}_{J-1}-\tilde\psi_{1IJ}\right)+\frac{c}{\Delta{y}}\left(\psi_{1RJ}-\psi_{1RJ+1}\cos^{y}_{J}-\psi_{1IJ+1}\sin^{y}_{J}\right)\nonumber\\
& &+\frac{c}{\Delta{z}}\left(\tilde\psi_{2IJ}-\tilde\psi_{2IJ-1}\cos^{z}_{J-1}-\tilde\psi_{2RJ-1}\sin^{z}_{J-1}\right),\label{eq:d76}
\end{eqnarray}

\begin{eqnarray}
\dot{\bm{A}}_{J}&=&\left\{\bm{A}_{J},H^{(1)}_{d}\right\}=0,\label{eq:77}
\end{eqnarray}
\begin{eqnarray}
\dot{Y}_{xJ}&=&\left\{Y_{xJ},H^{(1)}_{d}\right\}\nonumber\\
&=&\frac{e}{2\hbar}\left[\left(\psi_{1RJ+1}\psi_{4RJ}+\psi_{1IJ+1}\psi_{4IJ}+\psi_{2RJ}\psi_{3RJ+1}+\psi_{2IJ}\psi_{3IJ+1}\right.\right.\nonumber\\
& &\left.+\tilde\psi_{1RJ}\tilde\psi_{4RJ+1}+\tilde\psi_{1IJ}\tilde\psi_{4IJ+1}+\tilde\psi_{2RJ+1}\tilde\psi_{3RJ}+\tilde\psi_{2IJ+1}\tilde\psi_{3IJ}\right)\cos^{x}_{J}\nonumber\\
& &+\left(\psi_{1IJ+1}\psi_{4RJ}-\psi_{1RJ+1}\psi_{4IJ}-\psi_{2IJ}\psi_{3RJ+1}+\psi_{2RJ}\psi_{3IJ+1}\right.\nonumber\\
& &\left.\left.-\tilde\psi_{1IJ}\tilde\psi_{4RJ+1}+\tilde\psi_{1RJ}\tilde\psi_{4IJ+1}+\tilde\psi_{2IJ+1}\tilde\psi_{3RJ}-\tilde\psi_{2RJ+1}\tilde\psi_{3IJ}\right)\sin^{x}_{J}\right],\label{eq:78}
\end{eqnarray}
\begin{eqnarray}
\dot{Y}_{yJ}&=&\left\{Y_{yJ},H^{(1)}_{d}\right\}\nonumber\\
&=&\frac{e}{2\hbar}\left[\left(\psi_{1RJ+1}\tilde\psi_{4IJ}-\psi_{1IJ+1}\tilde\psi_{4RJ}-\tilde\psi_{2RJ}\psi_{3IJ+1}+\tilde\psi_{2IJ}\psi_{3RJ+1}\right.\right.\nonumber\\
& &\left.+\tilde\psi_{1RJ}\psi_{4IJ+1}-\tilde\psi_{1IJ}\psi_{4RJ+1}-\psi_{2RJ+1}\tilde\psi_{3IJ}+\psi_{2IJ+1}\tilde\psi_{3RJ}\right)\cos^{y}_{J}\nonumber\\
& &+\left(\psi_{1RJ+1}\tilde\psi_{4RJ}+\psi_{1IJ+1}\tilde\psi_{4IJ}+\tilde\psi_{2RJ}\psi_{3RJ+1}+\tilde\psi_{2IJ}\psi_{3IJ+1}\right.\nonumber\\
& &\left.\left.-\tilde\psi_{1RJ}\psi_{4RJ+1}-\tilde\psi_{1IJ}\psi_{4IJ+1}-\psi_{2RJ+1}\tilde\psi_{3RJ}-\psi_{2IJ+1}\tilde\psi_{3IJ}\right)\sin^{y}_{J}\right],\label{eq:79}
\end{eqnarray}
\begin{eqnarray}
\dot{Y}_{zJ}&=&\left\{Y_{zJ},H^{(1)}_{d}\right\}\nonumber\\
&=&\frac{e}{2\hbar}\left[\left(\psi_{1RJ+1}\psi_{3RJ}+\psi_{1IJ+1}\psi_{3IJ}-\psi_{2RJ}\psi_{4RJ+1}-\psi_{2IJ}\psi_{4IJ+1}\right.\right.\nonumber\\
& &\left.+\tilde\psi_{1RJ+1}\tilde\psi_{3RJ}+\tilde\psi_{1IJ+1}\tilde\psi_{3IJ}-\tilde\psi_{2RJ}\tilde\psi_{4RJ+1}-\tilde\psi_{2IJ}\tilde\psi_{4IJ+1}\right)\cos^{z}_{J}\nonumber\\
& &+\left(\psi_{1IJ+1}\psi_{3RJ}-\psi_{1RJ+1}\psi_{3IJ}-\psi_{2RJ}\psi_{4IJ+1}+\psi_{2IJ}\psi_{4RJ+1}\right.\nonumber\\
& &\left.\left.+\tilde\psi_{1IJ+1}\tilde\psi_{3RJ}-\tilde\psi_{1RJ+1}\tilde\psi_{3IJ}-\tilde\psi_{2RJ}\tilde\psi_{4IJ+1}+\tilde\psi_{2IJ}\tilde\psi_{4RJ+1}\right)\sin^{z}_{J}\right].\label{eq:80}
\end{eqnarray}
Where the Wilson line components $\cos/\sin^{\bm{x}}_{J}$ are defined as,
\begin{eqnarray}
\cos/\sin^{x}_{J}\triangleq\cos/\sin\left(\frac{e}{\hbar{c}}A_{xi+\frac{1}{2},j,k}\Delta{x}\right),\label{eq:81}\\
\cos/\sin^{y}_{J}\triangleq\cos/\sin\left(\frac{e}{\hbar{c}}A_{yi,j+\frac{1}{2},k}\Delta{y}\right),\label{eq:82}\\
\cos/\sin^{z}_{J}\triangleq\cos/\sin\left(\frac{e}{\hbar{c}}A_{zi,j,k+\frac{1}{2}}\Delta{z}\right).\label{eq:83}
\end{eqnarray}
In these equations, the translations of lattice index $J\pm1$ along the relevant gauge connections.

The subsystem generated by $H^{(2)}_{d}$ is given by,
\begin{eqnarray}
\dot{\psi}_{1RJ}/\dot{\tilde\psi}_{1RJ}=\left\{\psi_{1RJ}/\tilde\psi_{1RJ},H^{(2)}_{d}\right\}=\frac{1}{\hbar}\left(e\phi_{J}+mc^2\right)\psi_{1IJ}/\tilde\psi_{1IJ},\label{eq:84}
\end{eqnarray}
\begin{eqnarray}
\dot{\psi}_{2RJ}/\dot{\tilde\psi}_{2RJ}=\left\{\psi_{2RJ}/\tilde\psi_{2RJ},H^{(2)}_{d}\right\}=\frac{1}{\hbar}\left(e\phi_{J}+mc^2\right)\psi_{2IJ}/\tilde\psi_{2IJ},\label{eq:85}
\end{eqnarray}
\begin{eqnarray}
\dot{\psi}_{3RJ}/\dot{\tilde\psi}_{3RJ}=\left\{\psi_{3RJ}/\tilde\psi_{3RJ},H^{(2)}_{d}\right\}=\frac{1}{\hbar}\left(e\phi_{J}-mc^2\right)\psi_{3IJ}/\tilde\psi_{3IJ},\label{eq:86}
\end{eqnarray}
\begin{eqnarray}
\dot{\psi}_{4RJ}/\dot{\tilde\psi}_{4RJ}=\left\{\psi_{4RJ}/\tilde\psi_{4RJ},H^{(2)}_{d}\right\}=\frac{1}{\hbar}\left(e\phi_{J}-mc^2\right)\psi_{4IJ}/\tilde\psi_{4IJ},\label{eq:87}
\end{eqnarray}
\begin{eqnarray}
\dot{\psi}_{1IJ}/\dot{\tilde\psi}_{1IJ}=\left\{\psi_{1IJ}/\tilde\psi_{1IJ},H^{(2)}_{d}\right\}=-\frac{1}{\hbar}\left(e\phi_{J}+mc^2\right)\psi_{1RJ}/\tilde\psi_{1RJ},\label{eq:88}
\end{eqnarray}
\begin{eqnarray}
\dot{\psi}_{2IJ}/\dot{\tilde\psi}_{2IJ}=\left\{\psi_{2IJ}/\tilde\psi_{2IJ},H^{(2)}_{d}\right\}=-\frac{1}{\hbar}\left(e\phi_{J}+mc^2\right)\psi_{2RJ}/\tilde\psi_{2RJ},\label{eq:89}
\end{eqnarray}
\begin{eqnarray}
\dot{\psi}_{3IJ}/\dot{\tilde\psi}_{3IJ}=\left\{\psi_{3IJ}/\tilde\psi_{3IJ},H^{(2)}_{d}\right\}=-\frac{1}{\hbar}\left(e\phi_{J}-mc^2\right)\psi_{3RJ}/\tilde\psi_{3RJ},\label{eq:90}
\end{eqnarray}
\begin{eqnarray}
\dot{\psi}_{4IJ}/\dot{\tilde\psi}_{4IJ}=\left\{\psi_{4IJ}/\tilde\psi_{4IJ},H^{(2)}_{d}\right\}=-\frac{1}{\hbar}\left(e\phi_{J}-mc^2\right)\psi_{4RJ}/\tilde\psi_{4RJ},\label{eq:91}
\end{eqnarray}
\begin{eqnarray}
\dot{\bm{A}}_{J}&=&\left\{\bm{A}_{J},H^{(2)}_{d}\right\}=0,\label{eq:92}
\end{eqnarray}
\begin{eqnarray}
\dot{\bm{Y}}_{J}&=&\left\{\bm{Y}_{J},H^{(2)}_{d}\right\}=0.\label{eq:93}
\end{eqnarray}
The second subsystem can be solved exactly when $\phi_{J}(t)$ is given explicitly.

The subsystem generated by $H^{(3)}_{d}$ is given by,
\begin{eqnarray}
\dot{\psi}_{RJ}/\dot{\tilde\psi}_{RJ}=\left\{\psi_{RJ}/\tilde\psi_{RJ},H^{(3)}_{d}\right\}=0,\label{eq:94}
\end{eqnarray}
\begin{eqnarray}
\dot{\psi}_{IJ}/\dot{\tilde\psi}_{IJ}=\left\{\psi_{IJ}/\tilde\psi_{IJ},H^{(3)}_{d}\right\}=0,\label{eq:95}
\end{eqnarray}
\begin{eqnarray}
\dot{\bm{A}}_{J}=\left\{\bm{A}_{J},H^{(2)}_{d}\right\}=4\pi{c^2}\bm{Y}_{J}-c\left(\bigtriangledown_{d}\phi\right)_{J},\label{eq:96}
\end{eqnarray}
\begin{eqnarray}
\dot{\bm{Y}}_{J}=\left\{\bm{Y}_{J},H^{(2)}_{d}\right\}=-\frac{1}{4\pi}\left(\bigtriangledown^{T}_{d}\times\bigtriangledown_{d}\times\bm{A}\right)_{J}.\label{eq:97}
\end{eqnarray}
Where the discrete operator $\bigtriangledown^{T}_{d}\times\bigtriangledown_{d}\times$ is defined as,
\begin{eqnarray}
\left(\bigtriangledown^{T}_{d}\times\bigtriangledown_{d}\times\bm{A}\right)_{J}\triangleq\frac{1}{2}\frac{\partial}{\partial\bm{A}_{J}}\left[\sum^{M}_{K=1}\left(\bigtriangledown_{d}\times\bm{A}\right)^{2}_{K}\right].\label{eq:98}
\end{eqnarray}

Because the semi-discrete action functional admits $U(1)$ gauge symmetry, the semi-discrete LCFT for Dirac-Maxwell systems 
is gauge invariant. Because there are no explicit lattice coordinates in the semi-discrete action functional, it admits 
translation symmetry, then the semi-discrete LCFT for Dirac-Maxwell systems is translation invariant. We should emphasize 
that when the DEC lattice is fixed, the Lorentz boosts are forbidden, and there are only parity, time-reversal, and discrete 
rotations with angles $(l\pi/2,m\pi/2,n\pi/2)$ exsit in the discrete subgroup of $SO(3,1)$.

\section{Structure-preserving geometric algorithms}
\label{sec:4}

\subsection{Gauge invariant canonical symplectic algorithms}
\label{sec:4-1}
Based on the Hamiltonian splitting method, the three linear canonical subsystems can be solved independently. The solution 
maps of the subsystems will be combined in various ways to give desired structure-preserving geometric algorithms for 
the semi-discrete LCFT.

For the subsystem generated by $H^{(1)}_{d}$, the canonical equations \eqref{eq:69}-\eqref{eq:80} can be rewritten as,
\begin{eqnarray}
\left(\begin{array}{c}
\dot{\psi}_{iR}\\
\dot{\tilde\psi}_{iR}\\
\dot{\psi}_{iI}\\
\dot{\tilde\psi}_{iI}
\end{array}\right)=\Xi\left(\bm{A}\right)\left(\begin{array}{c}
\psi_{iR}\\
\tilde\psi_{iR}\\
\psi_{iI}\\
\tilde\psi_{iI}
\end{array}\right),\label{eq:99}
\end{eqnarray}
\begin{eqnarray}
\left(\begin{array}{c}
\dot{\bm{A}}\\
\dot{\bm{Y}}
\end{array}\right)=\left[\begin{array}{c}
0\\
\bm{\mathcal{J}}\left(\psi_{iR},\tilde\psi_{iR},\psi_{iI},\tilde\psi_{iI},\bm{A}\right)
\end{array}\right].\label{eq:100}
\end{eqnarray}
Where $\Xi(\bm{A})$ is an skew-symmetric matrix, which is also an infinitesimal generator of the symplectic group. 
To preserve the unitary property of the bispinor field, we adopt the symplectic mid-point method for this subsystem, and 
the one step map $M_{D}(\Delta{t}):(\psi_{iR},\tilde\psi_{iR},\bm{A},\psi_{iI},\tilde\psi_{iI},\bm{Y})^{n}\mapsto(\psi_{iR},\tilde\psi_{iR},\bm{A},\psi_{iI},\tilde\psi_{iI},\bm{Y})^{n+1}$ 
can be given by,
\begin{eqnarray}
\left(\begin{array}{c}
\psi_{iR}\\
\tilde\psi_{iR}\\
\psi_{iI}\\
\tilde\psi_{iI}
\end{array}\right)^{n+1}=\left(\begin{array}{c}
\psi_{iR}\\
\tilde\psi_{iR}\\
\psi_{iI}\\
\tilde\psi_{iI}
\end{array}\right)^{n}+\frac{\Delta{t}}{2}\Xi\left(\bm{A}^n\right)\left[\left(\begin{array}{c}
\psi_{iR}\\
\tilde\psi_{iR}\\
\psi_{iI}\\
\tilde\psi_{iI}
\end{array}\right)^{n}+\left(\begin{array}{c}
\psi_{iR}\\
\tilde\psi_{iR}\\
\psi_{iI}\\
\tilde\psi_{iI}
\end{array}\right)^{n+1}\right],\label{eq:101}
\end{eqnarray}
\begin{eqnarray}
\bm{A}^{n+1}=\bm{A}^{n},\label{eq:102}
\end{eqnarray}
\begin{eqnarray}
\bm{Y}^{n+1}=\bm{Y}^{n}+\Delta{t}\bm{\mathcal{J}}\left(\frac{\psi_{iR}^{n}+\psi_{iR}^{n+1}}{2},\frac{\tilde\psi_{iR}^{n}+\tilde\psi_{iR}^{n+1}}{2},\frac{\psi_{iI}^{n}+\psi_{iI}^{n+1}}{2},\frac{\tilde\psi_{iI}^{n}+\tilde\psi_{iI}^{n+1}}{2},\bm{A}^n\right).\label{eq:103}
\end{eqnarray}
Eq.~\eqref{eq:101} is a linear algebraic equation whose solution can be written as,
\begin{eqnarray}
\left(\begin{array}{c}
\psi_{iR}\\
\tilde\psi_{iR}\\
\psi_{iI}\\
\tilde\psi_{iI}
\end{array}\right)^{n+1}=\mathrm{Cay}\left[\Xi\left(\bm{A}^n\right)\frac{\Delta{t}}{2}\right]\left(\begin{array}{c}
\psi_{iR}\\
\tilde\psi_{iR}\\
\psi_{iI}\\
\tilde\psi_{iI}
\end{array}\right)^{n},\label{eq:104}
\end{eqnarray}
\begin{eqnarray}
\mathrm{Cay}\left[\Xi\left(\bm{A}^n\right)\frac{\Delta{t}}{2}\right]=\left[1-\Xi\left(\bm{A}^n\right)\frac{\Delta{t}}{2}\right]^{-1}\left[1+\Xi\left(\bm{A}^n\right)\frac{\Delta{t}}{2}\right].\label{eq:105}
\end{eqnarray}
Where $\mathrm{Cay}(S)$ denotes the Cayley transformation. It is well-known that $\mathrm{Cay}(S)$ is a symplectic rotation 
transformation when $S$ in the Lie algebra of the symplectic rotation group \cite{Hairer1}. As a result, the one step solution map 
$M_{D}(\Delta{t}):(\psi_{iR},\tilde\psi_{iR},\bm{A},\psi_{iI},\tilde\psi_{iI},\bm{Y})^{n}\mapsto(\psi_{iR},\tilde\psi_{iR},\bm{A},\psi_{iI},\tilde\psi_{iI},
\bm{Y})^{n+1}$ is symplectic and unitary for bispinor field. Once $\psi^{n+1}$ and $\tilde\psi^{n+1}$ are known, $\bm{Y}^{n+1}$ can be calculated explicitly 
via Eq.~\eqref{eq:103}. Thus, $M_{D}(\Delta{t}):(\psi_{iR},\tilde\psi_{iR},$ $\bm{A},\psi_{iI},\tilde\psi_{iI},\bm{Y})^{n}\mapsto(\psi_{iR},\tilde\psi_{iR},\bm{A},
\psi_{iI},\tilde\psi_{iI},\bm{Y})^{n+1}$ is a second order symplectic scheme, which also preserves the unitariness of the bispinor field.

For the subsystem generated by $H^{(2)}_{d}$, the canonical equations \eqref{eq:84}-\eqref{eq:93} can be solved exactly 
when $\phi_{J}(t)$ is given explicitly. Because the LCFT is gauge invariant, we can get an explicit $\phi_{J}(t)$ by 
adopting some gauge conditions, such as the temporal gauge $\phi=0$. Here, the one step solution map 
$M_{M}(\Delta{t}):(\psi_{iR},\tilde\psi_{iR},\bm{A},\psi_{iI},\tilde\psi_{iI},\bm{Y})^{n}\mapsto(\psi_{iR},\tilde\psi_{iR},\bm{A},\psi_{iI},\tilde\psi_{iI},
\bm{Y})^{n+1}$ can be given by,
\begin{eqnarray}
\psi^{n+1}_{1RJ}/\tilde\psi^{n+1}_{1RJ}=\psi^{n}_{1RJ}/\tilde\psi^{n}_{1RJ}\cos\left(\int^{\left(n+1\right)\Delta{t}}_{n\Delta{t}}\omega^{+}_{J}\left(\tau\right){\rm{d}}\tau\right)+\psi^{n}_{1IJ}/\tilde\psi^{n}_{1IJ}\sin\left(\int^{\left(n+1\right)\Delta{t}}_{n\Delta{t}}\omega^{+}_{J}\left(\tau\right){\rm{d}}\tau\right),\label{eq:106}
\end{eqnarray}
\begin{eqnarray}
\psi^{n+1}_{2RJ}/\tilde\psi^{n+1}_{2RJ}=\psi^{n}_{2RJ}/\tilde\psi^{n}_{2RJ}\cos\left(\int^{\left(n+1\right)\Delta{t}}_{n\Delta{t}}\omega^{+}_{J}\left(\tau\right){\rm{d}}\tau\right)+\psi^{n}_{2IJ}/\tilde\psi^{n}_{2IJ}\sin\left(\int^{\left(n+1\right)\Delta{t}}_{n\Delta{t}}\omega^{+}_{J}\left(\tau\right){\rm{d}}\tau\right),\label{eq:107}
\end{eqnarray}
\begin{eqnarray}
\psi^{n+1}_{3RJ}/\tilde\psi^{n+1}_{3RJ}=\psi^{n}_{3RJ}/\tilde\psi^{n}_{3RJ}\cos\left(\int^{\left(n+1\right)\Delta{t}}_{n\Delta{t}}\omega^{-}_{J}\left(\tau\right){\rm{d}}\tau\right)+\psi^{n}_{3IJ}/\tilde\psi^{n}_{3IJ}\sin\left(\int^{\left(n+1\right)\Delta{t}}_{n\Delta{t}}\omega^{-}_{J}\left(\tau\right){\rm{d}}\tau\right),\label{eq:108}
\end{eqnarray}
\begin{eqnarray}
\psi^{n+1}_{4RJ}/\tilde\psi^{n+1}_{4RJ}=\psi^{n}_{4RJ}/\tilde\psi^{n}_{4RJ}\cos\left(\int^{\left(n+1\right)\Delta{t}}_{n\Delta{t}}\omega^{-}_{J}\left(\tau\right){\rm{d}}\tau\right)+\psi^{n}_{4IJ}/\tilde\psi^{n}_{4IJ}\sin\left(\int^{\left(n+1\right)\Delta{t}}_{n\Delta{t}}\omega^{-}_{J}\left(\tau\right){\rm{d}}\tau\right),\label{eq:109}
\end{eqnarray}
\begin{eqnarray}
\psi^{n+1}_{1IJ}/\tilde\psi^{n+1}_{1IJ}=\psi^{n}_{1IJ}/\tilde\psi^{n}_{1IJ}\cos\left(\int^{\left(n+1\right)\Delta{t}}_{n\Delta{t}}\omega^{+}_{J}\left(\tau\right){\rm{d}}\tau\right)-\psi^{n}_{1RJ}/\tilde\psi^{n}_{1RJ}\sin\left(\int^{\left(n+1\right)\Delta{t}}_{n\Delta{t}}\omega^{+}_{J}\left(\tau\right){\rm{d}}\tau\right),\label{eq:110}
\end{eqnarray}
\begin{eqnarray}
\psi^{n+1}_{2IJ}/\tilde\psi^{n+1}_{2IJ}=\psi^{n}_{2IJ}/\tilde\psi^{n}_{2IJ}\cos\left(\int^{\left(n+1\right)\Delta{t}}_{n\Delta{t}}\omega^{+}_{J}\left(\tau\right){\rm{d}}\tau\right)-\psi^{n}_{2RJ}/\tilde\psi^{n}_{2RJ}\sin\left(\int^{\left(n+1\right)\Delta{t}}_{n\Delta{t}}\omega^{+}_{J}\left(\tau\right){\rm{d}}\tau\right),\label{eq:111}
\end{eqnarray}
\begin{eqnarray}
\psi^{n+1}_{3IJ}/\tilde\psi^{n+1}_{3IJ}=\psi^{n}_{3IJ}/\tilde\psi^{n}_{3IJ}\cos\left(\int^{\left(n+1\right)\Delta{t}}_{n\Delta{t}}\omega^{-}_{J}\left(\tau\right){\rm{d}}\tau\right)-\psi^{n}_{3RJ}/\tilde\psi^{n}_{3RJ}\sin\left(\int^{\left(n+1\right)\Delta{t}}_{n\Delta{t}}\omega^{-}_{J}\left(\tau\right){\rm{d}}\tau\right),\label{eq:112}
\end{eqnarray}
\begin{eqnarray}
\psi^{n+1}_{4IJ}/\tilde\psi^{n+1}_{4IJ}=\psi^{n}_{4IJ}/\tilde\psi^{n}_{4IJ}\cos\left(\int^{\left(n+1\right)\Delta{t}}_{n\Delta{t}}\omega^{-}_{J}\left(\tau\right){\rm{d}}\tau\right)-\psi^{n}_{4RJ}/\tilde\psi^{n}_{4RJ}\sin\left(\int^{\left(n+1\right)\Delta{t}}_{n\Delta{t}}\omega^{-}_{J}\left(\tau\right){\rm{d}}\tau\right),\label{eq:113}
\end{eqnarray}
\begin{eqnarray}
\left(\begin{array}{c}
\bm{A}\\
\bm{Y}
\end{array}\right)^{n+1}=\left(\begin{array}{c}
\bm{A}\\
\bm{Y}
\end{array}\right)^{n}.\label{eq:114}
\end{eqnarray}
Where $\omega^{\pm}_{J}(\tau)=[e\phi_{J}(\tau)\pm{m}c^2]/\hbar$ are eigen-frequencies of the fermion and anti-fermion. 
Because Eqs.~\eqref{eq:106}-\eqref{eq:114} are rigorous solutions of the second subsystem in one step, the map 
$M_{M}(\Delta{t}):(\psi_{iR},\tilde\psi_{iR},\bm{A},\psi_{iI},\tilde\psi_{iI},\bm{Y})^{n}\mapsto(\psi_{iR},\tilde\psi_{iR},\bm{A},\psi_{iI},\tilde\psi_{iI},
\bm{Y})^{n+1}$ is a symplectic scheme, which also preserves the unitariness of the bispinor field.

For the subsystem generated by $H^{(3)}_{d}$, the canonical equations \eqref{eq:94}-\eqref{eq:97} can be rewritten as,
\begin{eqnarray}
\left(\begin{array}{c}
\dot{\psi}_{iR}\\
\dot{\tilde\psi}_{iR}\\
\dot{\psi}_{iI}\\
\dot{\tilde\psi}_{iI}
\end{array}\right)=0,\label{eq:115}
\end{eqnarray}
\begin{eqnarray}
\left(\begin{array}{c}
\dot{\bm{A}}\\
\dot{\bm{Y}}
\end{array}\right)=Q\left(\begin{array}{c}
\bm{A}\\
\bm{Y}
\end{array}\right).\label{eq:116}
\end{eqnarray}
Where $Q$ is a constant matrix which belongs to the Lie algebra of the symplectic group. We also use the second order 
symplectic mid-point rule for this subsystem, and the one step map $M_{G}(\Delta{t}):(\psi_{iR},\tilde\psi_{iR},\bm{A},\psi_{iI},\tilde\psi_{iI},
\bm{Y})^{n}\mapsto(\psi_{iR},\tilde\psi_{iR},\bm{A},\psi_{iI},\tilde\psi_{iI},\bm{Y})^{n+1}$ can be given by,
\begin{eqnarray}
\left(\begin{array}{c}
\psi_{iR}\\
\tilde\psi_{iR}\\
\psi_{iI}\\
\tilde\psi_{iI}
\end{array}\right)^{n+1}=\left(\begin{array}{c}
\psi_{iR}\\
\tilde\psi_{iR}\\
\psi_{iI}\\
\tilde\psi_{iI}
\end{array}\right)^{n},\label{eq:117}
\end{eqnarray}
\begin{eqnarray}
\left(\begin{array}{c}
\bm{A}\\
\bm{Y}
\end{array}\right)^{n+1}=\mathrm{Cay}\left(Q\frac{\Delta{t}}{2}\right)\left(\begin{array}{c}
\bm{A}\\
\bm{Y}
\end{array}\right)^{n}.\label{eq:118}
\end{eqnarray}
$M_{G}(\Delta{t}):(\psi_{iR},\tilde\psi_{iR},\bm{A},\psi_{iI},\tilde\psi_{iI},\bm{Y})^{n}\mapsto(\psi_{iR},\tilde\psi_{iR},\bm{A},\psi_{iI},
\tilde\psi_{iI},\bm{Y})^{n+1}$ is symplectic. Because the bispinor field does not evolution in the third subsystem, the scheme is unitary.

Because the one step solution maps $M_{D}(\Delta{t})$ and $M_{G}(\Delta{t})$ generated by the Cayley transformation are 
time-symmetric symplectic schemes \cite{Hairer1}, they do not break the $U(1)$ gauge symmetry admitted by the semi-discrete 
LCFT for the Dirac-Maxwell systems. Obviously, the solution map $M_{M}(\Delta{t})$ does not break the symmetries, for it is 
locally rigorous. Given $M_{D}(\Delta{t})$, $M_{M}(\Delta{t})$, and $M_{G}(\Delta{t})$ for $H^{(1)}_{d}$, 
$H^{(2)}_{d}$, and $H^{(3)}_{d}$ subsystems respectively, a first order algorithm for the LCFT can be obtained by composition,
\begin{eqnarray}
M\left(\Delta{t}\right)=M_{G}\left(\Delta{t}\right){\circ}M_{D}\left(\Delta{t}\right){\circ}M_{M}\left(\Delta{t}\right).\label{eq:119}
\end{eqnarray}
A second order symplectic symmetric method can be constructed by the following symmetric composition,
\begin{eqnarray}
M^{2}\left(\Delta{t}\right)=M_{M}\left(\frac{\Delta{t}}{2}\right){\circ}M_{D}\left(\frac{\Delta{t}}{2}\right){\circ}M_{G}\left(\Delta{t}\right){\circ}M_{D}\left(\frac{\Delta{t}}{2}\right){\circ}M_{M}\left(\frac{\Delta{t}}{2}\right).\label{eq:120}
\end{eqnarray}
From a $2l$-th order symplectic symmetric method $M^{2l}(\Delta{t})$, a $2(l+1)$-th order symplectic symmetric method can be 
constructed as \cite{Hairer1},
\begin{eqnarray}
M^{2\left(l+1\right)}\left(\Delta{t}\right)=M^{2l}\left(a_{l}\Delta{t}\right){\circ}M^{2l}\left(b_{l}\Delta{t}\right){\circ}M^{2l}\left(a_{l}\Delta{t}\right),\label{eq:121}
\end{eqnarray}
\begin{eqnarray}
a_{l}=\left(2-2^{1/\left(2l+1\right)}\right)^{-1},~~b_{l}=1-2a_{l}.\label{eq:122}
\end{eqnarray}
Obviously, the high order algorithms for the LCFT are symplectic and unitary structure-preserving, which is also gauge invariant.

\subsection{Doubler, chirality and numerical stability}
\label{sec:4-2}
Fermion doubling problem is induced by a bad numerical dispersion of the free Dirac equation, which introduces pseudo 
cones in the Brillouin zone (BZ) of the lattice \cite{Gourdeau1,Gourdeau3}. As a result, pseudo-fermion modes are excited 
on the lattice and the fermion velocity can faster than $c$. We can prove that the nonphysical doublers are suppressed in 
this work and there are only two degenerate fermion flavors exist on the DEC lattice. There are several equivalent approaches 
to achieve the doubler modes, such as the Poincar\'e-Hopf theorem based topological methd used for searching the chiral 
fermion doublers \cite{Karsten}. Here we directly derive the numerical dispersion or inverse Dirac propagator of the lattice fermion 
constructed in this work. A plane wave mode of the bipsinor field on the DEC lattice can be given by,
\begin{eqnarray}
\psi^{n}_{J}=\frac{1}{\sqrt{2\hbar}}\left[\begin{array}{c}
\check{\psi}_{1R}+i\check{\psi}_{1I}\\
\left(\check{\psi}_{2R}+i\check{\psi}_{2I}\right){\rm{e}}^{i\left(k_{x}\frac{\Delta{x}}{2}+k_{z}\frac{\Delta{z}}{2}\right)}\\
\left(\check{\psi}_{3R}+i\check{\psi}_{3I}\right){\rm{e}}^{ik_{z}\frac{\Delta{z}}{2}}\\
\left(\check{\psi}_{4R}+i\check{\psi}_{4I}\right){\rm{e}}^{ik_{x}\frac{\Delta{x}}{2}}\\
\left(\check{\tilde\psi}_{1R}+i\check{\tilde\psi}_{1I}\right){\rm{e}}^{i\left(k_{x}\frac{\Delta{x}}{2}+k_{y}\frac{\Delta{y}}{2}\right)}\\
\left(\check{\tilde\psi}_{2R}+i\check{\tilde\psi}_{2I}\right){\rm{e}}^{i\left(k_{y}\frac{\Delta{y}}{2}+k_{z}\frac{\Delta{z}}{2}\right)}\\
\left(\check{\tilde\psi}_{3R}+i\check{\tilde\psi}_{3I}\right){\rm{e}}^{i\left(k_{x}\frac{\Delta{x}}{2}+k_{y}\frac{\Delta{y}}{2}+k_{z}\frac{\Delta{z}}{2}\right)}\\
\left(\check{\tilde\psi}_{4R}+i\check{\tilde\psi}_{4I}\right){\rm{e}}^{ik_{y}\frac{\Delta{y}}{2}}
\end{array}\right]{\rm{e}}^{i\left(k_{x}x_{i}+k_{y}y_{j}+k_{z}z_{k}-\omega{t}_{n}\right)}.\label{eq:123}
\end{eqnarray}
Where the superscript $\vee$ means amplitude. By substituting Eq.~\eqref{eq:123} into the $H^{(1)}_{d}$ subsystem, and 
setting the $U(1)$ gauge field $A^{\mu}=0$, we obtain the numerical dispersion of a mass free fermion as,
\begin{eqnarray}
\frac{\tan^{2}\left(\frac{\omega\Delta{t}}{2}\right)}{\left(\frac{\Delta{t}}{2}\right)^2}=c^2\left[\frac{\sin^{2}\left(\frac{k_{x}\Delta{x}}{2}\right)}{\left(\frac{\Delta{x}}{2}\right)^2}+\frac{\sin^{2}\left(\frac{k_{y}\Delta{y}}{2}\right)}{\left(\frac{\Delta{y}}{2}\right)^2}+\frac{\sin^{2}\left(\frac{k_{z}\Delta{z}}{2}\right)}{\left(\frac{\Delta{z}}{2}\right)^2}\right].\label{eq:124}
\end{eqnarray}
Once $(\Delta{t},\Delta\bm{x})\to(0,0)$, the continuum limit of Eq.~\eqref{eq:124} can be obtained as,
\begin{eqnarray}
\omega=\pm{c}\sqrt{k^{2}_{x}+k^{2}_{y}+k^{2}_{z}},\label{eq:125}
\end{eqnarray}
which is the rigorous dispersion of mass free fermions. With the de Broglie relation $E=\hbar\omega$, we obtain the 
energy spectrum,
\begin{eqnarray}
E=\pm\frac{2\hbar}{\Delta{t}}\arctan\left[\frac{c\Delta{t}}{2}\sqrt{\frac{\sin^{2}\left(\frac{k_{x}\Delta{x}}{2}\right)}{\left(\frac{\Delta{x}}{2}\right)^2}+\frac{\sin^{2}\left(\frac{k_{y}\Delta{y}}{2}\right)}{\left(\frac{\Delta{y}}{2}\right)^2}+\frac{\sin^{2}\left(\frac{k_{z}\Delta{z}}{2}\right)}{\left(\frac{\Delta{z}}{2}\right)^2}}\right].\label{eq:126}
\end{eqnarray}
Where the momentum $\bm{k}$ is restricted in the lattice BZ $\bm{k}\in[-\pi/\Delta\bm{x},\pi/\Delta\bm{x}]$. 
Eq.~\eqref{eq:126} is also the numerical dispersion of the staggered fermion \cite{Doel}. It shows that there is only one cone 
centered at $\bm{k}=0$ in the lattice BZ, and the Kramers and time-reversal symmetries are preserved. We should emphasis 
that due to the Dirac field is double sampled into eight complex field components, there are two degenerate fermion flavors, 
and the doubler degree is two. For free Dirac field these two flavors are uncoupled. When the gauge field is non-trivial, 
flavor mixing emerges, as the two flavors interact with different-valued gauge field on the DEC lattice. In this work, the 
strong gauge field will be treated as a classical field without fluctuation, then the flavor mixing is tolerable. The 
contribution of flavor mixing to gauge field is reflected in the $\psi\tilde\psi$ and $\tilde\psi\psi$ terms in 
Eqs.~\eqref{eq:78}-\eqref{eq:80}.

The Nielsen-Ninomiya no-go theorem forbids perfect fermions on a regular lattice. The reduce of fermion doublers in our 
work is achieved by giving up on partial chiral symmetry. In the Dirac representation, the chiral operator can be given 
by $\gamma^{5}=i\gamma^{0}\gamma^{1}\gamma^{2}\gamma^{3}$, and the chiral symmetry of free fermions can be defined by an 
anti-commutator $\{\slashed{D},\gamma^{5}\}=0$. Based on the Hamiltonian \eqref{eq:66} of semi-discrete lattice field 
theory, the spatial Dirac contraction of discrete gauge covariant derivative is obtained as,
\begin{eqnarray}
\slashed{D}_{d}=\left(\begin{array}{cccccccc}
0 & 0 & D^{>}_{z} & D^{>}_{x} & 0 & 0 & 0 & -iD^{>}_{y}\\
0 & 0 & D^{<}_{x} & -D^{<}_{z} & 0 & 0 & iD^{>}_{y} & 0\\
-D^{<}_{z} & -D^{>}_{x} & 0 & 0 & 0 & iD^{>}_{y} & 0 & 0\\
-D^{<}_{x} & D^{>}_{z} & 0 & 0 & -iD^{>}_{y} & 0 & 0 & 0\\
0 & 0 & 0 & -iD^{<}_{y} & 0 & 0 & D^{>}_{z} & D^{<}_{x}\\
0 & 0 & iD^{<}_{y} & 0 & 0 & 0 & D^{>}_{x} & -D^{<}_{z}\\
0 & iD^{<}_{y} & 0 & 0 & -D^{<}_{z} & -D^{<}_{x} & 0 & 0\\
-iD^{<}_{y} & 0 & 0 & 0 & -D^{>}_{x} & D^{>}_{z} & 0 & 0
\end{array}\right).\label{eq:chiral1}
\end{eqnarray}
Then we can derive the chiral anti-commutator with discrete derivative as,
\begin{eqnarray}
\left\{\slashed{D}_{d},\left(\begin{array}{cc}
\gamma^{5} & 0_{4}\\
0_{4} & \gamma^{5}\end{array}\right)\right\}\left(\begin{array}{c}
\psi_{J}\\
\tilde\psi_{J}\end{array}\right)=\left(D^{>}_{z}-D^{<}_{z}\right)\left(\begin{array}{c}
\psi_{J}\\
\tilde\psi_{J}\end{array}\right).\label{eq:chiral2}
\end{eqnarray}
The residual term in Eq.~\eqref{eq:chiral2} shows that the break of chiral symmetry is only induced by the non-balance 
between z-direction discrete gauge covariant derivatives. This partial chirality is gauge dependent and the complete chiral 
symmetry can naturally recover in the continuous limit. We indicate that the broken symmetry comes from the non-commutation 
between discrete derivative and continuous chiral operator, and we can define a complete discrete chirality on the DEC lattice 
by introducing a pair of discrete chiral operators,
\begin{eqnarray}
\Gamma^{\pm}=CS^{\pm}\circ\left(\begin{array}{cc}
\gamma^{5} & 0_{4}\\
0_{4} & \gamma^{5}\end{array}\right),~~\left\{\slashed{D}_{d},\Gamma^{\pm}\right\}\left(\begin{array}{c}
\psi_{J}\\
\tilde\psi_{J}\end{array}\right)=0.\label{eq:chiralneo1}
\end{eqnarray}
Where the shifting operators $CS^+$ translates lattice indexes of the 2nd, 3rd, 6th and 7th bispinor components as $J{\mapsto}J+1$ 
along z-direction, and $CS^-$ translates lattice indexes of the 1st, 4th, 5th and 8th bispinor components as $J{\mapsto}J-1$ 
along z-direction. Then $\Gamma^{\pm}$ are inverse operators.

By using the discrete chiral operators, we can define the discrete chiral fermions as,
\begin{eqnarray}
\left(\begin{array}{c}
\psi_{J}\\
\tilde\psi_{J}\end{array}\right)^{L}=P_{L}\left(\begin{array}{c}
\psi_{J}\\
\tilde\psi_{J}\end{array}\right),~\left(\begin{array}{c}
\psi_{J}\\
\tilde\psi_{J}\end{array}\right)^{R}=P_{R}\left(\begin{array}{c}
\psi_{J}\\
\tilde\psi_{J}\end{array}\right),\label{eq:chiral3}
\end{eqnarray}
\begin{eqnarray}
\left(\bar{\psi}_{J},\bar{\tilde\psi}_{J}\right)^{L}=\left(\bar{\psi}_{J},\bar{\tilde\psi}_{J}\right)P_{R},~\left(\bar{\psi}_{J},\bar{\tilde\psi}_{J}\right)^{R}=\left(\bar{\psi}_{J},\bar{\tilde\psi}_{J}\right)P_{L}.\label{eq:chiral4}
\end{eqnarray}
Where the chiral projection operators are given by $P_{L}=\frac{1}{2}(1-\Gamma^+)$ and $P_{R}=\frac{1}{2}(1+\Gamma^-)$. 
By substituting Eqs.~\eqref{eq:chiral3}-\eqref{eq:chiral4} into Eq.~\eqref{eq:66}, we obtain,
\begin{eqnarray}
H^{(1)}_{d}=-i\hbar{c}\sum^{M}_{J=1}\left[\left(\bar{\psi}_{J},\bar{\tilde\psi}_{J}\right)^{L}\slashed{D}_{d}\left(\begin{array}{c}
\psi_{J}\\
\tilde\psi_{J}\end{array}\right)^{L}+\left(\bar{\psi}_{J},\bar{\tilde\psi}_{J}\right)^{R}\slashed{D}_{d}\left(\begin{array}{c}
\psi_{J}\\
\tilde\psi_{J}\end{array}\right)^{R}\right]\frac{\Delta{V}}{2}.\label{eq:chiral5}
\end{eqnarray}
Then the discrete dynamical equations of free chiral fermions are obtained by the discrete variational principle,
\begin{eqnarray}
\left(\begin{array}{c}
\dot{\psi}_{J}\\
\dot{\tilde\psi}_{J}\end{array}\right)^{L}=-c\slashed{D}_{d}\left(\begin{array}{c}
\psi_{J}\\
\tilde\psi_{J}\end{array}\right)^{L},~~\left(\begin{array}{c}
\dot{\psi}_{J}\\
\dot{\tilde\psi}_{J}\end{array}\right)^{R}=-c\slashed{D}_{d}\left(\begin{array}{c}
\psi_{J}\\
\tilde\psi_{J}\end{array}\right)^{R}.\label{eq:chiral6}
\end{eqnarray}
We should emphasize that the complete discrete chirality is only a lattice analogue, and the free chiral fermions defined by 
continuous chiral operator are coupled on the DEC lattice, where the nonphysical coupling will vanish in the continuous limit.

The numerical dispersion \eqref{eq:124} shows that the energy $\hbar\omega$ is always real for arbitrary lattice 
periods $(\Delta{t},\Delta\bm{x})$, which means that the solution map $M_{D}(\Delta{t})$ of $H^{(1)}_{d}$ subsystem 
is locally unconditional stable. The $H^{(2)}_{d}$ subsystem is solved exactly, which means the solution map 
$M_{M}(\Delta{t})$ is unconditional stable. The solution map $M_{G}(\Delta{t})$ of $H^{(3)}_{d}$ subsystem equals to 
the Crank-Nicolson FDTD method for Maxwell's equations, which is a well-known unconditional stable CED scheme \cite{GSun}. 
As a result, the high order solution map $M(\Delta{t})$ for the LCFT is locally unconditional stable. 

To implement the algorithms, Jacobian inversions are needed, as the Cayley transformation brings several linear algebraic 
equations. The Krylov subspace theory provides us with many efficient linear solvers, such as the generalized minimum 
residual (GMRES) method, the incomplete Cholesky conjugate gradient (ICCG) method, and the biconjugate gradient stabilized 
(BICGSTAB) method, which can be used to solve the large sparse matrix equation. Based on these efficient linear solvers, 
the algorithms can be conveniently implemented via standard parallel strategies.

\subsection{Field quantization}
\label{sec:4-3}
In classical statistic regime, the quantization of Dirac field can be simulated via a statistically quantization-equivalent 
ensemble, which can reconstruct pairs of anticommuting creation and annihilation operators in a statistic sense. The low-cost 
fermions strategy is such an ensemble model widely used in real-time LGT simulations \cite{Borsanyi,Hebenstreit1}. Along this 
approach, we introduce a unified ensemble model of vacuum and plasmas into the LCFT to realize real-time LCFT simulations for 
abundant SFQED and RQP phenomena. Based on the standard canonical quantization procedure, the Dirac bispinor field can be 
quantized as \cite{Weinberg,Peskin,Greiner},
\begin{eqnarray}
\psi\left(\bm{x},t\right)=\frac{1}{\left(2\pi\hbar\right)^3}\int^{\infty}_{-\infty}\sum_{s}\left[a^{s}_{\bm{p}}u_{s}\left(\bm{p}\right){\rm{e}}^{-i\frac{Et}{\hbar}}+b^{s+}_{-\bm{p}}v_{s}\left(-\bm{p}\right){\rm{e}}^{i\frac{Et}{\hbar}}\right]{\rm{e}}^{i\frac{\bm{p}\cdot\bm{x}}{\hbar}}{\rm{d}}^{3}p,\label{eq:127}
\end{eqnarray}
\begin{eqnarray}
\bar{\psi}\left(\bm{x},t\right)=\frac{1}{\left(2\pi\hbar\right)^3}\int^{\infty}_{-\infty}\sum_{s}\left[b^{s}_{\bm{p}}\bar{v}_{s}\left(\bm{p}\right){\rm{e}}^{-i\frac{Et}{\hbar}}+a^{s+}_{-\bm{p}}\bar{u}_{s}\left(-\bm{p}\right){\rm{e}}^{i\frac{Et}{\hbar}}\right]{\rm{e}}^{i\frac{\bm{p}\cdot\bm{x}}{\hbar}}{\rm{d}}^{3}p.\label{eq:128}
\end{eqnarray}
Where $\{a^{s}_{\bm{p}},a^{s'+}_{\bm{p'}}\}=(2\pi\hbar)^3\delta(\bm{p}-\bm{p'})\delta_{s,s'}$ and 
$\{b^{s}_{\bm{p}},b^{s'+}_{\bm{p'}}\}=(2\pi\hbar)^3\delta(\bm{p}-\bm{p'})\delta_{s,s'}$ are creation and annihilation 
operators for fermions and antifermions respectively, with spin index $s=\pm\frac{1}{2}$. $u_{s}\left(\bm{p}\right)$ and 
$v_{s}\left(\bm{p}\right)$ are relevant eigen spinors of free particles, which can be normalized as,
\begin{eqnarray}
u_{\frac{1}{2}}\left(\bm{p}\right)=\sqrt{\frac{E+mc^2}{2E}}\left(\begin{array}{c}
U^{\uparrow}\\
\frac{c\left(\hat\sigma\cdot\bm{p}\right)}{E+mc^2}U^{\uparrow}
\end{array}\right),u_{-\frac{1}{2}}\left(\bm{p}\right)=\sqrt{\frac{E+mc^2}{2E}}\left(\begin{array}{c}
U^{\downarrow}\\
-\frac{c\left(\hat\sigma\cdot\bm{p}\right)}{E+mc^2}U^{\downarrow}
\end{array}\right),\label{eq:129}
\end{eqnarray}
\begin{eqnarray}
v_{\frac{1}{2}}\left(\bm{p}\right)=\sqrt{\frac{E+mc^2}{2E}}\left(\begin{array}{c}
\frac{c\left(\hat\sigma\cdot\bm{p}\right)}{E+mc^2}U^{\uparrow}\\
U^{\uparrow}
\end{array}\right),v_{-\frac{1}{2}}\left(\bm{p}\right)=\sqrt{\frac{E+mc^2}{2E}}\left(\begin{array}{c}
-\frac{c\left(\hat\sigma\cdot\bm{p}\right)}{E+mc^2}U^{\downarrow}\\
U^{\downarrow}
\end{array}\right).\label{eq:130}
\end{eqnarray}
Where $U^{s}$ are normalized Pauli spinors $U^{s+}U^{s'}=\delta_{s,s'}$. $E$ is the positive definite energy norm of an on 
shell fermion. Then the orthogonality relations $u_{s}^{+}(\bm{p})u_{s'}(\bm{p})=v_{s}^{+}(\bm{p})v_{s'}(\bm{p})=\delta_{s,s'}$ 
and $u_{s}^{+}(\bm{p})v_{s'}(-\bm{p})=0$ can be obtained directly.

To achieve the Fermi-Dirac statistics via a classical Dirac field ensemble, we replace the creation and annihilation 
operators with a class of stochastic variables and reconstruct a pair of stochastic Dirac spinors as,
\begin{eqnarray}
\psi_{M}\left(\bm{x},0\right)=\frac{1}{\left(2\pi\hbar\right)^3}\int^{\infty}_{-\infty}\sum_{s}\frac{1}{\sqrt{2}}\left[\xi^{s}\left(\bm{p}\right)u_{s}\left(\bm{p}\right)+\eta^{s}\left(\bm{p}\right)v_{s}\left(-\bm{p}\right)\right]{\rm{e}}^{i\frac{\bm{p}\cdot\bm{x}}{\hbar}}{\rm{d}}^{3}p,\label{eq:131}
\end{eqnarray}
\begin{eqnarray}
\psi_{F}\left(\bm{x},0\right)=\frac{1}{\left(2\pi\hbar\right)^3}\int^{\infty}_{-\infty}\sum_{s}\frac{1}{\sqrt{2}}\left[\xi^{s}\left(\bm{p}\right)u_{s}\left(\bm{p}\right)-\eta^{s}\left(\bm{p}\right)v_{s}\left(-\bm{p}\right)\right]{\rm{e}}^{i\frac{\bm{p}\cdot\bm{x}}{\hbar}}{\rm{d}}^{3}p.\label{eq:132}
\end{eqnarray}
Where the gender subscripts indicate (M) male and (F) female. $\xi^{s}\left(\bm{p}\right)$ and $\eta^{s}\left(\bm{p}\right)$ are 
stochastic variables which are sampled according to the ensemble average relations $\left<\xi^{s}(\bm{p})\xi^{s'*}(\bm{p'})\right>=
(2\pi\hbar)^3(1-2n^{s+}_{\bm{p}})\delta(\bm{p}-\bm{p'})\delta_{s,s'}$ and $\left<\eta^{s}(\bm{p})\eta^{s'*}(\bm{p'})\right>=
(2\pi\hbar)^3(1-2n^{s-}_{\bm{p}})\delta(\bm{p}-\bm{p'})\delta_{s,s'}$, where the other correlators vanish. 
\footnote{$\left<\cdot\right>\equiv\frac{1}{N_{e}}\sum\cdot$, where the ensemble capacity $N_{e}$ is the number of systems in a 
given ensemble. The minimum $N_{e}$ should be larger than the lattice degree of freedom, which guarantees all the lattice modes 
can be sampled.} The ensemble model reconstructs the anticommutation relations of the fermion ladder operators in classical statistic 
regime. To describe the Dirac vacuum, we can assume these stochastic variables admit the same amplitude distribution 
$(\sqrt{(2\pi\hbar)^3-\sigma^{2}},\sigma^{2})$ and uniform phase distribution ${\rm{U}}(-\pi,\pi]$ in the momentum space. 
To describe a single specie and spin polarized plasma background, we can set distributions of the stochastic variables 
$\xi(\bm{p})$ and $\eta(\bm{p})$ admit the Pauli blocking density. Then the ensemble averaged bilinear covariant gives 
rise to the background plasma density. Eqs.~\eqref{eq:131}-\eqref{eq:132} ensure the statistically equivalence between 
ensemble model and field quantization for the Dirac vacuum and non-trivial plasma backgrounds \cite{Hebenstreit1,YShi2}. 
On a DEC lattice, the stochastic Dirac spinors are discreted as,
\begin{eqnarray}
\psi_{MJ}^{0}=\frac{1}{V}\sum_{\bm{p}}\sum_{s}\frac{1}{\sqrt{2}}\left[\xi^{s}\left(\bm{p}\right)u_{s}\left(\bm{p}\right)+\eta^{s}\left(\bm{p}\right)v_{s}\left(-\bm{p}\right)\right]{\rm{e}}^{i\frac{\bm{p}\cdot\bm{x}_{J}}{\hbar}},\label{eq:dss1}
\end{eqnarray}
\begin{eqnarray}
\psi_{FJ}^{0}=\frac{1}{V}\sum_{\bm{p}}\sum_{s}\frac{1}{\sqrt{2}}\left[\xi^{s}\left(\bm{p}\right)u_{s}\left(\bm{p}\right)-\eta^{s}\left(\bm{p}\right)v_{s}\left(-\bm{p}\right)\right]{\rm{e}}^{i\frac{\bm{p}\cdot\bm{x}_{J}}{\hbar}}.\label{eq:dss2}
\end{eqnarray}
Then the non-trivial correlators are $\left<\xi^{s}(\bm{p})\xi^{s'*}(\bm{p'})\right>=V(1-2n^{s+}_{\bm{p}})\delta(\bm{p}-\bm{p'})\delta_{s,s'}$ 
and $\left<\eta^{s}(\bm{p})\eta^{s'*}(\bm{p'})\right>=V(1-2n^{s-}_{\bm{p}})\delta(\bm{p}-\bm{p'})\delta_{s,s'}$, where the 
discrete momentum space is given by $\bm{p}\in[-\pi\hbar/\Delta\bm{x},\pi\hbar/\Delta\bm{x}]$ with lattice spacing $2\pi\hbar/\Delta\bm{x}N_{\bm{x}}$.

Based on this ensemble model, the Lagrangian density Eq.~\eqref{eq:1} on $TG=(\psi_{MR},\psi_{MI},$ $\psi_{FR},\psi_{FI},\bm{A},
\phi,\dot{\psi}_{MR},\dot{\psi}_{MI},\dot{\psi}_{FR},\dot{\psi}_{FI},\dot{\bm{A}},\dot{\phi})$ can be rewritten as,
\begin{eqnarray}
\mathcal{L}=-\frac{1}{2}:\left<\bar{\psi}_{M}\left(i\hbar{c}\slashed{D}-mc^2\right)\psi_{F}+g.c.\right>:-\frac{1}{16\pi}\mathcal{F}^{\mu\nu}\mathcal{F}_{\mu\nu}.\label{eq:133}
\end{eqnarray}
Where the gender conjugate $g.c.$ means commutation between the stochastic bispinor pairs, and $:\cdot:$ means normal product. 
Then the Hamiltonian functional on $T^{*}G=(\psi_{MR},\psi_{MI},$ $\psi_{FR},\psi_{FI},\bm{A},-\psi_{FI}/4N_{e},
\psi_{FR}/4N_{e},-\psi_{MI}/4N_{e},\psi_{MR}/4N_{e},\bm{Y})$ can be obtained via the Legendre transformation,
\begin{eqnarray}
H=\int_{V}\left\{-\frac{1}{2}:\left<\psi_{M}^{+}\hat{H}\psi_{F}+g.c.\right>:+\frac{1}{8\pi}\left[16\pi^2c^2\bm{Y}^2+\left(\bigtriangledown\times\bm{A}\right)^2-8\pi{c}\bm{Y}\cdot\bigtriangledown\phi\right]\right\}{\rm{d}}^{3}x.\label{eq:134}
\end{eqnarray}

The Poisson algebra of this ensemble model can be constructed as,
\begin{eqnarray}
\left\{F,G\right\}&=&\int_{V}\left[\left(\frac{\delta{F}}{\delta\psi_{MR}}\right)^{T},\left(\frac{\delta{F}}{\delta\psi_{FR}}\right)^{T},\frac{\delta{F}}{\delta\bm{A}},\left(\frac{\delta{F}}{\delta\psi_{FI}}\right)^{T},\left(\frac{\delta{F}}{\delta\psi_{MI}}\right)^{T},\frac{\delta{F}}{\delta\bm{Y}}\right]\Omega^{-1}\nonumber\\
& &\cdot\left[\left(\frac{\delta{G}}{\delta\psi_{MR}}\right)^{T},\left(\frac{\delta{G}}{\delta\psi_{FR}}\right)^{T},\frac{\delta{G}}{\delta\bm{A}},\left(\frac{\delta{G}}{\delta\psi_{FI}}\right)^{T},\left(\frac{\delta{G}}{\delta\psi_{MI}}\right)^{T},\frac{\delta{G}}{\delta\bm{Y}}\right]^{T}{\rm{d}}^{3}x\nonumber\\
&=&\int_{V}\left[2N^{2}_{e}\sum^{4}_{i=1}\left<\frac{\delta{F}}{\delta\psi_{FiI}}\frac{\delta{G}}{\delta\psi_{MiR}}-\frac{\delta{G}}{\delta\psi_{FiI}}\frac{\delta{F}}{\delta\psi_{MiR}}+g.c.\right>+\frac{\delta{F}}{\delta\bm{A}}\cdot\frac{\delta{G}}{\delta\bm{Y}}-\frac{\delta{G}}{\delta\bm{A}}\cdot\frac{\delta{F}}{\delta\bm{Y}}\right]{\rm{d}}^{3}x,\label{eq:136}
\end{eqnarray}
where $N_{e}$ is the ensemble capacity, and the canonical symplectic 2-form field is given by,
\begin{eqnarray}
\Omega=\frac{1}{2}\sum^{4}_{i=1}\left<\bm{{\rm{d}}}\psi_{MiR}\wedge\bm{{\rm{d}}}\psi_{FiI}+g.c.\right>+\sum^{3}_{i=1}\bm{{\rm{d}}}Y_{i}\wedge\bm{{\rm{d}}}A_{i}.\label{eq:135}
\end{eqnarray}
As a result of the general variational principle, the stochastic Dirac spinors $\psi_{F}$ and $\psi_{M}$ admit the same 
dynamical equations \eqref{eq:23}-\eqref{eq:30}, and the Dirac current density $\bm{J}$ in Eq.~\eqref{eq:32} is replaced 
by the ensemble current form,
\begin{eqnarray}
\left<\bm{J}\right>=-\frac{ec}{2}:\left<\psi_{M}^{+}\bm{\alpha}\psi_{F}+g.c.\right>:.\label{eq:137}
\end{eqnarray}

The canonical symplectic structure-preserving geometric algorithms constructed in this work can be equipped with the 
ensemble model directly. It means that the calculations of the stochastic bispinors $\psi_{F}$ and $\psi_{M}$ using the 
updating schemes of $\psi$, and the discrete ensemble current density used for updating gauge field is given by,
\begin{eqnarray}
\left<\mathcal{J}_{x}\right>&=&-\frac{ec}{4\hbar}:\left<\left[\left(\psi_{M1RJ+1}\psi_{F4RJ}+\psi_{M1IJ+1}\psi_{F4IJ}+\psi_{M2RJ}\psi_{F3RJ+1}+\psi_{M2IJ}\psi_{F3IJ+1}\right.\right.\right.\nonumber\\
& &\left.+\tilde\psi_{M1RJ}\tilde\psi_{F4RJ+1}+\tilde\psi_{M1IJ}\tilde\psi_{F4IJ+1}+\tilde\psi_{M2RJ+1}\tilde\psi_{F3RJ}+\tilde\psi_{M2IJ+1}\tilde\psi_{F3IJ}\right)\cos^{x}_{J}\nonumber\\
& &+\left(\psi_{M1IJ+1}\psi_{F4RJ}-\psi_{M1RJ+1}\psi_{F4IJ}-\psi_{M2IJ}\psi_{F3RJ+1}+\psi_{M2RJ}\psi_{F3IJ+1}\right.\nonumber\\
& &\left.\left.\left.-\tilde\psi_{M1IJ}\tilde\psi_{F4RJ+1}+\tilde\psi_{M1RJ}\tilde\psi_{F4IJ+1}+\tilde\psi_{M2IJ+1}\tilde\psi_{F3RJ}-\tilde\psi_{M2RJ+1}\tilde\psi_{F3IJ}\right)\sin^{x}_{J}\right]+g.c.\right>:,\label{eq:138}
\end{eqnarray}
\begin{eqnarray}
\left<\mathcal{J}_{y}\right>&=&-\frac{ec}{4\hbar}:\left<\left[\left(\psi_{M1RJ+1}\tilde\psi_{F4IJ}-\psi_{M1IJ+1}\tilde\psi_{F4RJ}-\tilde\psi_{M2RJ}\psi_{F3IJ+1}+\tilde\psi_{M2IJ}\psi_{F3RJ+1}\right.\right.\right.\nonumber\\
& &\left.+\tilde\psi_{M1RJ}\psi_{F4IJ+1}-\tilde\psi_{M1IJ}\psi_{F4RJ+1}-\psi_{M2RJ+1}\tilde\psi_{F3IJ}+\psi_{M2IJ+1}\tilde\psi_{F3RJ}\right)\cos^{y}_{J}\nonumber\\
& &+\left(\psi_{M1RJ+1}\tilde\psi_{F4RJ}+\psi_{M1IJ+1}\tilde\psi_{F4IJ}+\tilde\psi_{M2RJ}\psi_{F3RJ+1}+\tilde\psi_{M2IJ}\psi_{F3IJ+1}\right.\nonumber\\
& &\left.\left.\left.-\tilde\psi_{M1RJ}\psi_{F4RJ+1}-\tilde\psi_{M1IJ}\psi_{F4IJ+1}-\psi_{M2RJ+1}\tilde\psi_{F3RJ}-\psi_{M2IJ+1}\tilde\psi_{F3IJ}\right)\sin^{y}_{J}\right]+g.c.\right>:,\label{eq:139}
\end{eqnarray}
\begin{eqnarray}
\left<\mathcal{J}_{z}\right>&=&-\frac{ec}{4\hbar}:\left<\left[\left(\psi_{M1RJ+1}\psi_{F3RJ}+\psi_{M1IJ+1}\psi_{F3IJ}-\psi_{M2RJ}\psi_{F4RJ+1}-\psi_{M2IJ}\psi_{F4IJ+1}\right.\right.\right.\nonumber\\
& &\left.+\tilde\psi_{M1RJ+1}\tilde\psi_{F3RJ}+\tilde\psi_{M1IJ+1}\tilde\psi_{F3IJ}-\tilde\psi_{M2RJ}\tilde\psi_{F4RJ+1}-\tilde\psi_{M2IJ}\tilde\psi_{F4IJ+1}\right)\cos^{z}_{J}\nonumber\\
& &+\left(\psi_{M1IJ+1}\psi_{F3RJ}-\psi_{M1RJ+1}\psi_{F3IJ}-\psi_{M2RJ}\psi_{F4IJ+1}+\psi_{M2IJ}\psi_{F4RJ+1}\right.\nonumber\\
& &\left.\left.\left.+\tilde\psi_{M1IJ+1}\tilde\psi_{F3RJ}-\tilde\psi_{M1RJ+1}\tilde\psi_{F3IJ}-\tilde\psi_{M2RJ}\tilde\psi_{F4IJ+1}+\tilde\psi_{M2IJ}\tilde\psi_{F4RJ+1}\right)\sin^{z}_{J}\right]+g.c.\right>:.\label{eq:140}
\end{eqnarray}
To keep the physical constraints, the sampling of gauge field configuration should adapt the self-consistent field 
condition at initial time. We should emphasize that the gauge field can be treated as a classical field without quantum 
fluctuation only valid in very high occupation states. The SFQED and RQP phenomena always satisfy this condition. When 
it comes to weak field problems, e.g. some quantum optics and quantum electronics phenomena, the strong-field condition 
breaks and the quantization of gauge field should also be taken into consideration.

\section{Real-Time LCFT Simulations}
\label{sec:5}

\subsection{Energy spectra}
\label{sec:5-1}
To verify the canonical symplectic structure-preserving geometric algorithms constructed in this work, we implement the 
code to obtain a class of numerical energy spectra of the Dirac-Maxwell theory based LCFT. As benchmarks, the analytical 
dispersion relations of linearized scalar QED are introduced to compare with these numerical energy spectra \cite{Eliasson,YShi1}. 
The dispersion relation of free Dirac fermions are given by \cite{Weinberg},
\begin{eqnarray}
\omega=\pm\sqrt{c^2\bm{k}^2+\frac{m^2c^4}{\hbar^2}},\label{eq:141}
\end{eqnarray}
which means that there are two fermion modes sharing a gap $2mc^2/\hbar$. The Dirac double-cone of positive and negative 
states is a basic property of relativistic particles. If there is no strong background magnetic field, e.g. the vacuum 
and unmagnetized plasmas, the Klein-Gordon-Maxwell (KGM) theory based scalar QED is a good toy model of the Dirac-Maxwell 
fields theory, both of which admit the same branches of linearized dispersion relations. The $1/2$-spin effects only modify 
the mode structures. The tree-level electromagnetic mode dispersion relation of the scalar QED can be given by \cite{YShi1},
\begin{eqnarray}
\omega^2=c^2\bm{k}^2+\omega_{p}^{2}.\label{eq:142}
\end{eqnarray}
Where $\omega_{p}$ is the plasma frequency of background fermions. When it comes to the vacuum, $\omega_{p}=0$ and 
Eq.~\eqref{eq:142} reduces to the light cone. The tree-level dispersion relation of electrostatic mode can be given 
by \cite{Eliasson},
\begin{eqnarray}
\left(\omega^2-c^2\bm{k}^2\right)\left(\omega^2-c^2\bm{k}^2-\omega^{2}_{p}\right)-4\frac{m^2c^4}{\hbar^2}\left(\omega^2-\omega^{2}_{p}\right)=0.\label{eq:143}
\end{eqnarray}
Eq.~\eqref{eq:143} shows that the electrostatic mode consists of four branches. In vacuum, two gapless branches relate to 
the fermions moving with self gauge fields, which are known as Langmuir modes. The other two gapped branches are pair modes, 
and the half gap $2mc^2/\hbar$ means that if the photon energy $\hbar\omega>2mc^2$, the fermion pairs will be generated and 
the quanta of these pair plasmas have finite group velocities. The pair mode is also known as \emph{Zitterbewegung} effect 
in relativistic quantum mechanics, which is described as the interference between positive and negative states of a fermion 
on the Compton space-time scale.

\begin{figure}[htbp]
\centerline{\includegraphics[width=9cm,height=6.9cm]{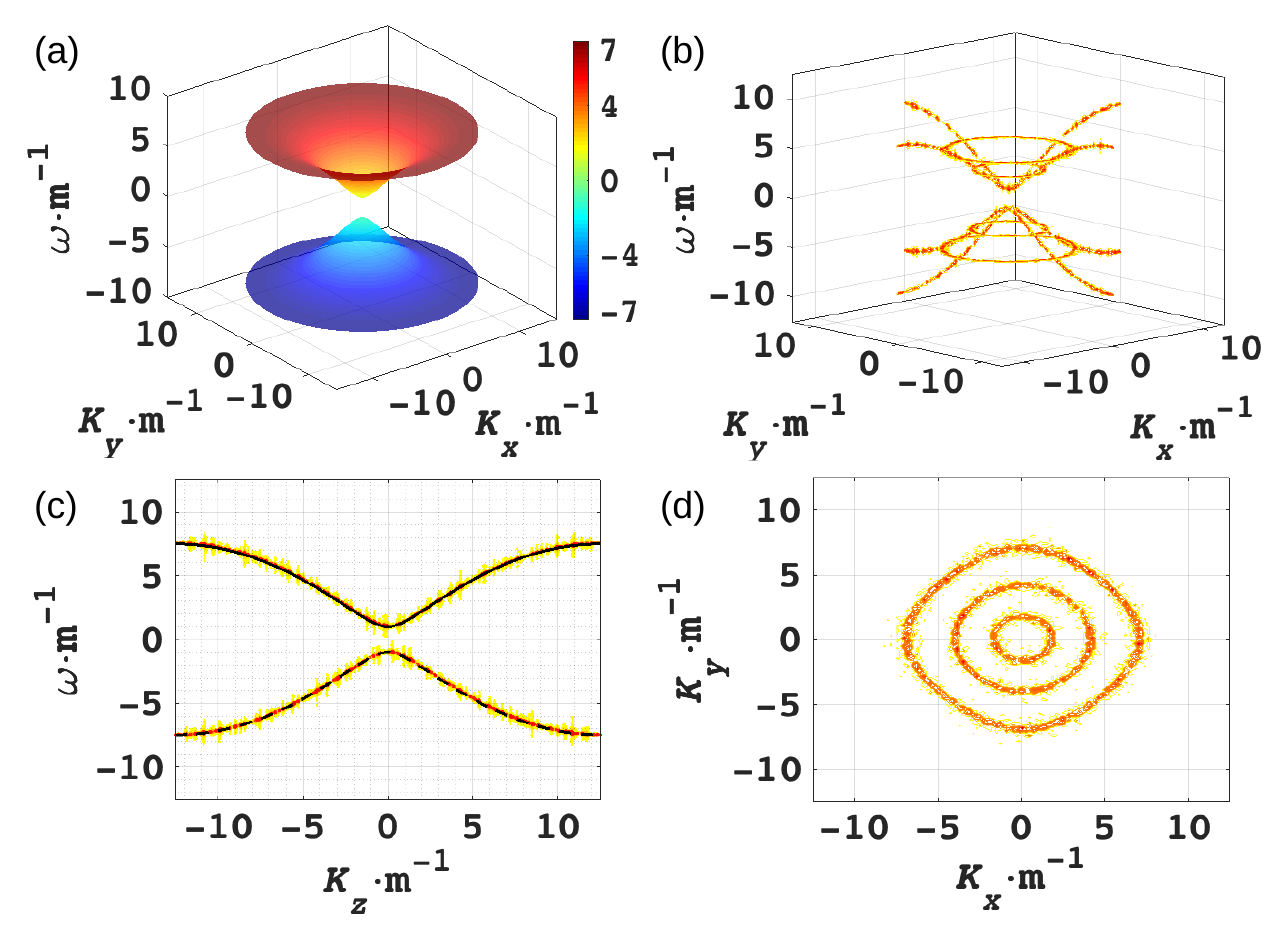}}
\caption{Numerical dispersion relation of a free fermion. (a) The Dirac double-cone in the BZ of a DEC lattice. 
Different from the Dirac double-cone in continuous space-time, the exact dispersion relation of the LCFT shown here is 
modified near the BZ boundary. It can be found that each branch of the spectra has only one cone centered at the origin of 
BZ. As a result, there are only two degenerate fermion flavors exist. (b) The numerical dispersion relation (contour slice) 
on $k_{z}=0$ plane. (c) The numerical dispersion relation (contour plot) along $k_{z}$ axis. These free fermion dispersion 
relations are obtained by a real-time LCFT simulation, which show good consistency with the analytical Dirac double-cone of 
the LCFT (solid line for positive state and dashed line for negative state). The energy gap $2m$ between positive and negative 
states is perfectly recovered. (d) The numerical isoenergic surfaces (contour plot) on $k_{z}=0$ plane. $\omega{\cdot}m^{-1}=2,4,6$ 
from the inside out. ($m=0.25, \Delta{t}=0.5, \Delta{x_{i}}=1, N_{t}=512, N_{x_{i}}=256$)}
\label{fig:2}
\end{figure}

\begin{figure}[htbp]
\centerline{\includegraphics[width=8.2cm,height=6.3cm]{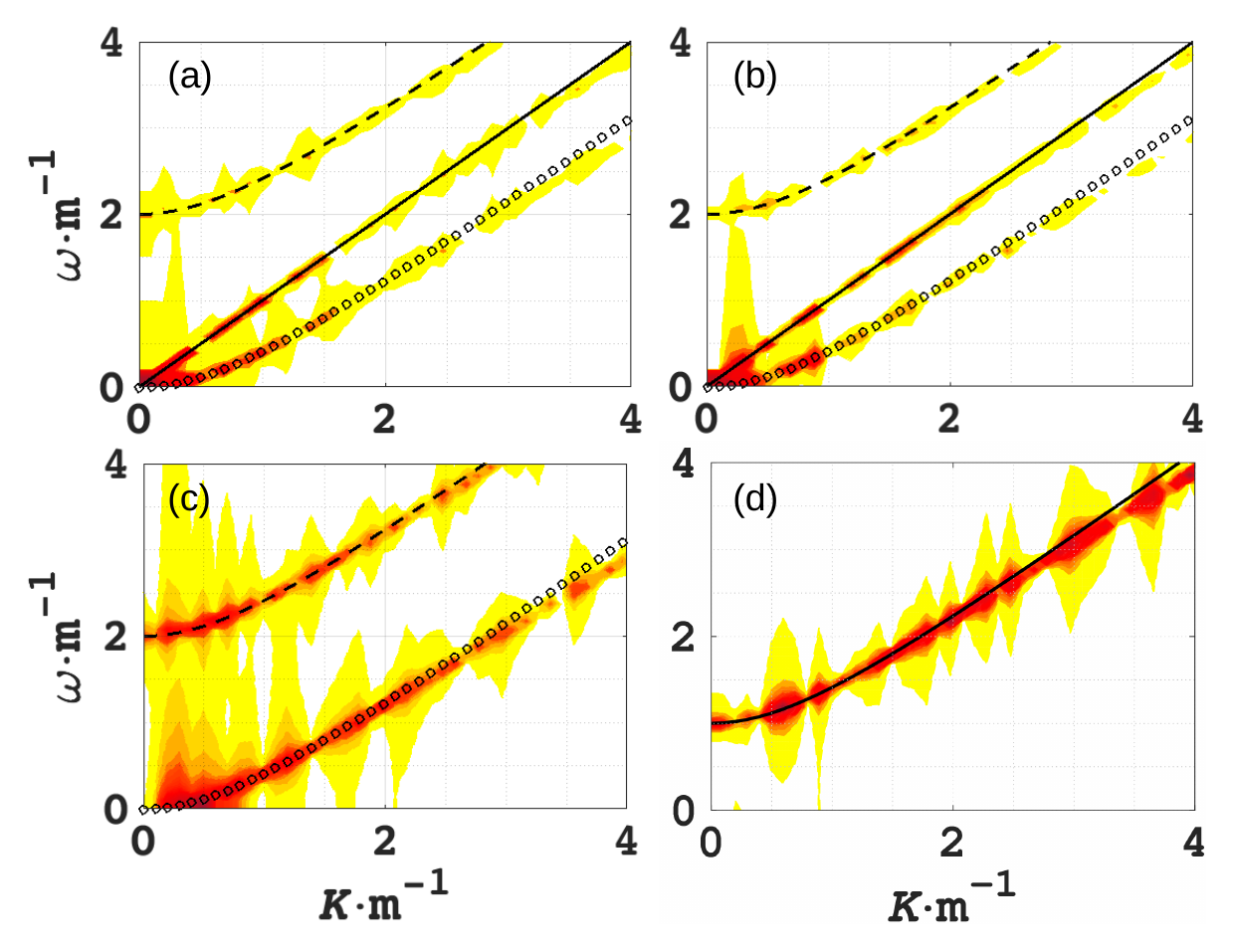}}
\caption{Numerical energy spectra (colour-filled contour plot) along $k_{z}$ axis obtained by a real-time LCFT simulation. (a)-(b) 
Energy spectra of $E_{x}$ and $E_{y}$, where the light cone (solid line) is well traced by the analytical electromagnetic mode 
dispersion. Distinguished from the 0-spin scaler QED model, the existence of pair and Langmuir modes (dashed and circle lines) 
implies the field structures of these two modes are polarization hybrid. (c) Energy spectrum of $E_{z}$, where the gapped pair 
mode (dashed line) and gapless Langmuir mode (circle line) are well traced by the analytical electrostatic mode dispersion. 
The energy gap $2m$ is a threshold beyond which pair plasmas will be generated. (d) Energy spectrum of $\psi_{1R}$. The 
weak self gauge field dressed bispinor spectrum is well traced by a free fermion dispersion (solid line). ($m=0.25, 
\Delta{t}=0.5, \Delta{x_{i}}=1, N_{t}=512, N_{x_{i}}=256$)}
\label{fig:3}
\end{figure}

To implement real-time LCFT simulations, the natural units are used, where the constants $\hbar=c=1$, the elementary charge 
$e=0.0854$, so that the fine structure constant $\alpha=e^{2}/\hbar{c}\approx1/137$ is physically correct. Then all physical 
variables can be normalized by a unified dimension, such as [M]. Here we employ the fermion mass $m=0.25$, the energy 
dimension is [M], and the dimension of length and time is [M]$^{-1}$. To calculate the energy spectra, a uniform 
$256\times256\times256 (N_{x}{\times}N_{y}{\times}N_{z})$ DEC lattice is introduced, where the spatial lattice periods 
$\Delta{x}=\Delta{y}=\Delta{z}=1$, and the temporal lattice period $\Delta{t}=0.5$. In all directions, the periodic boundary 
is used to naturally introduce an infrared truncation. To initialize the simulations so that a broad spectrum of linear waves 
are excited, the bispinor field is given using small amplitude unbiased white noise with standard deviation $\sigma=1\times10^{-6}$, 
and the temporal gauge is adopted explicity. After a $N_{t}=512$ steps simulation, the numerical spectra of the LCFT can be read out 
from simulation results by taking multi-dimensional fast Fourier transforms (FFT) of bispinor and electric field components. 
By abandoning the gauge field in simulation, we can obtain the dispersion relation of a free fermion via the same procedure. 

Fig.~\ref{fig:2} illustrates the numerical dispersion relation of a free fermion in the LCFT. The Dirac double-cone shown 
in Fig.~\ref{fig:2} (a) is an exact result derived from the linearized LCFT, which can be seen as a massive fermion extension 
of Eq.~\eqref{eq:124}. Different from the bispinor field in continuous spac-time manifold, the shape of numerical Dirac 
double-cone is modified near the boundary of lattice BZ. The dispersion relations (contour slice) shown in Fig.~\ref{fig:2} (b)-(c) 
are obtained by a simulation. Fig.~\ref{fig:2} (b) illustrates the dispersion on $k_{z}=0$ plane and Fig.~\ref{fig:2} (c) shows 
the dispersion along $k_{z}$ axis. The benchmark lines given in Fig.~\ref{fig:2} (c) are projections of the exact Dirac double-cone 
of the LCFT. It can be found that the dispersion relation and the energy gap $2m$ of a free fermion is perfectly recovered. The 
numerical results also show that there is only one Dirac double-cone centered at the origin of BZ. The contour plot shown in 
Fig.~\ref{fig:2} (d) illustrates the isoenergic surfaces on $k_{z}=0$ plane, where the energy circles are single valued and labeled 
as $\omega{\cdot}m^{-1}=2,4,6$ from the inside out. It proofs that the algorithms involve two degenerate fermion flavors.

The dispersion relations along $k_{z}$ axis obtained by a simulation shown in Fig.~\ref{fig:3} provide us with a complete numerical 
energy spectra of the Dirac-Maxwell theory based LCFT. The momentum in these plots is cut off at $k=4m$ ($k\Delta{z}\approx1$) where 
the continuous approximation is no longer sufficient. The numerical energy spectrum (colour-filled contour plot) of $E_{z}$ shown in 
Fig.~\ref{fig:3} (c) is well traced by the analytical electrostatic mode dispersion, where the dashed line indicates gapped pair mode 
(energy gap $2m$) and the circle line indicates gapless Langmuir mode. The numerical energy spectra (colour-filled contour plot) of 
$E_{x}$ and $E_{y}$ plotted in Fig.~\ref{fig:3} (a)-(b) show three different branches, where the well traced light cone (solid line) 
is the typical dispersion of a tree-level electromagnetic mode, and the distinct pair (dashed line) and Langmuir (circle line) modes 
imply that the field structures of these two modes are polarization hybrid. The hybrid polarization originates from 1/2-spin induced 
polarization currents hybridization, which is distinguished from the 0-spin KGM theory. Fig.~\ref{fig:3} (d) shows a weak self gauge 
field dressed fermion mode, which can be well traced by the free fermion dispersion. That the analytical dispersion relations are 
recovered by numerical spectra indicates that our solutions faithfully capture the propagation of linear waves up to the lattice resolution.

\begin{figure}[htbp]
\centerline{\includegraphics[width=8.15cm,height=6.45cm]{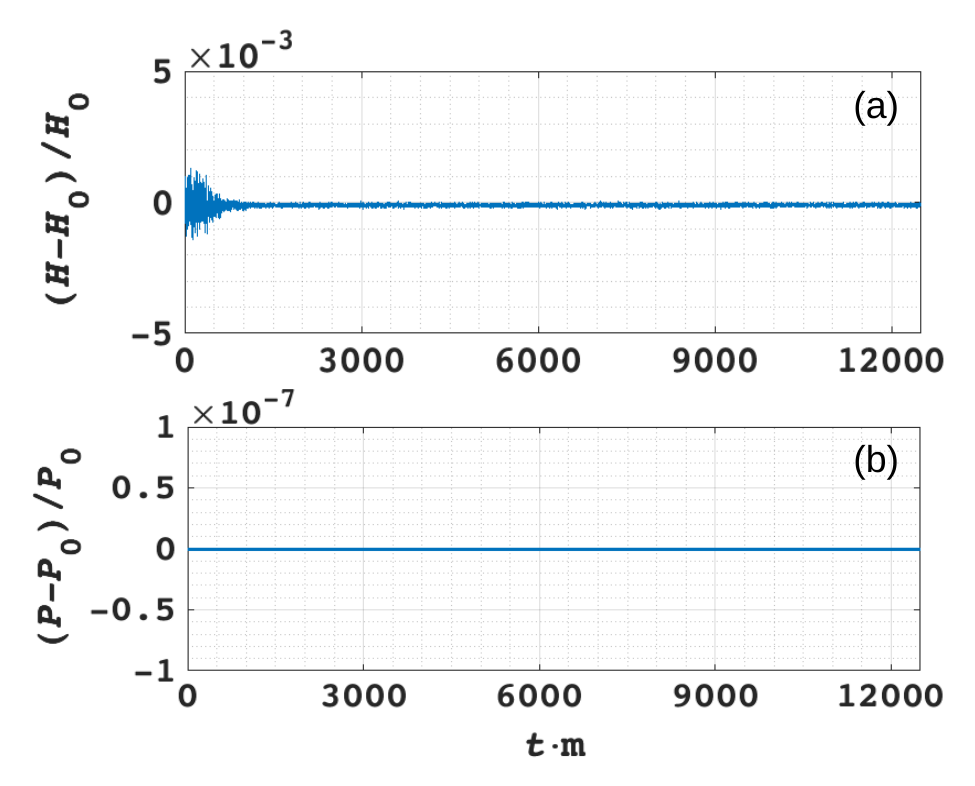}}
\caption{Numerical errors of the conserved quantities. (a) Relative error of the total Hamiltonian. (b) Relative error 
of the total probability. Where the subscripts $0$ of the total Hamiltonian $H$ and probability $P$ indicate the initial values. 
($m=0.25, \Delta{t}=0.05, \Delta{x_{i}}=1, N_{t}=1\times10^{6}, N_{x_{i}}=10$)}
\label{fig:4}
\end{figure}

To illustrate the advantages of our algorithms, we implement a long-term simulation and record the numerical errors of the 
conserved quantities. The simulation domain is a uniform $10\times10\times10$ DEC lattice, and the the periodic 
boundary is used in all directions. $\Delta{x}=\Delta{y}=\Delta{z}=1$, and $\Delta{t}=0.05$. At initial time, a unbiased 
white noise with standard deviation $\sigma=3\times10^{-2}$ is introduced into the bispinor field. After a million steps 
simulation, the relative numerical errors of the total Hamiltonian and total probability are plotted in Fig.~\ref{fig:4}. 
We find that after a extremely long-term simulation, the numerical errors of conserved quantities are bounded by small values 
without coherent accumulation. The excellent conservation property in these numerical solutions comes from the preservation 
of geometric structures and symmetries by using our algorithms. The conservation is a footstone to implement secular 
simulations for nonlinear multi-scale SFQED and RQP phenomena.

\subsection{Schwinger mechanism induced $e$-$e^+$ pairs creation}
\label{sec:5-2}
\begin{figure}[htbp]
\centerline{\includegraphics[width=8cm,height=2.8cm]{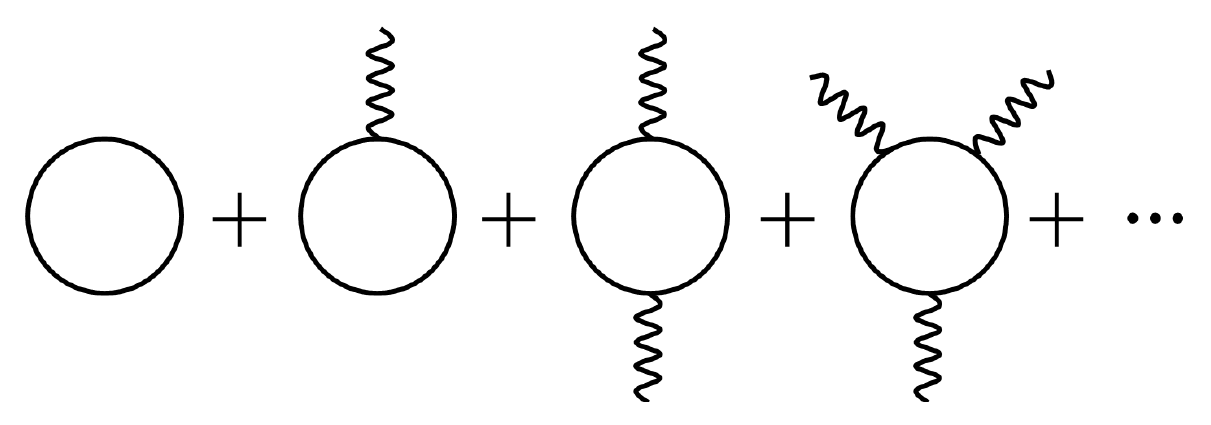}}
\caption{The Feynman diagram of Schwinger effect. In order to calculate the Schwinger pair production, one should sum 
the infinite set of diagrams, each of which contains one electron loop and any number of external photon legs.}
\label{fig:5}
\end{figure}

The Schwinger mechanism induced creation and annihilation of electron and positron pairs are genuine phenomena in SFQED, 
which can not be described via classical theories \cite{BSXie}. In Sec.\ref{sec:5-1}, the numerical spectra show that the 
pair mode can be found once the energy of the $\gamma$ photon exceeds double electron rest energy. Schwinger effect states 
that when the photon wavelength is not very short, the $e$-$e^+$ pair can also be generated once the gauge field strength 
is extremely strong \cite{Schwinger1,Weinberg}. The typical electrostatic field strength of the Schwinger limit is 
$E_{S}=1.32\times10^{16}$ V/cm, and the equivalent magnetic field strength and laser intensity are of orders $10^9$ T and 
$10^{29}$ W$\cdot$cm$^{-2}$ respectively \cite{Schwinger1}. Beyond the Schwinger threshold, the virtual $e$-$e^+$ pairs 
can be pulled apart from quantum fluctuations on the Compton space-time scale and large on shell $e$-$e^+$ pairs can be 
created from the vacuum. Although there are some other QED mechanisms can create on shell $e$-$e^+$ pairs in the vacuum, 
e.g. the Breit-Wheeler process, the Schwinger effect becomes the dominate process once the $U(1)$ gauge field becomes a 
low frequency and extremely strong field. The Feynman diagram of Schwinger effect is shown in Fig.\ref{fig:5}.

To simulate the Schwinger mechanism induced $e$-$e^+$ pair creation, we set a longitudinal quasi-static electric field 
whose strength is normalized by the Schwinger limit $E_{S}=m^2/e$. Numerical experiments are implemented on a 
$1\times1\times256$ uniform DEC lattice, and the periodic boundary is used in all directions. $e/m=0.2$, $\Delta{x}=\Delta{y}=
\Delta{z}=0.05/m$, and $\Delta{t}=0.5\Delta{x_{i}}$. At initial time, an ensemble model based Dirac vacuum state is 
introduced by sampling a class of stochastic Dirac spinors $\psi_{M}(\bm{x},0)$ and $\psi_{F}(\bm{x},0)$ as 
$n^{+}_{\bm{p}}=n^{-}_{\bm{p}}=0$, where the ensemble capacity $N_{e}=512$. The gauge field $A^{\mu}$ is sampled in 
the temporal gauge, where $\bm{A}(\bm{x},0)=\bm{0}$ and $\bm{Y}(\bm{x},0)=(0,0,-E_{S}/4\pi)$ are given as an initial 
condition. By setting $\bm{Y}(\bm{x},0)=\bm{0}$, we can also simulate the quantum fluctuations of Dirac vacuum. After 
a 20000 steps simulation, the numerical results are recorded, which include the electric field evolution, pair production 
rate, Hamiltonian transfer, and spectral density.

\begin{figure}[htbp]
\centerline{\includegraphics[width=7.5cm,height=5.6cm]{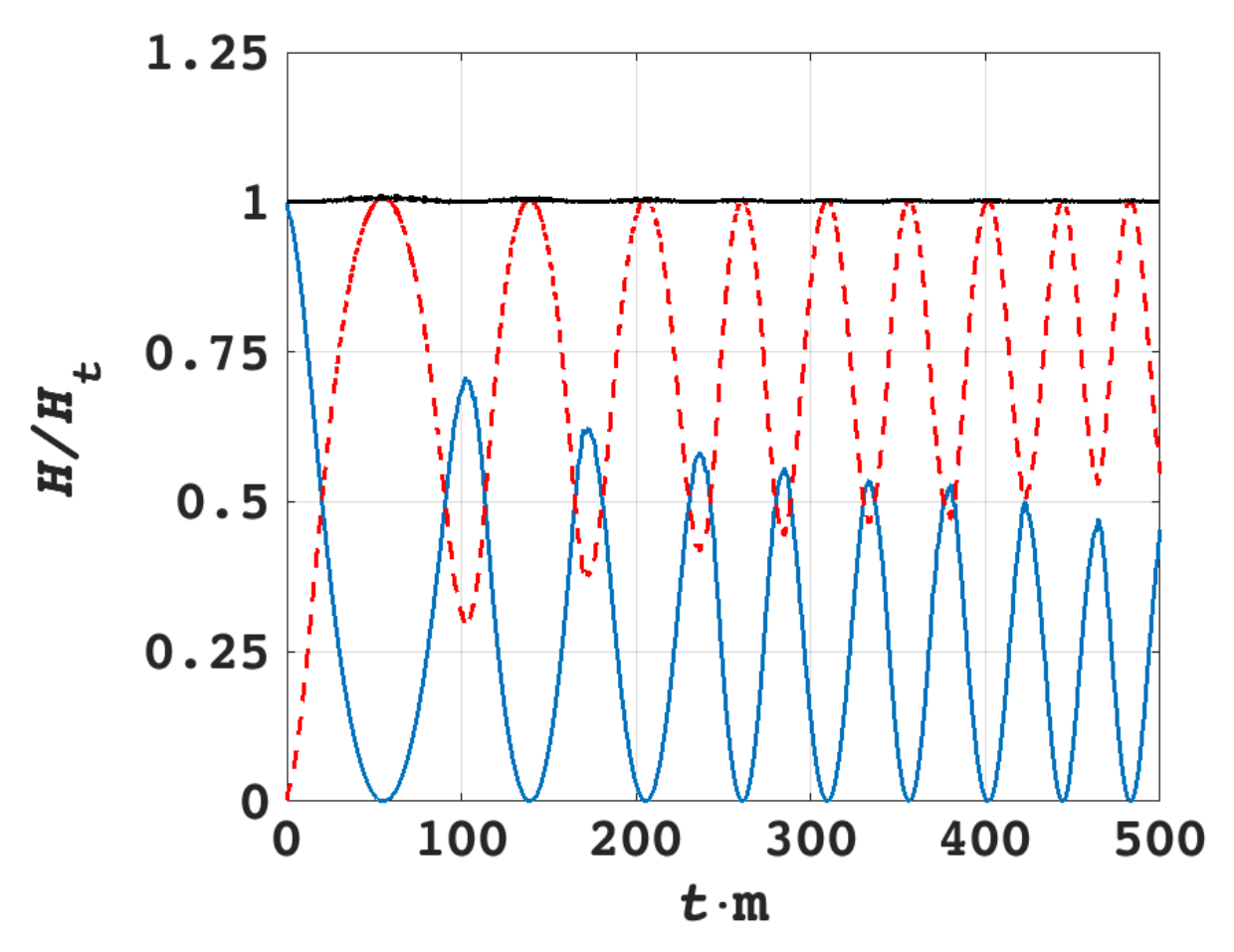}}
\caption{Numerical evolution of different parts of the QED Hamiltonian (normalized by total Hamiltonian). The total 
Hamiltonian $H_{t}$ (black solid line) of the interacting fermion-boson system is perfectly conserved in this long-term 
simulation, which is a result of the preservation of the geometric structures. The energy transfer between $e$-$e^+$ pair 
plasmas and $U(1)$ gauge field can be read out from the fermionic sector $H_{F}$ (red dashed line) and gauge sector $H_{G}$ 
(blue solid line) of the total Hamiltonian. The nonlinear envelopes of these curves demonstrate that the pair production 
experienced a process from rapid growth to saturation, where the nonlinear suppression comes from the nonpertubative 
field backreaction and Pauli exclusion effect. With the increase of fermion density, the plasma frequency 
will exhibit blue shift, which can be found as the chirp features in these Hamiltonian curves. ($e/m=0.2, \Delta{x_{i}}=0.05/m, 
\Delta{t}=0.5\Delta{x_{i}}, N_{t}=2\times10^4, N_{x}=N_{y}=1, N_{z}=256, N_{e}=512$)}
\label{fig:6}
\end{figure}

Fig.~\ref{fig:6} illustrates the numerical evolution of normalized Hamiltonians, where the vacuum energy has been renormalized 
by the normal product. The blue solid line show us the decaying oscillation of the $U(1)$ gauge field. During this process, 
on shell $e$-$e^+$ pairs are continuously created and driven, and the energy of photons is continuously transfered into the 
fermions energy. With the growth of fermion density, the pair plasma frequency increases and the chirp feature of the 
plasma oscillation can be distinctly recognized from the Hamiltonian of gauge field. This is a nonlinear phenomenon, 
as the production of $e$-$e^+$ pair can be effectively suppressed by the radiation reaction, which means that the energy 
of gauge field will be absorbed by self generated pair plasmas and the pair production will reach a saturation level. 
This nonlinearity can also be read out from the fermion Hamiltonian (red dashed line). The lower envelope of this curve 
told us that the production rate of the $e$-$e^+$ pair in the simulation domain has a very high level in the early time 
of the vacuum breakdown ($t{\cdot}m\sim50$), and then the pair production rate is saturated with a relatively stable plasma 
frequency after a long time evolution ($t{\cdot}m\sim500$). During this nonlinear Schwinger process, the total Hamiltonian 
in the simulation domain is perfectly conserved (black solid line). In summary, the photon energy is continuously absorbed 
by new on shell $e$-$e^+$ pairs, and the energy between ptoton and pair plasmon is exchanged cycle by cycle with a chirped 
plasma frequency.

\begin{figure}[htbp]
\centerline{\includegraphics[width=8.8cm,height=5.5cm]{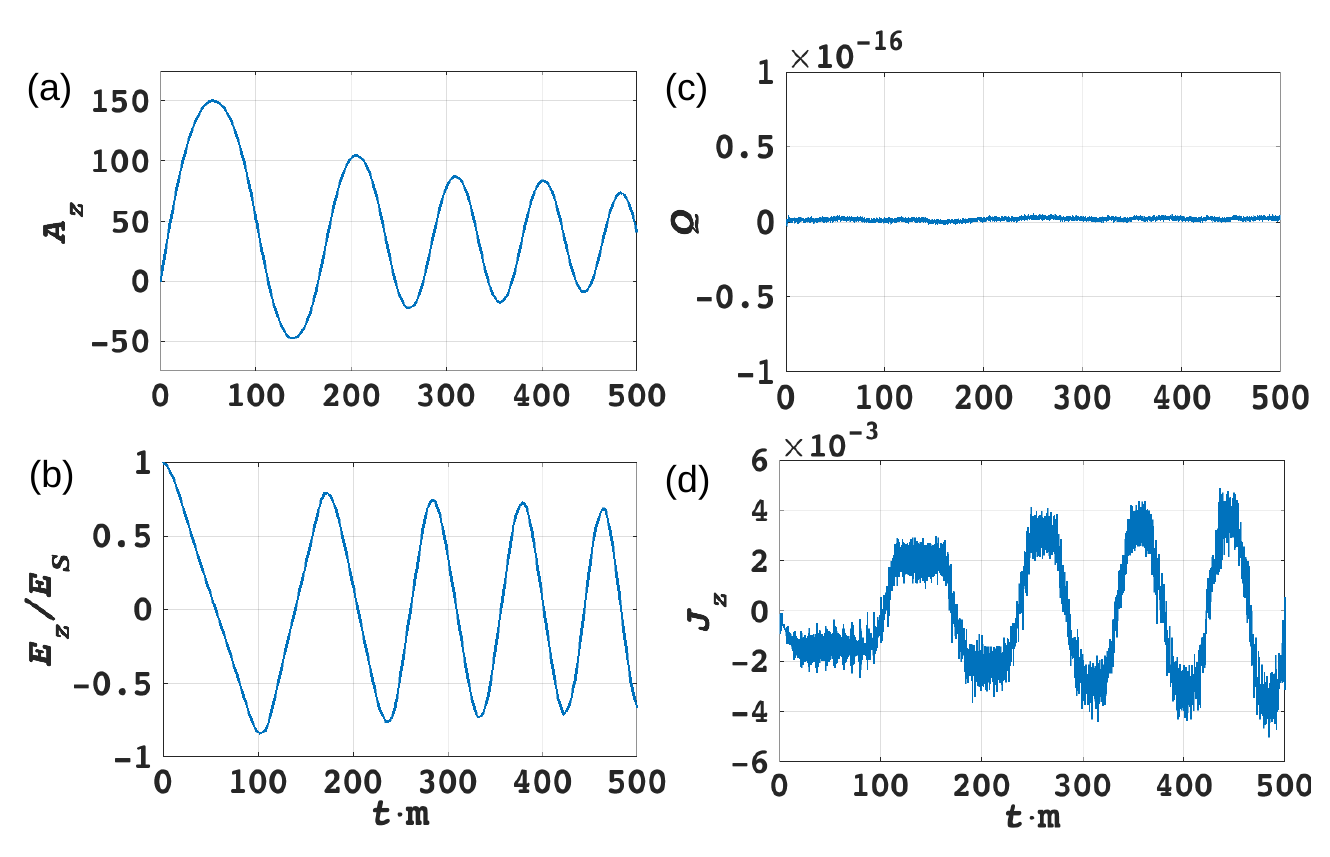}}
\caption{Dynamical properties of $U(1)$ gauge field and fermions. (a)-(b) Numerical evolution of gauge connection $A_{z}$ 
and associated electric field $E_{z}$ (normalized by $E_{S}$). The chirped electric field exhibits dissipative anharmonic 
feature, which can be used as a probe to diagnose the state of $e$-$e^+$ pair oscillators. In this simulation ($t\sim500/m$), 
the electric field amplitude approaches 0.68$E_{S}$, which means that more than half of the gauge field energy has been 
consumed to create $e$-$e^+$ pairs. All QED phenomena admit the charge conservation law, which means that the net charge 
$Q$ in Schwinger process must keep 0. It can be found in subfigure (c), where a numerical noise induced extremely small 
net charge is well conserved in the simulation. (d) shows the pair plasma oscillation induced ensemble current density. 
With the nonlinear increase of fermion density, The current amplitude experienced a growth process with a decay rate. 
Obviously, the chirped oscillations of current and elctric field are well matched with a $\pi/2$ phase difference. 
($e/m=0.2, \Delta{x_{i}}=0.05/m, \Delta{t}=0.5\Delta{x_{i}}, N_{t}=2\times10^4, N_{x}=N_{y}=1, N_{z}=256, N_{e}=512$)}
\label{fig:7}
\end{figure}

Moreover, we plot the guage connection $A_{z}$, electric field $E_{z}$, net charge $Q$ and Dirac current density 
$\mathcal{J}_{z}$ in Fig.~\ref{fig:7} to demonstrate the dynamical properties of $U(1)$ gauge field and fermions. The 
numerical evolution of gauge field shown in Fig.~\ref{fig:7} (a)-(b) illustrates the dissipative anharmonic effects 
induced by the separation and recombination of nonlinear $e$-$e^+$ pair oscillators. From the Hamiltonian curves plotted 
in Fig.~\ref{fig:6}, we already know that after hundreds of Compton periods $1/m$, the pair production will be effectively 
suppressed by the nonpertubative field backreaction. In Fig.~\ref{fig:7} (b), we find a slowly varying electric field 
amplitude $E_{z}\approx0.68E_{S}$ at the end of this simulation. During this time, more than half of the gauge field 
energy has been consumed to create $e$-$e^+$ pairs, and the consumed energy is transferred and stored in the pair plasmas. 
Gauge symmetry induced charge conservation law can be found in Fig.~\ref{fig:7} (c), where an extremely small net charge 
$Q$ in the simulation domain is well conserved in this simulation. The fermion and antifermion are created and annihilated 
in pairs during the Schwinger process. As a result, the net charge in whole space must keep 0. The well bounded weak 
numerical noise of $Q$ exhibits advanced long-term performance of our structure-preserving algorithms in simulating 
complicated relativistic quantum effects. The ensemble current density of pair plasmon shown in Fig.~\ref{fig:7} (d) 
provide us with an intuitive picture of the pair plasma motion. Just as a classical plasma oscillator, the oscillations 
of current and elctric field are well matched with a $\pi/2$ phase difference. Different from the classical oscillator, 
the nonlinear increase of pair plasma density gives rise to a nonlinear current amplitude growth and an oscillation frequency 
blue shift.

\begin{figure}[htbp]
\centerline{\includegraphics[width=7.5cm,height=4.7cm]{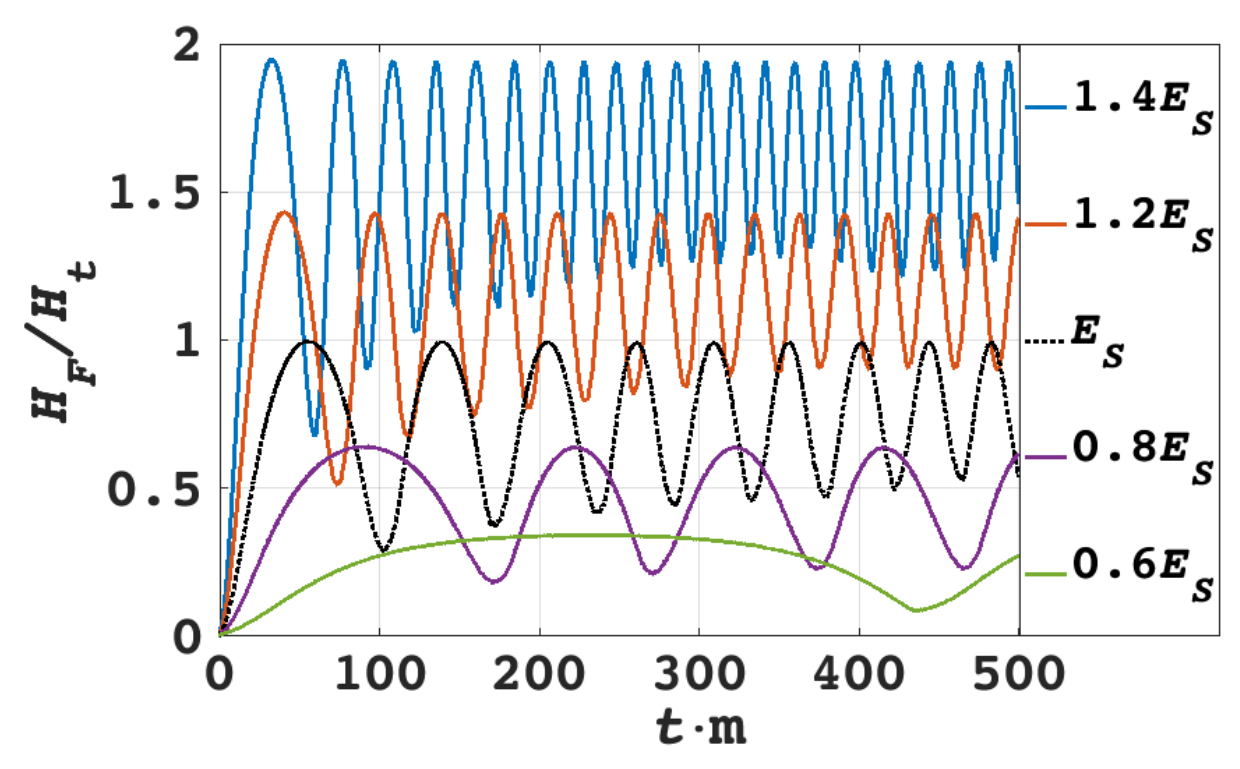}}
\caption{Numerical evolution of fermion Hamiltonians (normalized by total Hamiltonian with background electric field $E_{S}$) 
with different background gauge field strengths. The lower envelopes of these Hamiltonian curves can be used as an indicator 
of the pair production rate. It shows that when the background electric field approaches 0.5$E_{S}$, the Schwinger effect 
becomes negligible. With a background electric field stronger than $E_{S}$, the pair plasmas can be quickly generated and 
then saturated by the nonlinear radiation backreaction in a relatively short period. ($e/m=0.2, \Delta{x_{i}}=0.05/m, 
\Delta{t}=0.5\Delta{x_{i}}, N_{t}=2\times10^4, N_{x}=N_{y}=1, N_{z}=256, N_{e}=512$)}
\label{fig:8}
\end{figure}

To compare the Schwinger effect with different background gauge field strengths, we implement a class of simulations 
under different initial electric fields (normalized by $E_{S}$), and plot the associated fermion Hamiltonian curves in 
Fig.~\ref{fig:8}. The lower envelope of the curve with $E_{z}=0.6$ demonstrate that the Schwinger effect can be effectively 
cut off when the background electric field is lower than half $E_{S}$. In this situation, the extremely tenuous fermion 
density gives rise to a very low plasma frequency, and the energy transfer from photons to fermions is very slow. On the 
contrary, the lower envelope of the curve with $E_{z}=1.4$ shows us that the nonlinear suppression effect in Schwinger 
process can be significantly enhanced when the background electric field is far stronger than $E_{S}$.

Finally, we can make a brief summary that the Schwinger mechanism induced fermion pairs production is inherently a 
nonlinear and non-perturbative phenomenon, which exhibits abundant anharmonic, non-equilibrium and self-modulation features. 
To illustrate the complete physics of this process, real-time LGT simulation or other non-perturbative methods are needed. 
Due to the symmetric and geometric structure-preserving nature, Our algorithm provide an efficient, accurate, stable and 
conservative approach to implement real-time LCFT simulations to study this kind of complicated SFQED phenomena.

\subsection{Vacuum Kerr effect}
\label{sec:5-3}
\begin{figure}[htbp]
\centerline{\includegraphics[width=6.3cm,height=6cm]{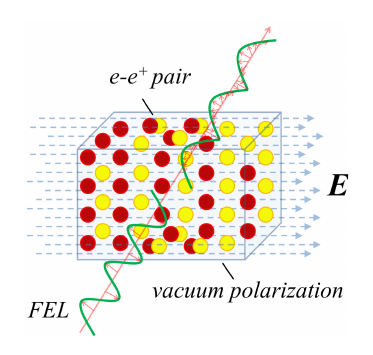}}
\caption{Schematic of the vacuum Kerr effect. The Dirac vacuum is strongly polarized under an extreme electric field 
whose strength approaches the Schwinger threshold. Just as the birefraction in a polarized dielectric medium, a linear 
polarized FEL beam can transfer into an elliptical polarized beam after it propagating through the polarized vacuum.}
\label{fig:9}
\end{figure}

The Schwinger mechanism induced pair plasmas are strongly polarized, which means that the Dirac vacuum is strongly 
polarized under an extreme electric field. The Kerr effect states that when a dielectric medium is polarized by an 
external electric field, it will exhibit birefraction property, for the refractive index parallel to external electric 
field is modulated. Be treated as an QED analogue of the classical polarized dielectric medium, the polarized Dirac 
vacuum may also exhibit birefraction property, and a Kerr-like effect can be expected to be observed in the SFQED regime. 
A schematic of the vacuum Kerr effect is shown in Fig.~\ref{fig:9}.

To simulate the vacuum Kerr effect, we set a transverse quasi-static electric field whose strength approaches $E_{S}$, 
and then introduce a weak linear polarized free electron laser (FEL) beam as incident wave. Numerical experiment is 
implemented on a $40\times40\times20$ uniform DEC lattice, and the periodic boundary for both fermion and gauge field is 
used in $x$ and $y$ directions. To cut off the longitudinal radiations, we introduce the second order Mur's boundary in 
$z$ direction to simulate a open space. $e/m=0.2$, $\Delta{x}=\Delta{y}=\Delta{z}=0.1/m$, and $\Delta{t}=0.5\Delta{x_{i}}$. 
At initial time, an ensemble model based pair plasma state is introduced by sampling the stochastic bispinors 
$\psi_{M}(\bm{x},0)$ and $\psi_{F}(\bm{x},0)$ as $n^{+}_{\bm{p}}=n^{-}_{\bm{p}}=0.1$, where the ensemble capacity 
$N_{e}=3200$. The gauge field $A^{\mu}$ is sampled in the temporal gauge, where $\bm{A}(\bm{x},0)=\bm{0}$ and 
$\bm{Y}(\bm{x},0)=(0,0,-E_{S}/4\pi)$ are given as an initial condition. To excite an incident FEL plane wave, we set a 
total-scattered fields boundary in the $z=5\Delta{z}$ (source) plane, and the laser frequency is given by $\omega=0.2{\pi}m$. 
After a 400 steps simulation, the numerical results are recorded, where the trace of magnetic field vector in the $z=15\Delta{z}$ 
(target) plane demonstrates the polarization state of the beam.

\begin{figure}[htbp]
\centerline{\includegraphics[width=8.8cm,height=9.75cm]{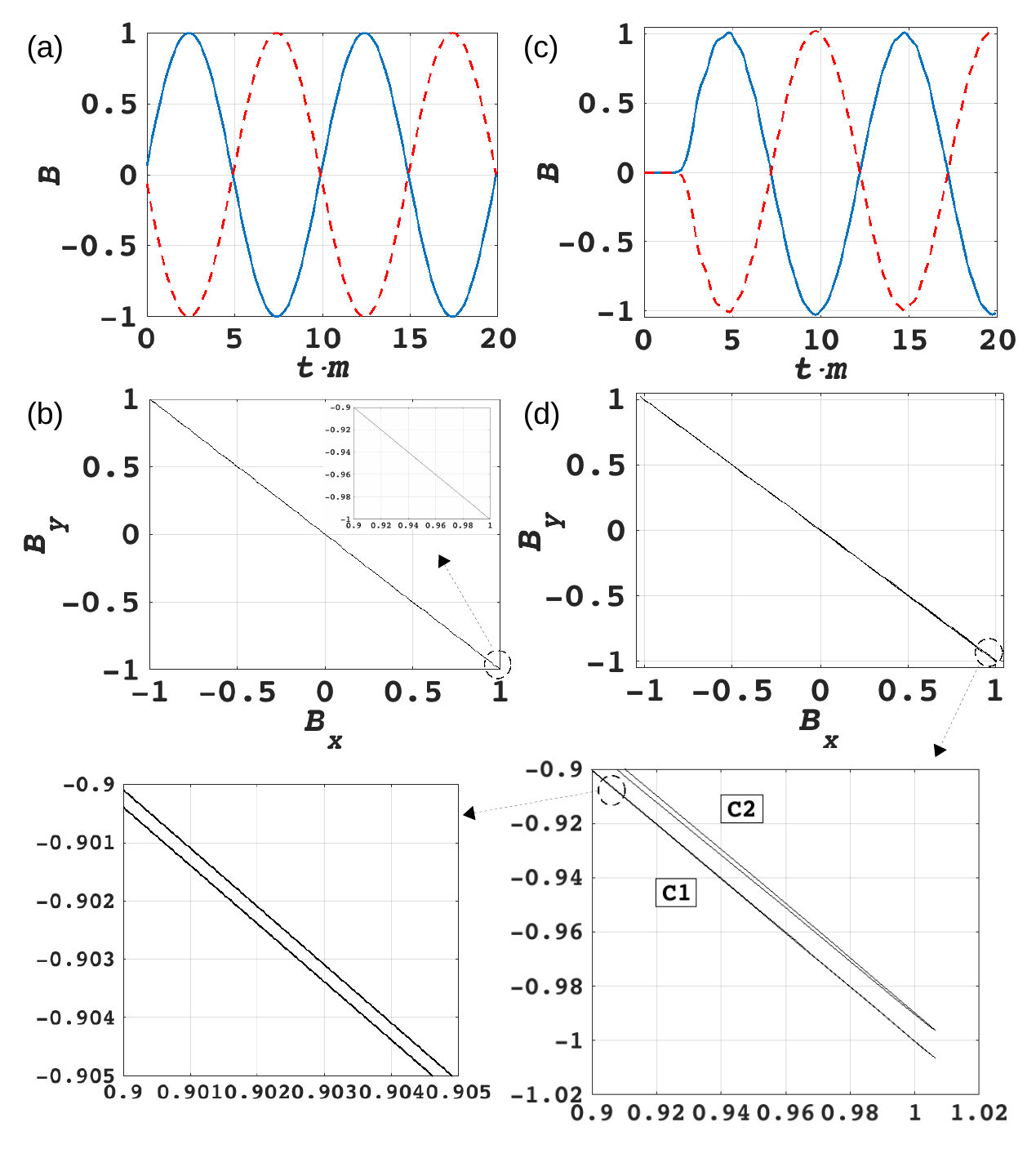}}
\caption{Vacuum Kerr effect induced FEL polarization conversion from linear to elliptical-like states. (a) The magnetic 
field components in the source plane, where the bule solid line denotes $B_{x}$ and the red dashed line denotes $B_{y}$. 
(b) The trace of magnetic field vector in the source plane is a line segment, which means that the incident FEL plane wave 
is linear polarized (fine structure of the trace can be found in the inset). (c) The magnetic field components in the target 
plane, where the bule solid line denotes $B_{x}$ and the red dashed line denotes $B_{y}$. (d) The trace of magnetic field 
vector in the target plane exhibits elliptical shape with varying parameters (insets illustrate fine structures of the trace, 
where C1 and C2 denote the first and the second cycles of the beam respectively), which means that the Dirac vacuum exhibits 
birefraction property and the birefraction index is time dependent. The magnetic field vectors plotted here are normalized 
by the incident wave. ($e/m=0.2, \Delta{x_{i}}=0.1/m, \Delta{t}=0.5\Delta{x_{i}}, \omega=0.2{\pi}m, N_{t}=400, N_{x}=N_{y}=40, 
N_{z}=20, N_{e}=3600$)}
\label{fig:10}
\end{figure}

Fig.~\ref{fig:10} illustrates the conversion of FEL polarization states in the source and target planes. The incident FEL 
beam is a $3\pi/4$ linear polarized plane wave, which can be found in Fig.~\ref{fig:10} (b), where the trace of magnetic field 
vector in the source plane draws a perfect line segment. When propagating throw the strongly polarized Dirac vacuum, the 
$x$ and $y$ components of the FEL beam have different phase velocities. Then there is a phase difference between $x$ and 
$y$ modes in the target plane, and the beam polarization will transferred into a elliptical-like state, which can be found 
in Fig.~\ref{fig:10} (d). From the numerical results of Schwinger mechanism induced $e$-$e^+$ pairs creation, we know that 
the pair plasma density and the background electric field strength are time dependent. As a result, the Vacuum Kerr effect 
admits a varying coefficient and the phase difference is time dependent. The insets of Fig.~\ref{fig:10} (d) demonstrate 
this feature, where the second cycle (C2) of the beam admits a larger semi-major axis than the first cycle (C1), and the 
trace gyrocenter is drifting perpendicular to the major axis. If the pair plasma is dense enough, more significant 
polarization conversion features can be expected to be observed.

In summary, the numerical experiments illustrated in Sec.\ref{sec:5} cover relativistic QED wave structures, fermion pairs 
creation and annihilation effects, self-consistent interactions between fermion plasmas and $U(1)$ gauge field, nonlinear 
and non-perturbative nature of strong-field physics. The good properties of these numerical solutions ensure the 
structure-preserving real-time LCFT simulation method is expected to be a unified first-principle based theoretical tool 
in studying SFQED and RQP phenomena.

\section{Conclusion and Outlook}
\label{sec:6}

In this paper, we developed a class of high-order canonical symplectic structure-preserving geometric algorithms for 
simulating the quantized Dirac-Maxwell theory based SFQED and RQP. We constructed a canonical field theory of the Dirac-Maxwell 
systems, and obtained the canonical symplectic form and Poisson algebra admitted by this field theory. Based on the 
Noether's theorem, this field theory admits charge, energy-momentum and angular momentum conservation laws via the gauge 
and Poincar\'e symmetries. In DEC framework, we constructed a LCFT which is a good semi-discrete analogue of the continuous 
canonical field theory. The $U(1)$ gauge field is discreted to form a cochain complex which guarantees the Bianchi identities 
of the $U(1)$ gauge theory. With the Hodge dual relations, the bispinor field components are discreted as eight different 
differential forms, which naturally generate a staggered checkerboard-like lattice. Two kinds of discrete gauge covariant 
derivatives, i.e. pull-back and push-forward, are used to construct a gauge invariant semi-discrete action. A well-defined 
discrete Poisson bracket is constructed, which admits bilinearity, anticommutativity, product rule, and Jacobi identity. 
With the previous numerical techniques, the semi-discrete LCFT is gauge invariant, which also preserves the canonical 
symplectic and unitary structures. By using the Hamiltonian splitting method, we obtained three linear subsystems which 
can be solved independently, and constructed a class of high-order structure-preserving geometric algorithms via the Cayley 
transformation and symmetric composition technique. The algorithms preserve the gauge symmetry and geometric structures of 
the semi-discrete LCFT. We proved that the numerical dispersion of the mass free fermions subsystem has only one Dirac 
double-cone centered at the origin of lattice BZ, and there are only two degenerate fermion flavors in our algorithms. 
The locally unconditional stable property can also be obtained from the numerical dispersion. The structure-preserving and 
unconditional stable properties make the scheme superior to conventional Wilson and staggered fermions. To simulate the quantization 
of Dirac field to achieve a correct Fermi-Dirac statistics, we introduce an unified statistically quantization-equivalent 
ensemble model to describe the Dirac vacuum and non-trivial plasma backgrounds. Although the algorithms are unconditional 
stable, it does not means that the lattice periods can be chosen arbitrary large values. On the one hand, some basic physics 
can not the captured if the lattice periods exceed the typical space-time scales, such as the Compton scale. On the other 
hand, large lattice periods will lead to sparse matrices with very large condition numbers, which are very expensive for 
matrix inversion. Additionally, the topology of the $U(1)$ gauge field is changed into a torus by the Wilson lines, which 
means that the lattice periods should not be too large to avoid topological modes.

The numerical energy spectra of the LCFT were calculated and compared with the analytical dispersion relations of linearized 
scalar QED. Simulation results show that the relativistic quantum wave dynamics and the vacuum responses can be captured 
perfectly. The gapless lower branches (Langmuir) of electrostatic mode relate to the fermions moving with self gauge fields. 
The gaped higher branches of electrostatic mode are QED pair modes, where the virtual fermion pairs in quantum fluctuations 
are generated. As the quanta of pair plasmas, the pair plasmons have finite group velocities, which means the virtual pairs 
created and annihilated on the Compton space-time scale are very different from the classical plasmas. The nonlinear Schwinger 
effect was also simulated to illustrate the power of our algorithms. To simulate the quantum fluctuations, we introduced 
an ensemble of statistically quantization-equivalent initial conditions via random momentum and phase, which can be used 
as a statistical model of the quantized Dirac vacuum. With a uniform strong field, the pair production rate is obtained with 
a nonlinear suppression, which means that the energy of the gauge field will be absorbed by self generated pair plasmas and 
the pair creation will reach a saturation level. This nonlinear property of Schwinger mechanism can only be resolved by 
non-perturbative methods, such as the quantum particle-in-cell (PIC) and real-time LQED methods \cite{Nerush,Hebenstreit1,Ridgers1}. 
Our algorithms provide a more accurate and efficient solver for simulating these SFQED and RQP problems because of the 
advanced conservation performance in secular simulations and good unconditional stable property. We also simulated the 
vacuum Kerr effect, where the vacuum response can be resolved. After propagating through a strongly polarized vacuum area, 
a linear polarized FEL beam transferred into an elliptical polarized one, where the vacuum birefraction property was well 
traced. Because the vacuum polarization state is strongly affected by the evolution of background electric field, the 
birefraction index of polarized Dirac vacuum is time dependent. This dynamical property can only be resolved by nonlinear 
non-perturbative methods. All simulations implemented in this work show a common property that the numerical errors of 
conserved quantities, e.g. total Hamiltonian and charge, are bounded by a very small value after a long-term simulation. 
This advantage enables us to simulate nonlinear multi-scale problems dominated by relativistic quantum effects, such as a 
high energy FEL beam interacting with RQP and the the magnetosphere of an X-ray pulsar.

In summary, the gauge invariant canonical symplectic structure-preserving geometric algorithms constructed 
in this work provide us with a powerful first-principle based theoretical tool to implement quantized Dirac-Maxwell theory 
based real-time LCFT simulations. Because of the nonlinear and non-perturbative nature of this approach, it can bring 
abundant physics from the interacting fields. With well-designed field quantization models, this method opens a new door 
toward high-quality simulations in SFQED and RQP fields.

\appendix
\section{Wigner Function Based Pseudo Distributions and Observables}
\label{sec:app}

The observables of electron and positron, e.g. particle numbers and Hamiltonian densities, are hard to read out from the 
bispinor field directly. To get these important observables, the Dirac-Heisenberg-Wigner (DHW) theory can be introduced 
as an auxiliary tool, where the Wigner function is recognized as a linear map from real space to phase space \cite{Hebenstreit0}. 
The Wigner function transforms operators in real space into pseudo distributions in phase space. Taking number density and 
Hamiltonian for example, the map can be given by,
\begin{eqnarray}
\rho\left(\bm{x},\bm{p}\right)\triangleq\int^{\infty}_{-\infty}-\frac{1}{2}\left<\psi^{+}_{M}\left(\bm{x}-\frac{\bm{x'}}{2}\right)\psi_{F}\left(\bm{x}+\frac{\bm{x'}}{2}\right)+g.c.\right>{\rm{e}}^{i\frac{\bm{p}\cdot\bm{x'}}{\hbar}}{\rm{d}}^{3}x',\label{eq:a1}
\end{eqnarray}
\begin{eqnarray}
H\left(\bm{x},\bm{p}\right)\triangleq\int^{\infty}_{-\infty}-\frac{1}{2}\left<\psi^{+}_{M}\left(\bm{x}-\frac{\bm{x'}}{2}\right)\hat{H}\psi_{F}\left(\bm{x}+\frac{\bm{x'}}{2}\right)+g.c.\right>{\rm{e}}^{i\frac{\bm{p}\cdot\bm{x'}}{\hbar}}{\rm{d}}^{3}x'.\label{eq:a2}
\end{eqnarray}
We should emphasize that Eqs.~\eqref{eq:a1}-\eqref{eq:a2} are nonphysical, for they contradict Heisenberg's uncertainty 
principle. But these pseudo distributions can help us to construct useful physical observables. The pseudo distribution 
of electron and positron can be given by,
\begin{eqnarray}
\rho^{\pm}\left(\bm{x},\bm{p}\right)=\frac{H\left(\bm{x},\bm{p}\right)-H_{vac}\left(\bm{x},\bm{p}\right)\pm\hbar\omega_{\bm{p}}\rho\left(\bm{x},\bm{p}\right)}{2\hbar\omega_{\bm{p}}}.\label{eq:a3}
\end{eqnarray}
Where the spectral energy $\hbar\omega_{\bm{p}}=\sqrt{\left(c\bm{p}-e\bm{A}\right)^2+m^2c^4}$, and $H_{vac}$ means pseudo 
distribution of vacuum Hamiltonian. Eq.~\eqref{eq:a3} can be used as an approximate evaluation of the local spectral pair 
production. The physically correct local and total pair productions are obtained as,
\begin{eqnarray}
\rho^{\pm}\left(\bm{x}\right)=\frac{1}{\left(2\pi\hbar\right)^3}\int^{\infty}_{-\infty}\rho^{\pm}\left(\bm{x},\bm{p}\right){\rm{d}}^{3}p,\label{eq:a4}
\end{eqnarray}
\begin{eqnarray}
n^{\pm}=\int^{\infty}_{-\infty}\rho^{\pm}\left(\bm{x}\right){\rm{d}}^{3}x.\label{eq:a5}
\end{eqnarray}
The other observables of electron and positron can be obtained via the same procedure.

\acknowledgments
This work is supported by the National Nature Science Foundations of China (NSFC-11805273, 11905220, 12005141). Numerical 
simulations were implemented on the SongShan supercomputer at National Supercomputing Center in Zhengzhou, the TH-3 Prototype 
and TH-1A supercomputers at National Super Computer Center in Tianjin and the ShenMa high performance computing cluster at 
Institute of Plasma Physics, Chinese Academy of Sciences.


\providecommand{\noopsort}[1]{}\providecommand{\singleletter}[1]{#1}%








\end{document}